\documentclass[11pt]{article}

\usepackage{jheppub}
\usepackage{epsfig}
\usepackage{amssymb}
\usepackage{amsmath}
\usepackage{epstopdf}
\usepackage{extarrows}
\usepackage{tikz}
\usetikzlibrary{arrows.meta, positioning}
\usepackage{dsfont,bbm}
\usepackage{subfig}
\usepackage{multirow}

\usepackage{shuffle}
\usepackage{booktabs}
\usepackage{array}
\usepackage{slashed}

\usepackage{mathrsfs}

\graphicspath{{./fig/}}

\newcommand{\trphitwo}{\text{tr}(\phi^2)}
\newcommand{\tr}{\text{tr}}

\newcommand{\bp}{\bar{\psi}}
\newcommand{\psibarpsi}{\bp\psi}
\newcommand{\psiGabarpsi}{\bp\gamma^{\mu}\psi}
\newcommand{\trpp}{\psibarpsi}
\newcommand{\trpGp}{\psiGabarpsi}

\newcommand{\ff}{\boldsymbol{F}}
\newcommand{\cf}{\mathcal{F}}

\newcommand{\amp}{\boldsymbol{A}}
\newcommand{\ca}{\mathcal{A}}

\newcommand{\gA}{\boldsymbol{M}}

\newcommand{\La}{\mathcal{L}}
\newcommand{\LaYMS}{\La^{\text{YMs}}}
\newcommand{\LaYMSH}{\La^{\text{YMsH}}}

\newcommand{\CK}{\text{CK}}
\newcommand{\klt}{\mathbf{S}}

\newcommand{\clsA}{\mathrm{I}}
\newcommand{\clsB}{\mathrm{II}}

\newcommand{\rank}{\text{Rank }}

\newcommand{\trs}{\tr_{\text{s}}}
\newcommand{\sk}{\tilde{\mathcal{S}}}

\newcommand{\soa}[3]{\mathscr{A}^{[#1]}_{#2,\, #3}}
\newcommand{\soga}[3]{\mathscr{M}^{[#1]}_{#2,\, #3}}

\newcommand{\idd}{\mathds{1}}

\newcommand{\tf}{f}
\newcommand{\td}{d}

\newcommand{\blc}{\mathcal{B}}

\newcommand{\lns}{\tilde{\mathcal{N}}_{\text{L}}}
\newcommand{\ns}{\tilde{\mathcal{N}}}

\newcommand{\co}{\tilde{\mathcal{C}}}

\newcommand{\colorxy}[2]{[\, #1 \,|\, #2 \,]}

\newcommand{\set}[1]{{\{#1\}}}

\newcommand{\ampbas}{\mathscr{B}}
\newcommand{\campbas}{\mathscr{B}}

\newcommand{\ampnh}[2]{\amp_{#2}^{[#1]}}
\newcommand{\campnh}[2]{\ca_{#2}^{[#1]}}
\newcommand{\ganh}[2]{\gA_{#2}^{[#1]}}
\newcommand{\ffnh}[2]{\ff_{#2}^{[#1]}}
\newcommand{\cffnh}[2]{\cf_{#2}^{[#1]}}

\newcommand{\clrBlue}{\color{cyan!70!black}}
\newcommand{\clrOrange}{\color{orange!87!black}}
\newcommand{\clrGreen}{\color{green!70!black}}
\newcommand{\clrPurple}{\color{violet!100!black}}
\newcommand{\clrRed}{\color{red!80!black}}

\newcommand{\bH}{\bar{H}}
\newcommand{\bHBlue}{{\clrBlue \bH}}
\newcommand{\bHBlueIdx}[1]{{\clrBlue \bH_{#1}}}
\newcommand{\bHOrangeIdx}[1]{{\clrOrange \bH_{#1}}}
\newcommand{\bHGreenIdx}[1]{{\clrGreen \bH_{#1}}}
\newcommand{\bHPurpleIdx}[1]{{\clrPurple \bH_{#1}}}

\newcommand{\FV}{\tilde{V}}

\newcommand{\photon}{\mathfrak{p}}

\newcommand{\FC}{\tilde{V}}
\newcommand{\bC}{\bar{V}}
\newcommand{\bCPurple}{{\clrRed \bC}}

\newcommand{\ip}{D}

\newcommand{\Nt}[4]{N_{#1}(#2,\, #3,\, #4)}

\newcommand{\Pbas}{\mathbb{P}}

\newcommand{\cint}{\mathcal{T}}

\newcommand{\spa}[2]{\langle #1 #2 \rangle}
\newcommand{\spb}[2]{[#1 #2]}
 
\preprint{USTC-ICTS/PCFT-26-36}
\title{Dyeing form factors as amplitudes}

\author[a,b,c]{Xinyue Li,}
\emailAdd{lixinyue251@mails.ucas.ac.cn}
\author[a,b,c,d]{Gang Yang,}
\emailAdd{yangg@itp.ac.cn}
\author[b,c]{Guorui Zhu}
\emailAdd{zhuguorui@itp.ac.cn}
\affiliation[a]{School of Fundamental Physics and Mathematical Sciences, \\
Hangzhou Institute for Advanced Study, UCAS, Hangzhou 310024, China}
\affiliation[b]{Institute of Theoretical Physics, Chinese Academy of
Sciences, Beijing 100190, China}
\affiliation[c]{School of Physical Sciences, University of Chinese
Academy of Sciences, Beijing 100049, China}
\affiliation[d]{Peng Huanwu Center for Fundamental Theory, Hefei,
Anhui 230026, China}

\abstract{
The double copy of form factors has revealed a striking feature: poles that are spurious from the gauge-theory perspective become physical propagators in gravity. At the same time, form factors obey hidden factorization relations on the kinematics of these poles. We explain both phenomena by introducing a ``dyeing'' procedure, which promotes the color-singlet operator, or the Higgs particle representing it, to an adjoint massive state. The original form factor is recovered by the inverse ``bleaching'' operation, realized as a $U(1)$ decoupling of the dyed leg.
In the dyed theory, these apparent spurious poles turn into ordinary physical propagators of colored amplitudes, and the hidden factorization relations follow from standard BCJ relations. Applying this framework to multiple operator insertions gives a systematic double-copy construction for multi-Higgs amplitudes and, as a byproduct, reveals scalar-ordering sectors. We also discuss higher-length scalar operators and fermionic operators, including the dyed vector construction for $\trpGp$, as well as a loop-level example.
}

\begin{document}

\maketitle


\section{Introduction}

The double copy has revealed a deep relation between gauge and gravity
theories. It first appeared in the Kawai--Lewellen--Tye (KLT) relations for
string amplitudes, and later became a field-theory framework through
Bern--Carrasco--Johansson (BCJ) color-kinematics duality and the
Cachazo--He--Yuan (CHY) representation
\cite{Kawai:1985xq,Bern:2008qj,Bern:2010ue,Cachazo:2013hca,Cachazo:2014xea}.
In the BCJ form, a gauge-theory amplitude is written as a sum over cubic
graphs whose color factors and kinematic numerators obey the same
algebraic relations. Replacing the color factors by another copy of the
kinematic numerators then gives gravity amplitudes. This idea has led to
many results for scattering amplitudes; see reviews in 
\cite{Bern:2019prr,Bern:2022wqg,Adamo:2022dcm}.

It is natural to ask whether such a structure extends beyond ordinary
scattering amplitudes. A basic class of observables with local
gauge-invariant operator insertions is given by form factors (FFs)
\cite{Maldacena:2010kp, Brandhuber:2010ad,Bork:2010wf}:
\begin{equation}
F_{\mathcal O,n}(1,\ldots,n)
=
\int d^D x\, e^{i q\cdot x}
\langle 1,\ldots,n|\mathcal O(x)|0\rangle
=
\delta^D\!\left(\sum_{i=1}^n p_i + q\right)
\langle 1,\ldots,n|\mathcal O(0)|0\rangle .
\label{eq:FF_def}
\end{equation}
Here all momenta are taken as outgoing, so the operator carries the off-shell momentum $q$ with $\sum_i p_i + q = 0$.
FFs are therefore close to amplitudes, but they also contain a color-singlet off-shell insertion.
This makes them a useful testing ground for extending on-shell methods
to local gauge-invariant observables.

In applying the double copy for FFs, the main difficulty stems from the color-singlet nature of the operator. 
In particular, Feynman diagrams in which the operator is inserted at different positions can have different propagator structures while sharing the same color factors, thereby imposing strong constraints on CK duality.  Moreover, local observables break diffeomorphism invariance in gravity, so the
double copy of a FF cannot be interpreted as
the insertion of a local gravitational operator at a fixed spacetime
point. A  useful alternative viewpoint is to regard the gauge-invariant operator as an effective interaction vertex coupling to 
a color-singlet Higgs particle:
${\cal O} \rightarrow H {\cal O}$, in analogy with the Higgs effective theory considered in \cite{Wilczek:1977zn, Shifman:1979eb, Dawson:1990zj, Kniehl:1995tn, Chetyrkin:1997un}.
This converts the form factors into scattering amplitudes
containing a color-singlet massive external state carrying momentum $q$.

The double copy for tree-level FFs was initiated and developed in
\cite{Lin:2021pne,Lin:2022jrp,Lin:2023rwe}. For the FF of
${\rm tr}(\phi^2)$ in Yang--Mills-scalar theory, the basic CK and KLT
constructions were established in \cite{Lin:2021pne,Lin:2022jrp}, together with an
analysis of the emergence of ``spurious'' poles in CK-dual numerators.
The construction was extended in \cite{Lin:2023rwe} to higher-length
scalar operators ${\rm tr}(\phi^m)$, where new color identities were
identified, closed constructions of CK-dual numerators were given, and
the vectors governing hidden factorization relations and KLT-kernel
factorization were studied.
These works uncovered several features that do not occur in
ordinary amplitudes in the same way:
\begin{itemize}
\item
CK-dual FF numerators contain poles that cancel in the full
gauge level sum, and are therefore spurious for the FF itself.
After the double copy, however, these poles survive and become
physical propagators in gravity:
\begin{equation}
\text{spurious poles in gauge theory}
\;\xrightarrow{\ \text{double copy}\ }\;
\text{physical propagators in gravity}.
\notag 
\end{equation}
\item
FFs obey hidden factorization relations on these spurious poles,
\begin{equation}
\vec v\cdot \vec F_n\big|_{s_{\rm sp}=0}
=
F_m \times A_{n+2-m},
\end{equation}
where the vectors $\vec v$ also appear in the factorization of the KLT
kernel \cite{Lin:2022jrp,Lin:2023rwe}.
The left-hand-side structure of the equation is reminiscent of the BCJ relations for pure amplitudes,
suggesting an underlying amplitude-like structure hidden behind the FF.
\end{itemize}

The goal of this paper is to identify this structure. We introduce a
systematic operation that we call ``dyeing''. Starting from the color-singlet
Higgs particle that represents the operator insertion, dyeing promotes
this auxiliary particle to an adjoint massive state,\footnote{This
dyed Higgs is distinct from the usual colored
Higgs in spontaneously broken gauge theories, whose double copy has been
studied in~\cite{Chiodaroli:2015rdg,Naculich:2015coa}.
Here, the dyed Higgs couples typically through $d$-symmetric color structures, as discussed
in later sections.}
\begin{equation}
H \quad \xrightarrow{\ \text{dyeing}\ } \quad \bar H = H^a T^a .
\end{equation}
For the operator ${\rm tr}(\phi^2)$, this produces a colored parent
theory in which the auxiliary Higgs can participate in ordinary
gauge level color relations. One can also consider an inverse operation, called ``bleaching'':
it removes the artificial adjoint color of the Higgs by a $U(1)$
decoupling limit, thereby projecting the colored parent amplitude back
to the original FF. The precise Lagrangian and bleaching
identities are given in Section~\ref{sec:dyedphi2}.

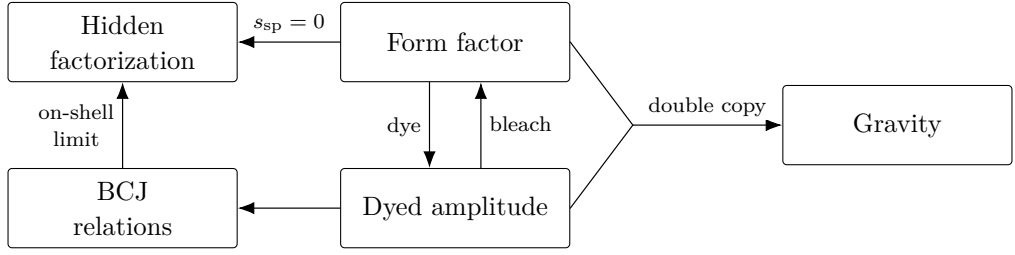
\begin{figure}[t]
  \centering
  \begin{tikzpicture}[
    font=\small,
    line cap=round,
    line join=round,
    box/.style={
      draw,
      rounded corners=1.5pt,
      align=center,
      inner sep=4pt,
      minimum height=1.05cm,
      text width=2.75cm
    },
    op/.style={
      -{Latex[length=2.2mm]},
      line width=0.45pt
    },
    edgelab/.style={
      font=\scriptsize,
      fill=white,
      inner sep=1pt
    },
    every node/.style={inner sep=2pt}
  ]
    \node[box] (hid) at (0,0) {Hidden\\factorization};
    \node[box] (bcj) at (0,-2.2) {BCJ\\relations};
    \node[box] (ff) at (4.4,0) {Form factor};
    \node[box] (dyed) at (4.4,-2.2) {Dyed amplitude};
    \node[box] (grav) at (10.25,-1.1) {Gravity};
    \coordinate (dcjoin) at (6.75,-1.1);

    \draw[op] ([xshift=-0.34cm]ff.south) -- node[pos=0.52, left=2pt, edgelab] {dye} ([xshift=-0.34cm]dyed.north);
    \draw[op] ([xshift=0.34cm]dyed.north) -- node[pos=0.52, right=2pt, edgelab] {bleach} ([xshift=0.34cm]ff.south);
    \draw[line width=0.45pt] (ff.east) -- (dcjoin);
    \draw[line width=0.45pt] (dyed.east) -- (dcjoin);
    \draw[op] (dcjoin) -- node[above=2pt, edgelab] {double copy} (grav.west);
    \draw[op] (dyed.west) -- (bcj.east);
    \draw[op] (bcj.north) -- node[left=2pt, edgelab, align=center] {on-shell\\limit} (hid.south);
    \draw[op] (ff.west) -- node[above=2pt, edgelab] {$s_{\rm sp}=0$} (hid.east);
  \end{tikzpicture}
  \caption{Schematic map of the dyeing and bleaching framework.}
  \label{fig:dyeing_map}
\end{figure}

Our main results, whose relations are sketched in Figure~\ref{fig:dyeing_map}, are as follows:
\begin{itemize}
\item
The dyed framework provides a direct interpretation of the spurious poles. In the dyed theory, these poles are ordinary physical
propagators of the massive adjoint Higgs. They disappear in the original FF because the corresponding color factors vanish after
bleaching. Their kinematic numerators, however, remain part of the
CK-dual structure and are retained by the double copy. Thus the dyed
theory explains why such poles are invisible in FFs but physical in gravity. Equivalently, it supplies a systematic
origin for auxiliary topologies whose color factors vanish after the
projection, but whose numerators are still needed by the FF
double copy.

\item
The same picture explains the hidden factorization relations. The dyed
object is a fully colored amplitude with massive adjoint matter, and
its color identities lead to the amplitude-level BCJ relations. 
We use the on-shell limit of a physical pole inside a BCJ
relation to extract the hidden factorization relation,
\begin{equation}
\text{BCJ relations in dyed amplitudes}
\;\xrightarrow{\ \text{on-shell limit}\ }\;
\text{hidden factorization relations of FFs}.
\notag
\end{equation}

\item
For multiple operator insertions, the dyeing construction
gives a systematic treatment in terms of multi-Higgs amplitudes, where the CK construction and double copy can be performed similarly. 
We also find that these amplitudes decompose into gauge-invariant sectors corresponding to different ordering of scalar particles. 
The corresponding propagator matrix is block
diagonal, and the same block structure is inherited by the KLT kernel.
Each scalar-ordered gauge-theory sector double copies to a
scalar-ordered gravity sector.

\item
The framework extends to a general class of operators including higher-length scalar operators ${\rm tr}(\phi^m)$ and the bi-fermion operator $\trpp$.
Moreover, we apply the same idea to the vector current
$\trpGp$. This example shows that dyeing is
not limited to scalar operator insertions: it can also accommodate
Lorentz indices carried by the operator. It therefore points toward a
more general use of the dyed construction for double-copy problems
involving higher-spin operator insertions.

\item
We also present a simple one-loop example, which suggests that the dyed
viewpoint can be generalized beyond tree level: before bleaching, the
operator insertion can be treated as part of a dyed integrand
with a massive internal propagator.
\end{itemize}

A shorter account of the main ideas presented here appeared in \cite{Li:2026cia}.

The rest of the paper is organized as follows. In Section~\ref{sec:reviewCK}
we review CK duality, KLT kernels, propagator matrices, and the basic
FF double-copy examples. In Section~\ref{sec:dyedphi2} we
introduce the dyeing and bleaching procedure for the primitive
${\rm tr}(\phi^2)$ operator and derive hidden factorization relations
from on-shell limits of BCJ relations. In Section~\ref{sec:multi_higgs}
we generalize the
construction to multiple dyed Higgs particles and develop the
scalar-ordering decomposition. In Section~\ref{sec:dyedgeneral} we
discuss general operators, including higher-length scalar operators and
fermionic operators. Section~\ref{sec:discussion} contains the summary
and outlook. A preliminary one-loop example is presented in
Appendix~\ref{sec:loop}.


 \section{Review}
\label{sec:reviewCK}
In this section, we review CK duality and the double-copy construction.
We first use a four-point gluon amplitude to illustrate the basic concepts,
and then use a three-point FF example to demonstrate the double copy of FFs.
This section is mainly served as a review of basic concepts and a setup of useful notation used throughout the paper.

\subsection{Amplitudes}

\newcommand{\afgppm}{\Theta_{2\times2}^{\amp_4^g}}

\begin{figure}[t]
  \centerline{\includegraphics[height=2.5cm]{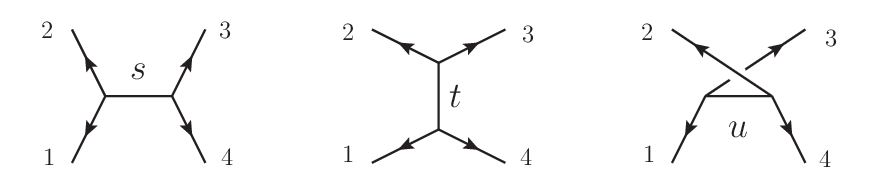} }
  \caption{Trivalent graphs for the four-point tree amplitude. }
  \label{treeA4}
\end{figure}

A fundamental example of CK duality is the four-gluon tree
amplitude, which can be represented by a sum over trivalent topologies
shown in Figure~\ref{treeA4},
\begin{equation}
    \amp_4^g(1^g,2^g,3^g,4^g) = \frac{C_sN_s}{s} + \frac{C_tN_t}{t} + \frac{C_uN_u}{u}\,,
    \label{eq:fullcolortreeA4}
\end{equation}
where $C_{\set{s,t,u}}$ and $N_{\set{s,t,u}}$ are color factors and kinematic numerators
of the corresponding channels.
$s = (p_1 + p_2)^2$, $t = (p_2 + p_3)^2$ and $u = (p_1 + p_3)^2$ are Mandelstam variables.
The color factors are given by
\begin{equation}\label{color_def}
  C_s={f}^{a_1a_2s}{f}^{sa_3a_4},\; \,
  C_t={f}^{a_2a_3t}{f}^{ta_4a_1}, \;\,
  C_u={f}^{a_1a_3u}{f}^{ua_2a_4},
\end{equation}
where $f^{abc} = \tr(T^a T^b T^c) - \tr(T^a T^c T^b)$ is the structure constant\footnote{
    The symmetric tensor is defined as
    $d^{abc} = \tr(T^a T^b T^c) + \tr(T^a T^c T^b)$
    in this paper.
}.
The color factors $C_{\set{s,t,u}}$ satisfy the Jacobi identity
\begin{equation}\label{stu_Jacobi_C}
    C_s= C_t + C_u \,.
\end{equation}
The CK duality requires the numerators to satisfy the same
linear relation
\begin{equation}\label{stu_Jacobi_N}
    N_s^\CK= N_t^\CK + N_u^\CK \,,
\end{equation}
which is referred to as a ``dual Jacobi relation'' or ``CK-dual relation''.
The superscript ``CK'' is used to denote CK-satisfying numerators.

\subsubsection*{DDM color basis}

To determine $N_\set{s,t,u}^\CK$,
a straightforward method is to compare (\ref{eq:fullcolortreeA4}) with
color-ordered amplitudes of a specific color basis.

A useful choice is the Del Duca-Dixon-Maltoni (DDM) color basis~\cite{DelDuca:1999sg},
whose elements can be represented by planar half-ladder diagrams.
These diagrams are generated systematically by fixing two external states
at the endpoints of a line,
and inserting the remaining states along this line in all possible permutations.

\begin{figure}
    \begin{center}
        \includegraphics[width=0.6\textwidth]{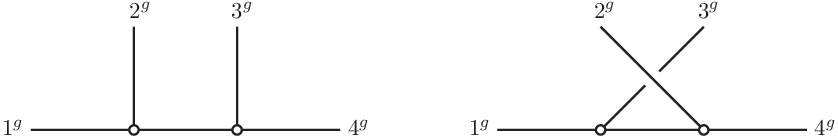}
    \end{center}
    \caption{Half-ladder diagrams for DDM color basis of $A_4^g$.}
    \label{fig:a4_ddm}
\end{figure}

For the four-point case $\amp_4^g$,
fixing $1^g$ and $4^g$ and permuting $2^g$ and $3^g$ gives two half-ladder diagrams
shown in Figure~\ref{fig:a4_ddm}.
Due to the planarity,
a half-ladder diagram can be uniquely specified by its planar ordering.
For instance,
the two diagrams correspond to the orderings $(1,2,3,4)$ and $(1,3,2,4)$.
We define an operator $\co$ to map these orderings to their corresponding color factors,
\begin{equation}
    \co(1,2,3,4) = f^{a_1 a_2 b} f^{b a_3 a_4} = C_s,\quad
    \co(1,3,2,4) = f^{a_1 a_3 b} f^{b a_2 a_4} = C_u\, .
\end{equation}
All other color factors can be expanded in terms of $C_{\set{s,u}}$ using color relations
such as the Jacobi identity.

\subsubsection*{Propagator matrix and BCJ relations}

The full-color amplitude $\amp_4^g$ can be decomposed into the DDM color basis as
\begin{equation}
    \amp_4^g(1^g, 2^g, 3^g, 4^g) = C_s \ca_{4,s}^{g} + C_u \ca_{4,u}^{g}
    = \vec{C}^T \cdot \vec{\ca}^{g}_4\, ,
    \label{eq:a4g_color_decompose}
\end{equation}
with
\begin{equation}
    \vec{C} = \begin{bmatrix} C_s \\ C_u \end{bmatrix}\, ,
    \quad
    \vec{\ca}^g_4 = \begin{bmatrix} \ca^g_{4,s} \\ \ca^g_{4,u} \end{bmatrix},
\end{equation}
where $\ca_{4, \set{s,u}}^g$ are color-ordered amplitudes.
In this case, these are equal to the partial amplitudes in the single-trace basis
$\tr(T^{a_1} T^{a_2} T^{a_3} T^{a_4})$ and $\tr(T^{a_1} T^{a_3} T^{a_2} T^{a_4})$, respectively.
Furthermore,
by projecting (\ref{eq:fullcolortreeA4}) into this basis,
it can be rearranged into a compact form
\begin{align}
    \amp_4^g(1^g,2^g,3^g,4^g) &=
\vec{C}^T \cdot \afgppm \cdot \vec{N}^\CK \, ,
    \label{eq:a4g_ctn}
\end{align}
with
\begin{equation}
    \afgppm =
    \begin{bmatrix}
        \frac{1}{s}+\frac{1}{t} & -\frac{1}{t} \\
        -\frac{1}{t} & \frac{1}{t}+\frac{1}{u}
    \end{bmatrix}\,,
    \quad
    \vec{N}^\CK= \begin{bmatrix} N_s^\CK \\ N_u^\CK \end{bmatrix}\, .
\end{equation}
The matrix $\afgppm$ is composed of propagators,
and therefore is referred to as the ``propagator matrix''.
Using the four-point massless kinematic condition $s + t + u = 0$, one can show that the matrix is singular with
\begin{equation}
    \rank \afgppm = 1 \,.
\end{equation}

Comparing (\ref{eq:a4g_color_decompose}) with (\ref{eq:a4g_ctn}) gives
\begin{equation}
    \vec{\ca}^g_4 = \afgppm \cdot \vec{N}^\CK \, .
    \label{eq:a4g_atn}
\end{equation}
The singularity of $\afgppm$ means that the elements of $\vec{\ca}^g_4$ are not linearly independent.
The resulting linear relations among these partial amplitudes are the well-known BCJ relations \cite{Bern:2008qj}.
A systematic approach to determine these relations involves analyzing the left null space of $\afgppm$,
\begin{equation}
    \lns(\afgppm) =
    \left\{\vec{v}_{\text{L}}\ |\ \vec{v}^T_{\text{L}} \cdot \afgppm = 0\right\} \, .
    \label{eq:left_null_space}
\end{equation}
Consequently, the BCJ relations are generated by
\begin{equation}
    \vec{v}^T_{\text{L}} \cdot \vec{\ca}^g_4 =
    \vec{v}^T_{\text{L}} \cdot \afgppm \cdot \vec{N}^\CK
    = 0 \,.
\end{equation}
For a symmetric matrix like $\afgppm$,
the left null space is the same as the right null space.
Thus we do not distinguish them and denote both by $\ns(\afgppm)$.
In this case,
$\ns(\afgppm)$ is spanned by a single vector
$
    \vec{v}_{\ns}^{\, T} = [s,\, -u]
$,
which gives the BCJ relation of $\amp_4^g$
\begin{equation}
    \vec{v}_{\ns}^T \cdot \vec{\ca}^g_4 = s \ca_{4,s}^g - u \ca_{4,u}^g = 0\,.
    \label{eq:a4g_bcj}
\end{equation}

\subsubsection*{Double copy and KLT form}

\newcommand{\afgklt}{\klt_{1 \times 1}^{\amp_4^g}}
Once the CK-satisfying numerators are determined,
the double copy procedure is implemented by replacing color factors with the corresponding numerators in
(\ref{eq:fullcolortreeA4}).
The result is a four-graviton tree-level amplitude,
\begin{equation}
    \gA_4(1^h, 2^h, 3^h, 4^h) = \frac{N_s^2}{s}+\frac{N_t^2}{t}+\frac{N_u^2}{u}\,.
    \label{eq:4pt_gluon_dc}
\end{equation}
$\gA_4$ is invariant under the following transformation
\begin{equation}
    \varepsilon_i^{\mu\nu}\to\varepsilon_i^{\mu\nu}+\alpha p_i^{(\mu}\xi_i^{\nu)}\,,
    \label{eq:diffeo_trans}
\end{equation}
as required by diffeomorphism invariance of gravity amplitudes.
Here
$\varepsilon_i^{\mu\nu} = \varepsilon_i^{(\mu}\varepsilon_i^{\nu)}$
is the polarization tensor of $i^h$,
and $\xi^\mu$ is a vector that satisfies $\xi \cdot p_i = 0$.

Alternatively, the double copy can be implemented using the KLT formula~\cite{Kawai:1985xq},
\begin{equation}
    \gA_4 =
    \ca^g_{4, s}
    \times
    \afgklt \colorxy{C_s}{C_s}
    \times
    \ca^g_{4, s}\, ,
\end{equation}
with the KLT kernel given by the inverse of the propagator matrix
sub-block~\cite{Mizera:2016jhj},
\begin{equation}
    \afgklt \colorxy{C_s}{C_s}
    = \left(\afgppm \colorxy{C_s}{C_s}\right)^{-1}
    = \left( \frac{1}{s} + \frac{1}{t} \right)^{-1} \,.
\end{equation}
Throughout this work,
a pair of color factors is often used to index elements of
the KLT kernel and propagator matrix.
The KLT kernel $\klt$ can generally be extracted
from the propagator matrix $\Theta$
by selecting a maximal full-rank submatrix of $\Theta$.

\subsubsection*{Bi-adjoint scalar}

\newcommand{\LaBAS}{\La^{\text{BAS}}}
\newcommand{\LaBASYMSH}{\LaBAS_{\text{YMsH}}}
\newcommand{\ba}{\bar{a}}
\newcommand{\bb}{\bar{b}}
\newcommand{\bc}{\bar{c}}
\newcommand{\pg}{\varphi_{g}}
\newcommand{\pp}{\varphi_{\phi}}
\newcommand{\pH}{\varphi_{H}}
\newcommand{\pdH}{\varphi_{\bHBlue}}

The matrix $\Theta^{\amp_n^g}$ can be interpreted as an amplitude of
bi-adjoint scalar (BAS)
theory~\cite{Du:2011js,Bjerrum-Bohr:2012kaa,Cachazo:2013iea}.
We review the BAS setup here
in order to build $\Theta_n$ systematically.

The BAS theory to match the propagator matrix of pure gluon scattering is
\begin{equation}
    \LaBAS_{g} = \LaBAS_{\pg} + \LaBAS_{ggg} \, ,
\end{equation}
with
\begin{gather}
    \LaBAS_{\pg} =
    \frac{1}{2}(\partial_\mu \pg^{a\ba}) (\partial^\mu \pg^{a\ba})\, ,
    \quad
    \LaBAS_{ggg} =
    \frac{\lambda_{ggg}}{3!} f^{abc}f^{\ba\bb\bc}
    \pg^{a\ba} \pg^{b\bb} \pg^{c\bc}\, .
    \label{eq:BAS_ggg}
\end{gather}
$\pg^{a\ba}$ is used to match the gluon $g$, and carries two adjoint indices $\set{a, \ba}$ of two different gauge groups, $SU(N)$ and $SU(\bar{N})$.
The subscript $g$ of $\varphi$ indicates the bi-adjoint field
is used to match gluons.
Each element of the propagator matrix of $\amp_n^g$ is given by
the corresponding double-partial color component of the $\LaBAS_{g}$ amplitude,
\begin{equation}
    \Theta^{\amp_n^g} \colorxy{C_1}{C_2} =
    \ampbas^g_n \colorxy{C_1}{C_2} \,.
\end{equation}
In practice,
the double partial amplitude $\campbas_n^g\colorxy{C_1}{C_2}$
can be built by summing over all propagators of overlapping
topologies that contribute to both color structures $\set{C_1, C_2}$, with the appropriate sign; see \cite{Cachazo:2013iea}.

 \subsection{Form factor}

We review the FF of $\tr(\phi^2)$ in Yang--Mills-scalar (YMs) theory
\begin{equation}
    \La^{\text{YMs}} =
    - \frac{1}{4} \tr(F_{\mu\nu}F^{\mu\nu})
    + \frac{1}{2} \tr(D^\mu\phi D_\mu\phi)\,
\end{equation}
as a basic example,
where the scalar field $\phi = \phi^a T^a$ is
in the adjoint representation of the gauge group.

\begin{figure}
    \centering
    \includegraphics[width=0.4\linewidth]{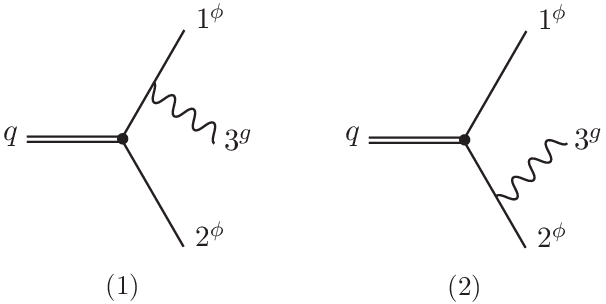}
    \caption{
        Feynman diagrams of $\ff_3$.
        The operator is represented by black double lines with a circular head,
        while $\phi$ is drawn with black straight lines
        and gluons are curly lines.
    }
    \label{fig:revtrphi2FF}
\end{figure}

The three-point FF $\ff_3(1^\phi,2^\phi,3^g)$ can be expressed as a sum of the two cubic diagrams in Figure~\ref{fig:revtrphi2FF},
\begin{equation}
    \ff_3(1^\phi,2^\phi,3^g)
    =
    \frac{C_1N_1}{s_{13}}+\frac{C_2N_2}{s_{23}}\,,
    \label{eq:ff3_fullcolor}
\end{equation}
with color factors
\begin{equation}
    C_1=C_2=f^{a_1a_3a_2}\,.
    \label{eq:rev_trphi2ff_col_jac_rel}
\end{equation}
Imposing CK duality gives
\begin{equation}
    N_{1}^\CK=N_{2}^\CK\,.
\end{equation}
We choose $C_1$ as the color basis.
$\ff_3$ can be decomposed as $\ff_3 = C_1 \cf_{3,(1)}$,
where $\cf_{3,(1)}$ is the partial FF of $C_1$.
In this case,
it equals the partial FF in the single-trace basis $\tr(T^{a_1}T^{a_3}T^{a_2})$,
\begin{equation}
    \cf_{3,(1)} =
    \cf_3(1^\phi, 3^g, 2^\phi) =
    2\left(-\frac{p_1 \cdot \varepsilon_3}{s_{13}} +
    \frac{p_2 \cdot \varepsilon_3}{s_{23}}\right)\, .
\end{equation}
By comparison, CK-satisfying numerators are obtained as
\begin{equation}
    N_1^\CK = N_2^\CK =
    \frac{s_{13}s_{23}}{s_{13}+s_{23}}\cf_3(1^\phi, 3^g, 2^\phi)\, ,
    \label{eq:revFF_NCK}
\end{equation}
where the spurious pole $(s_{13} + s_{23}) = -(s_{12} - q^2)$ appears.

\subsubsection*{From a spurious pole to a real pole}

After the double-copy replacement,
the spurious pole transforms into a real pole in the resulting gravity quantity,
\begin{equation}
    \mathcal{G}_3
    =
    \cf_3(1^\phi, 3^g, 2^\phi)
    \times
    \frac{s_{13}s_{23}}{s_{13}+s_{23}}
    \times
    \cf_3(1^\phi, 3^g, 2^\phi)\,.
    \label{eq:revtrphi2ff_dc}
\end{equation}
One verifies that $\mathcal{G}_3$ factorizes into the expected tree-level
gravity amplitudes on all of its poles,
including the new pole $(s_{12} - q^2)$,
\begin{equation}
\lim_{s_{12} \rightarrow q^2} (s_{12} - q^2) \times \mathcal{G}_3 =
    1 \times (2\varepsilon_3\cdot q)^2=
    \mathcal{G}_2(1^\phi,2^\phi)
    \times
    \gA_3(\mathbf{P}_{12}^S,q^S,3^h)\,.
    \label{eq:ff3_factorize}
\end{equation}
Here,
$\mathcal{G}_2(1^\phi, 2^\phi)$ is
the double copy of $\cf_2(1^\phi, 2^\phi)$,
and $\gA_3(\mathbf{P}_{12}^S,q^S, 3^h)$ is a three-point gravity amplitude,
where $S$ is a scalar field with mass $q^2$ coupled to gravity,
and $\mathbf{P}_{12} = p_1 + p_2$.
Therefore, the pole $(s_{12} - q^2)$ is a physical pole at the gravity level.
Together with the fact that $\mathcal{G}_3$ is diffeomorphism invariant,
$\mathcal{G}_3$ can be interpreted as a consistent four-point gravity amplitude
$\gA_4(1^\phi, 2^\phi, 3^h, q^S)$ of a certain theory.

\subsubsection*{Hidden factorization relations of FFs}
Taking the square root of (\ref{eq:ff3_factorize})
gives a hidden factorization relation of FF~\cite{Lin:2021pne,Lin:2022jrp},
\begin{equation}
    s_{13}\mathcal{F}_3(1^\phi,3^g,2^\phi)|_{s_{12}=q^2}
    =
    \mathcal{F}_2(1^\phi,2^\phi) \times \mathcal{A}_3(\mathbf{P}_{12}^{\bH},3^g,q^{\bH})\,,
    \label{eq:f3_st_hidden_fac}
\end{equation}
where $\ca_3(\mathbf{P}_{12}^{\bH},3^g,q^{\bH}) = (2 \varepsilon_3 \cdot q)$ is
a three-point gauge-theory amplitude.
$\bH$ is a scalar field under the adjoint representation.
This feature extends to higher-point cases.
For example, the four-point FF obeys
\begin{align}
    -\left[
        s_{24}\cf_4(1^{\phi},3^g,4^g,2^{\phi})+
        (s_{24}+s_{34})\cf_4(1^{\phi},4^g,3^g,2^{\phi})
    \right]_{s_{123}=q^2}
    &
    \notag \\
    &\hspace{-3cm}
    =\cf_3(1^\phi,3^g,2^\phi)
    \times \ca_3(\mathbf{P}_{123}^{\bH},4^g,q^{\bH})\,,
    \notag \\
    -\left[
        \frac{s_{13}s_{24}}{s_{13}+s_{23}}
        \cf_{4}(1^{\phi},3^g,4^g,2^{\phi})+
        \frac{s_{14}s_{23}}{s_{13}+s_{23}}
        \cf_{4}(1^{\phi},4^g,3^g,2^{\phi})
    \right]_{s_{12}=q^{2}}
    &
    \notag \\
    &\hspace{-3cm}
    =\cf_{2}(1^\phi,2^\phi)
    \times
    \ca_{4}(\mathbf{P}_{12}^{\bH},3^g,4^g,q^{\bH})\,.
    \label{eq:f4_hidden_fac}
\end{align}
The physical interpretation of these relations was not apparent in
previous works; it will be addressed in the next section.

\subsubsection*{Higgs-EFT}

By identifying the operator insertion as a color-singlet massive scalar,
these FFs can be interpreted as amplitudes of a Higgs effective field theory.

For example,
$\ff_{\tr(\phi^2), n}(1^\phi, 2^\phi, 3^g, \ldots)$ of YMs
can be considered as
$\amp_{n+1}(q^H, 1^\phi, 2^\phi, 3^g, \ldots)$
of the Yang-Mills-scalar-Higgs (YMsH) theory with a corresponding interaction term
\begin{equation}
    \La^{\text{YMsH}}_{\tr(\phi^2)} =
    \La^{\text{YMsH}} + \La^{H\phi\phi}\, ,
\end{equation}
with
\begin{gather}
    \La^{\text{YMsH}} = \La^{\text{YMs}} + \La^{\text{Higgs}},
    \notag \\
    \La^{\text{Higgs}} = \frac{1}{2}(\partial_\mu H)(\partial^\mu H) - \frac{1}{2} m_H^2 H^2,
    \quad
    \La^{H\phi\phi} = \lambda H \tr(\phi^2)\, ,
\end{gather}
where $\lambda$ is a coupling constant.
Note that the field $H$ here is a color-singlet scalar.

To build the propagator matrix of $\ff_n$,
we can again write down a BAS Lagrangian to match the YMsH theory,
where two additional scalar fields are introduced through the mapping
\begin{equation}
    \phi^a \rightarrow \varphi_\phi^{a\ba}, \quad
    H \rightarrow \varphi_H\,.
\end{equation}
Unlike $\set{\pg^{a\ba}, \pp^{a\ba}}$, $\pH$ is color singlet and matches the Higgs field.
The kinetic parts are
\begin{equation}
    \LaBAS_{\pp} =
    \frac{1}{2}(\partial_\mu \pp^{a\ba}) (\partial^\mu \pp^{a\ba})\, ,
    \quad
    \LaBAS_{\pH} =
    \frac{1}{2}(\partial_\mu \pH) (\partial^\mu \pH)
    - \frac{1}{2} m_H^2 \pH^2 \, .
\end{equation}
The BAS theory involves three types of interaction vertices: $g$-$g$-$g$, $\phi$-$\phi$-$g$, and $H$-$\phi$-$\phi$.
The first is defined in (\ref{eq:BAS_ggg}), while the latter two are given by
\begin{equation}
    \LaBAS_{\phi\phi g} =
    \frac{\lambda_{\phi\phi g}}{3!} f^{abc}f^{\ba\bb\bc}
    \pp^{a\ba} \pp^{b\bb} \pg^{c\bc}\,, \quad
    \LaBAS_{H\phi\phi} =
    \frac{\lambda_{H\phi\phi}}{3!} \delta^{ab}\delta^{\ba\bb}
    \pH \pp^{a\ba} \pp^{b\bb} \, .
\end{equation}
Using these components, the BAS Lagrangian corresponding to the YMsH theory is defined as
\begin{gather}
    \LaBAS_{\text{YMsH}_{\tr(\phi^2)}} =
    \LaBAS_{\text{YMsH}} + \LaBAS_{H\phi\phi}\, ,
\end{gather}
with
\begin{equation}
    \LaBASYMSH = \LaBAS_{\text{YMs}} + \LaBAS_{\pH} \,,
    \quad
    \LaBAS_{\text{YMs}} =
    \LaBAS_{\pg} + \LaBAS_{\pp} + \LaBAS_{ggg} + \LaBAS_{\phi\phi g}\,.
    \label{eq:BAS_YMSH}
\end{equation}
Each propagator matrix element of
$\ff_n$ is the corresponding double-partial BAS amplitude,
\begin{equation}
    \Theta^{\ff_n}\colorxy{C_1}{C_2}
    =
    \campbas_{n+1}^{\text{YMsH}_{\tr(\phi^2)}}
    \colorxy{C_1}{C_2}\, .
\end{equation}
External states of $\ff_n$ are matched to the BAS scalar fields defined above through the mapping
$g\rightarrow\pg$, $\phi\rightarrow\pp$, and the operator leg
$H\rightarrow\pH$. 


  \section{Dyed theory: primitive example} \label{sec:dyedphi2}

In this section,
we introduce basic examples to illustrate
the core concept and utility of ``dyed theories''.
By examining the scattering amplitudes of a corresponding dyed theory, we identify connections between amplitudes and FFs
at the level of CK duality and the double copy.
Furthermore,
the hidden factorization relations of FFs are shown to arise as direct
consequences of the BCJ relations within a dyed theory.

\subsection{From form factors to dyed amplitudes}

\newcommand{\fftabc}{\ff_3(1^\phi, 2^\phi, 3^g)}
\newcommand{\cftacb}{\cf_3(1^\phi, 3^g, 2^\phi)}
\newcommand{\YMSH}{\text{YMsH}}
\newcommand{\dYMSH}{\overline{\text{YMsH}}}
\newcommand{\dHiggs}{\overline{\text{Higgs}}}
\newcommand{\LadYMSH}{\La^{\dYMSH}}
\newcommand{\LadHiggs}{\La^{\dHiggs}}
\newcommand{\LadHpp}{\La^{{\clrBlue \bH} \phi \phi}}

\begin{figure}
    \begin{center}
        \includegraphics[width=0.14\linewidth]{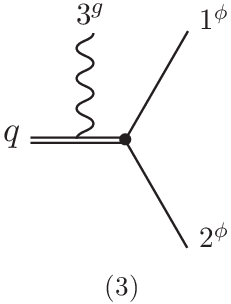}
    \end{center}
    \caption{The auxiliary topology for $\fftabc$.}
    \label{fig:f3_auxiliary}
\end{figure}

To demonstrate the physical intuition driving the formulation of dyed theories,
we start with the hidden factorization relation of the three-point FF.
This mathematical relation can be recast into a more transparent structural form
by hypothesizing the existence of an auxiliary trivalent topology of
$\ff_3(1^\phi, 2^\phi, 3^g)$ in Figure~\ref{fig:f3_auxiliary},
where the Higgs field is gauged and can therefore interact directly with gluons.
The squared mass of the Higgs introduced this way is $q^2$.

Since
\begin{equation}
    \cftacb = \frac{N_1}{s_{13}} + \frac{N_2}{s_{23}}\, ,
\end{equation}
the left-hand side of (\ref{eq:f3_st_hidden_fac}) is
\begin{equation}
    s_{13} \cftacb|_{s_{12} = q^2} = (N_1 - N_2)|_{s_{12} = q^2}\, ,
\end{equation}
where we used $s_{23} = -s_{13}$ under $s_{12} = q^2$.
The right-hand side of (\ref{eq:f3_st_hidden_fac}),
which is the product of two tree-level gauge blocks,
can be understood as the factorization of
the massive propagator of the auxiliary diagram.
Therefore, (\ref{eq:f3_st_hidden_fac}) can be written as
\begin{equation}
    (N_1 - N_2 - N_3)|_{s_{12} = q^2} = 0\, ,
\end{equation}
where $N_3$ is the numerator of the auxiliary diagram.

This suggests the existence of a set of numerators that satisfy CK duality,
and motivates considering a broader theory where gauge vectors couple
directly to the ``operator'', or equivalently, a generalized Higgs field.
We realize this by assigning
the Higgs field to the adjoint representation of the $SU(N)$ gauge group
\begin{equation}
    {\clrBlue \bH} = H^a T^a \, ,
  \label{eq:H_dyed}
\end{equation}
which we call the ``dyed Higgs''.
Starting from the Higgs-EFT $\LaYMSH$, the dyed theory is
\begin{equation}
    \LadYMSH_{\tr(\phi^2)} =
    \LadYMSH + \LadHpp \, ,
\end{equation}
with
\begin{gather}
    \LadYMSH = \LaYMS + \LadHiggs \,,
    \notag \\
    \LadHiggs = \frac{1}{2}(D_\mu H)^a(D^\mu H)^a
    - \frac{1}{2} m^2 H^a H^a \, ,
    \label{eq:La_dyed_higgs}
\end{gather}
and a Yukawa interaction
\begin{equation}
    \LadHpp = \lambda d^{abc} H^a \phi^b \phi^c\, .
    \label{eq:La_dyed_Hpp}
\end{equation}
$\LadYMSH$ is called the
dyed Yang-Mills-scalar-Higgs ($\dYMSH$) theory.\footnote{
	    An overline denotes a dyed object throughout this paper.
}

\subsubsection*{Bleaching}

\newcommand{\blcBH}{\blc_\bHBlue}

Partial amplitudes expressed in the single-trace basis satisfy
specific linear relations known as $U(1)$ decoupling identities,
which are derived by substituting the group generator of
a given external state with the identity matrix $\idd$.
Due to the presence of $d^{abc}$ vertices in the constructed dyed theory,
the $U(1)$ decoupling procedure has a similar role,
serving as a mathematical bridge between the dyed and non-dyed theories.
For example, it connects $\LadYMSH_{\tr(\phi^2)}$ and $\LaYMSH_{\tr(\phi^2)}$.

To clarify the notation,
the substitution of a color generator $T^q$ with the identity matrix $\idd$
for a specific external leg $q$ is denoted by the operator $\blc_q$,\footnote{
    $\blc$ stands for ``bleach'',
    which acts as the inverse operation to ``dye''.
    Therefore, the $U(1)$ decoupling operation is also called
	    the bleaching procedure in this context.
}
which acts on color tensors and is extended to amplitudes and Lagrangians.
Applied to the structure constants, one finds
\begin{equation}
    \blc_{c}(f^{abc}) =
    \tf^{ab\idd} \equiv \tr(T^a T^b \idd) - \tr(T^b T^a \idd) = 0\, ,
    \label{eq:f_u1_dec}
\end{equation}
This identity leads to the $U(1)$ decoupling relations among color-ordered amplitudes of pure gluon scattering.

With the normalization convention $\tr(T^aT^b) = \frac{1}{2} \delta^{ab}$, bleaching the symmetric structure constant $d^{abc}$ gives
\begin{equation}
    \blc_{c}(d^{abc}) =
    \td^{ab\idd} \equiv
    \tr(T^a T^b \idd) + \tr(T^b T^a \idd) = \delta^{ab}\,.
    \label{eq:d_u1_dec}
\end{equation}
Consequently,
\begin{equation}
    \blcBH(D_\mu H^a) = \partial_\mu H\,
    , \quad
    \blcBH\left(d^{abc} H^a \phi^b \phi^c\right) = H \delta^{bc} \phi^b \phi^c
    \, .
\end{equation}
Formally,
\begin{equation}
    \blcBH(\LadHiggs) = \La^{\text{Higgs}}\, ,
    \quad
    \blcBH(\LadHpp) = \La^{H\phi\phi}\, .
\end{equation}
Finally,
bleaching the dyed YMsH theory recovers the standard YMsH theory:
\begin{equation}
    \blcBH(\LadYMSH_{\tr(\phi^2)}) = \LaYMSH_{\tr(\phi^2)} \, .
\end{equation}

 \subsection{The four-point example}

\newcommand{\ampppgq}{\amp_{4}(1^\phi, 2^\phi, 3^g, q^{\clrBlue \bH})}
\newcommand{\gapphq}{\gA_{4}(1^\phi, 2^\phi, 3^h, q^{H})}

To parallel our earlier review of the three-point FF and
introduce the mechanics of the dyed theory,
we analyze the four-point amplitude $\ampppgq$ of the dyed theory
using the standard CK-duality method.

\begin{figure}
    \begin{center}
        \includegraphics[width=0.6\textwidth]{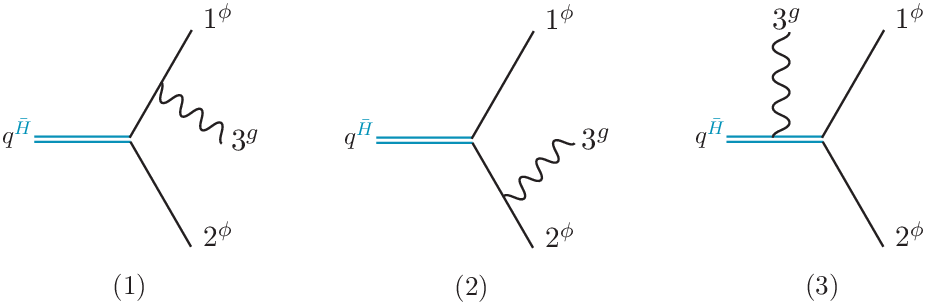}
    \end{center}
    \caption{
        Trivalent topologies of $\ampppgq$.
        Blue double lines are dyed Higgs $\bHBlue$.
    }
    \label{fig:a4_cubic}
\end{figure}

$\ampppgq$ has three trivalent topologies shown in Figure~\ref{fig:a4_cubic},
\begin{equation}
    \ampppgq =
    \frac{C_{1}^{\amp_{4}} N_{1}^{\amp_{4}}}{s_{13}} +
    \frac{C_{2}^{\amp_{4}} N_{2}^{\amp_{4}}}{s_{23}} +
    \frac{C_{3}^{\amp_{4}} N_{3}^{\amp_{4}}}{s_{12} - m^2}\, .
    \label{eq:a4_cubic_graphs}
\end{equation}
To distinguish the dyed Higgs $\bHBlue$ from the operator,
$\bHBlue$ is depicted as a blue double line in graphs without a black dot.

A key distinction between
$\ampppgq$ and $\fftabc$ lies in the color factors ---
dyed amplitudes utilize both the antisymmetric structure constant $f$ and
the symmetric structure constant $d$ as building blocks,
\begin{equation}
    C_{1}^{\amp_{4}} = f^{a_1 a_3 b} d^{b a_2 q},\
    C_{2}^{\amp_{4}} = f^{a_3 a_2 b} d^{b q a_1},\
    C_{3}^{\amp_{4}} = f^{q a_3 b} d^{b a_1 a_2}\, ,
    \label{eq:a4_cfs}
\end{equation}
where we use $q$ to denote the color index of the $q$-leg.
Analogous to the Jacobi relations of $ff$,
the color factors associated with the three interaction channels formed
by the product of $f$ and $d$ obey a linear color relation.
In our case, this relation is
\begin{equation}
    C_{3}^{\amp_{4}} = - C_{1}^{\amp_{4}} + C_{2}^{\amp_{4}} \, .
    \label{eq:a4_color_relation}
\end{equation}

Kinematic numerators generated via
Feynman rules automatically satisfy CK duality,
\begin{gather}
    N_{1}^{\amp_{4}} = -2 (p_1 \cdot \epsilon_3) = N_{1}^{\ff_3}\, ,
    \quad
    N_{2}^{\amp_{4}} = 2 (p_2 \cdot \epsilon_3) = N_{2}^{\ff_3}\, ,
    \notag \\
    N_{3}^{\amp_{4}} = 2 (p_1 + p_2) \cdot \epsilon_3\, .
    \label{eq:a4_numerators_feyn}
\end{gather}
By replacing $C_i^{\amp_4} \rightarrow N_i^{\amp_4}$ in (\ref{eq:a4_cubic_graphs}),
the double copy gives a four-point gravity amplitude
\begin{align}
    \gapphq &= 4 \left(
        \frac{(p_1 \cdot \epsilon_3)^2}{s_{13}} +
        \frac{(p_2 \cdot \epsilon_3)^2}{s_{23}} +
        \frac{((p_1 + p_2) \cdot \epsilon_3)^2}{s_{12} - m^2}
        \right) \notag \\
            &= -4 \frac{(s_{13}(p_2 \cdot \epsilon_3) - s_{23}(p_1 \cdot \epsilon_3))^2}
            {s_{13} s_{23} (s_{12} - m^2)} \, .
    \label{eq:m4_expression}
\end{align}
This result is identical to the one obtained from the double copy of $\ff_3$.
Unlike the FF case, there is no spurious pole in the numerators in this construction.

\subsubsection*{Propagator matrix, BCJ relations and double copy}

\newcommand{\afppmtr}{{\Theta_{2 \times 2}^{\amp_4}}}
\newcommand{\afklttr}{{\klt_{1 \times 1}^{\amp_4}}}
\newcommand{\catrabcq}{\ca_4(1^\phi, 2^\phi, 3^g, q^{\bHBlue})}
\newcommand{\catracbq}{\ca_4(1^\phi, 3^g, 2^\phi, q^{\bHBlue})}
\newcommand{\ppmAfour}{\Theta^{\amp_{4}}_{2 \times 2}}

To make this structure explicit,
we work in the color basis
\begin{equation}
    \left\{C_{\tr}^{\amp_4}\right\}
    = \left\{\tr(T^{a_1} T^{a_3} T^{a_2} T^{q}),\
    \tr(T^{a_1} T^{a_2} T^{a_3} T^{q})\right\}\, .
\end{equation}
The corresponding propagator matrix is
\begin{equation}
    \afppmtr =
    \begin{bmatrix}
    \frac{1}{s_{13}} + \frac{1}{s_{23}} & -\frac{1}{s_{23}} \\[8pt]
    -\frac{1}{s_{23}} & \frac{1}{s_{23}} + \frac{1}{s_{12} - m^2}
    \end{bmatrix}.
\end{equation}
The matrix is singular after imposing kinematic conditions, with
\begin{equation}
    \rank \afppmtr = 1 \, .
    \label{eq:a4_rank}
\end{equation}
The null space $\ns(\afppmtr)$ is spanned by
$\vec{v}_{\amp_4}^{\, \tr} = [- s_{13},\ s_{12} - m^2]^T$,
which gives the BCJ relation
\begin{equation}
    -s_{13} \catracbq + (s_{12} - m^2) \catrabcq = 0 \, .
    \label{eq:a4_bcj_tr}
\end{equation}
$\ca_4$ are partial amplitudes of corresponding color factors in $\set{C_{\tr}^{\amp_4}}$.

Since the rank of $\afppmtr$ is $1$,
the KLT kernel $\afklttr$ is extracted by
selecting a single color factor from $\set{C_{\tr}^{\amp_4}}$,
and therefore generated by taking the corresponding
$1 \times 1$ submatrix from $\afppmtr$.
For example, we choose $\tr(T^{a_1} T^{a_3} T^{a_2} T^{q})$,
\begin{equation}
    \afklttr \colorxy{1,3,2,q}{1,3,2,q} =
    \left(\afppmtr \colorxy{1,3,2,q}{1,3,2,q}\right)^{-1} =
    \left(\frac{1}{s_{13}} + \frac{1}{s_{23}}\right)^{-1}\,.
\end{equation}
The KLT-form double copy is then expressed as
\begin{equation}
    \gapphq =
    \catracbq \times
    \afklttr \colorxy{1,3,2,q}{1,3,2,q} \times
    \catracbq \,.
\end{equation}

\subsubsection*{$U(1)$ decoupling of the Higgs leg and revisit $\ff_3$}

\newcommand{\blcqBH}{\blc_{q^{\bHBlue}}}

The bleaching operation $\blcBH$ connects the two objects,
\begin{equation}
    \blcqBH(\ampppgq) = \ff_3(1^\phi, 2^\phi, 3^g) \, .
    \label{eq:a4_u1_decoup}
\end{equation}
We stress that $\blc$ acts on color factors $C_i$,
rather than kinematic numerators $N_i$ and propagators.
Therefore,
\begin{align}
    \blcqBH(\ampppgq)
&= \frac{\blcqBH(C_{1}^{\amp_{4}}) N_{1}^{\amp_{4}}}{s_{13}} +
       \frac{\blcqBH(C_{2}^{\amp_{4}}) N_{2}^{\amp_{4}}}{s_{23}} +
       \frac{\blcqBH(C_{3}^{\amp_{4}}) N_{3}^{\amp_{4}}}{s_{12} - m^2}\, .
    \label{eq:a4_blc_step_1}
\end{align}
With (\ref{eq:f_u1_dec}) and (\ref{eq:d_u1_dec}),
\begin{gather}
    \blcqBH(C_{1}^{\amp_{4}})
    = f^{a_1 a_3 b} \delta^{b a_2} = C_{1}^{\ff_3}\, ,
    \quad
    \blcqBH(C_{2}^{\amp_{4}})
    = f^{a_3 a_2 b} \delta^{b a_1} = C_{2}^{\ff_3}\, , \notag \\
    \blcqBH(C_{3}^{\amp_{4}}) = 0 \times d^{b a_1 a_2} = 0
    \equiv C_{3}^{\ff_3}\, ,
\end{gather}
which maps the color factor of the third diagram to $0$.
Together with (\ref{eq:a4_numerators_feyn}),
(\ref{eq:a4_blc_step_1}) can be rewritten as
\begin{equation}
    \blcqBH(\ampppgq)
    = \frac{C_1^{\ff_3} N_1^{\ff_3}}{s_{13}}
     + \frac{C_2^{\ff_3} N_2^{\ff_3}}{s_{23}}
     + \frac{0 \times N_3^{\ff_3}}{s_{12} - m^2}
    = \ff_3(1^\phi, 2^\phi, 3^g) \, .
\end{equation}
Since $\blc$ preserves all kinematics,
decoupling the dyed amplitude directly gives a set of valid
CK-satisfying FF numerators $\{N_{1,2,3}^{\ff_3}\}$ = $\{N_{1,2,3}^{\amp_4}\}$,
where a non-trivial $N_3^{\ff_3}$ is inherited from the dyed amplitude.
Moreover,
the bleaching procedure preserves color relations.
For $\amp_4$, this indicates
\begin{equation}
    \blcqBH \left(
        C_{3}^{\amp_{4}} = - C_{1}^{\amp_{4}} + C_{2}^{\amp_{4}}
    \right)\
    \Rightarrow
    \
    C_{3}^{\ff_{3}} = - C_{1}^{\ff_{3}} + C_{2}^{\ff_{3}}
    \, .
    \label{eq:a4_dec_color_relation}
\end{equation}
This motivates performing the double copy on the FF side using the newly generated numerators:
$\{N_{1,2,3}^{\ff_3}\}$ = $\{N_{1,2,3}^{\amp_4}\}$,
and the double-copy result is identical to (\ref{eq:m4_expression}) without
introducing any spurious poles like $(s_{12} - m^2)$ in FFs.

\newcommand{\cftracb}{\cf_3(1^\phi, 3^g, 2^\phi)}
\newcommand{\ftppmtr}{{\Theta_{1 \times 1}^{\ff_3}}}

We mention another interesting observation: the following dyed partial (color-ordered) amplitude is equivalent to the partial FF,
\begin{equation}
    \catracbq = \cftracb\, .
    \label{eq:a4_eqto_f3}
\end{equation}
This equivalence is non-trivial because
the decoupling procedure $\blcqBH$ typically annihilates certain color structures,
which would normally alter the partial amplitude.
However, a careful selection of topologies gives partial amplitudes that
remain unaffected by the bleaching operator.
Such amplitudes are termed ``bleaching-invariant amplitudes''.
Graphically, $\blcqBH$ annihilates any topology with a $q^{\bHBlue}$-$g$-$\bHBlue$ vertex due to the $f^{qg\bHBlue}$ factor. The partial amplitude $\catracbq$ is composed of only diagrams $(1)$ and $(2)$ in Figure~\ref{fig:a4_cubic}, and therefore equals $\cftacb$. In contrast, $\catrabcq$ contains the contribution from diagram $(3)$ and is not breaching invariant.

Finally,
the propagator matrix of the FF forms a submatrix
of the dyed amplitude's propagator matrix $\afppmtr$,
\begin{equation}
    \ftppmtr \colorxy{1,3,2,q}{1,3,2,q} =
    \frac{1}{s_{13}} + \frac{1}{s_{23}} =
    \afppmtr \colorxy{1,3,2,q}{1,3,2,q}\,.
\end{equation}
Since the rank of $\afppmtr$ is $1$,
this indicates that the FF's matrix acts equivalently as
the inverse KLT kernel of the four-point amplitude under
a specific color basis choice,
\begin{equation}
    \left( \klt_{1 \times 1}^{\ff_3} \right)^{-1}=
     \ftppmtr =
     \left( \afklttr \right)^{-1} \,.
\end{equation}
Together with (\ref{eq:a4_eqto_f3}),
this feature ensures that the double copy results for both
$\ff_3$ and $\amp_4$ are mathematically identical.

\subsubsection*{Hidden factorization relation}

Now we deduce the promised hidden factorization relation
(\ref{eq:f3_st_hidden_fac}).
Imposing the on-shell condition on the BCJ relation (\ref{eq:a4_bcj_tr}) at the pole $(s_{12} - m^2)$
factorizes the second term,
\begin{align}
    \lim_{s_{12} \rightarrow m^2}
    (s_{12} - m^2) \times \catrabcq
    &= \ca_3(1^{\phi}, 2^{\phi}, -\mathbf{P}_{12}^{\bHBlue})
    \times
    \ca_3(\mathbf{P}_{12}^{\bHBlue}, 3^g, q^{\bHBlue})
    \notag \\
    &= \cf_2(1^{\phi}, 2^{\phi}) \times
    \ca_3(\mathbf{P}_{12}^{\bHBlue}, 3^g, q^{\bHBlue})
    \, ,
\end{align}
where the equivalence $
    \ca_3(1^{\phi}, 2^{\phi}, -\mathbf{P}_{12}^{\bHBlue}) =
    \cf_2(1^{\phi}, 2^{\phi}) = 1
$ is used.
Hence the BCJ relation becomes
\begin{equation}
    \left. s_{13} \catracbq \right|_{s_{12} = m^2}
    =
    \cf_2(1^{\phi}, 2^{\phi}) \times
    \ca_3(\mathbf{P}_{12}^{\bHBlue}, 3^g, q^{\bHBlue})
    \, ,
    \label{eq:a4_bcj_to_hidfac}
\end{equation}
which reproduces the hidden factorization of $\ff_3$.
This gives the basic mechanism:
the hidden factorization relation follows from taking the on-shell limit
of a physical pole in the BCJ relation of the dyed theory.

 \subsection{The five-point example}
\label{sec:one_higgs_five_point_example}

\newcommand{\fffour}{\ff_{4}(1^{\phi}, 2^{\phi}, 3^{g}, 4^{g})}
\newcommand{\ampfive}{\amp_{5}(1^{\phi}, 2^{\phi}, 3^{g}, 4^{g}, q^{\bHBlue})}
\newcommand{\gapphhq}{\gA_{5}(1^\phi, 2^\phi, 3^h, 4^h, q^{H})}
\newcommand{\afiveddm}{\vec{{\ca_5}}}
\newcommand{\afiveddmc}{C^{\amp_{5}}}
\newcommand{\afivetr}{\vec{{\ca_5^{\tr}}}}
\newcommand{\Csone}{C_1}
\newcommand{\Cstwo}{C_2}
\newcommand{\Csthree}{C_3}
\newcommand{\Cuone}{C_{11}}
\newcommand{\Cutwo}{C_{12}}
\newcommand{\Cuthree}{C_{13}}
\newcommand{\afiveppm}{\Theta_{6 \times 6}^{\amp_5}}
\newcommand{\afiveklt}{\KLT_{2 \times 2}^{\amp_5}}
\newcommand{\afiveip}[1]{{\ip}^{\amp_5}_{{#1}}}
\newcommand{\catracdbq}{\ca_5(1^\phi, 3^g, 4^g, 2^\phi, q^{\bHBlue})}
\newcommand{\catracbdq}{\ca_5(1^\phi, 3^g, 2^\phi, 4^g,  q^{\bHBlue})}
\newcommand{\catradcbq}{\ca_5(1^\phi, 4^g, 3^g, 2^\phi, q^{\bHBlue})}
\newcommand{\catrabcdq}{\ca_5(1^\phi, 2^\phi, 3^g, 4^g, q^{\bHBlue})}

\begin{figure}[t]
    \centerline{\includegraphics[width=0.85\linewidth]{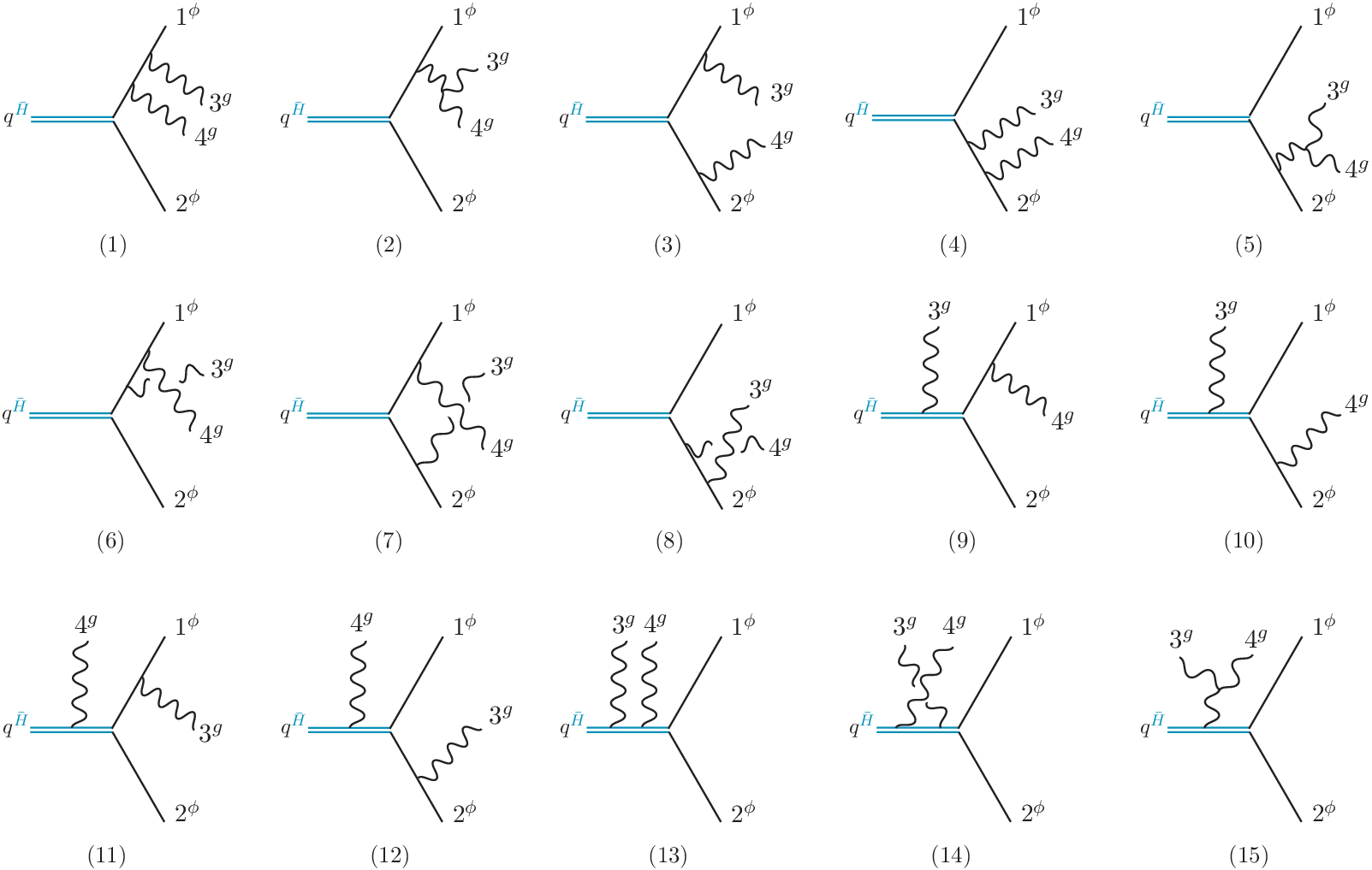} }
    \caption{
        Trivalent graphs of $\ampfive$.
        Cubic diagrams of $\fffour$ can be generated by replacing
        $q^{\bHBlue}$ with $q^{H}$ in the first eight diagrams.
    }
    \label{fig:trphi2_5pt}
\end{figure}

We next consider the five-point dyed amplitude $\ampfive$, which is promoted from the four-point $\fffour$. 
This extends the previous discussion and clarifies further the general structure.
The amplitude can be expanded into fifteen distinct cubic diagrams
plotted in Figure~\ref{fig:trphi2_5pt}.

Fixing $1^\phi$ and $q^{\bHBlue}$ and permuting
$\set{2^\phi, 3^g, 4^g}$,
we work in the color basis
\begin{equation}
    \label{eq:A5ddm1tr}
    \set{\afiveddmc_{\tr}}=
    \Big\{
        \begin{aligned}
        &\tr(T^{a_1}T^{a_3}T^{a_4}T^{a_2}T^{q})\,,\quad
        \tr(T^{a_1}T^{a_4}T^{a_3}T^{a_2}T^{q})\,,\quad
        \tr(T^{a_1}T^{a_3}T^{a_2}T^{a_4}T^{q})\,\\
        &\tr(T^{a_1}T^{a_4}T^{a_2}T^{a_3}T^{q})\,,\quad
        \tr(T^{a_1}T^{a_2}T^{a_3}T^{a_4}T^{q})\,,\quad
        \tr(T^{a_1}T^{a_2}T^{a_4}T^{a_3}T^{q})
        \end{aligned}
    \Big\}\, .
\end{equation}
The propagator matrix $\afiveppm$ has rank two after imposing kinematic conditions.
Restricting to the first two elements of $\set{\afiveddmc_{\tr}}$ gives the full-rank $2\times2$ block
\begin{align}
    \afiveppm \colorxy{1,3,4,2,q}{1,3,4,2,q}
    &=
    \frac{1}{\afiveip{1}} + \frac{1}{\afiveip{2}} + \frac{1}{\afiveip{3}} +
    \frac{1}{\afiveip{4}} + \frac{1}{\afiveip{5}}
    \notag \\
    &=
    \frac{1}{s_{13}} \left( \frac{1}{s_{24}} + \frac{1}{s_{134}} \right) +
    \frac{1}{s_{34}} \left( \frac{1}{s_{134}} + \frac{1}{s_{234}} \right)+
    \frac{1}{s_{24}s_{234}} \, ,
    \notag \\
    \afiveppm \colorxy{1,3,4,2,q}{1,4,3,2,q}
    &= \afiveppm \colorxy{1,4,3,2,q}{1,3,4,2,q}
    \notag \\
    &= - \left( \frac{1}{\afiveip{2}} + \frac{1}{\afiveip{5}} \right)
    = - \frac{1}{s_{34}} \left( \frac{1}{s_{134}} + \frac{1}{s_{234}} \right)\,,
    \notag \\
    \afiveppm \colorxy{1,4,3,2,q}{1,4,3,2,q}
    &=
    \frac{1}{\afiveip{2}} + \frac{1}{\afiveip{5}} + \frac{1}{\afiveip{6}} +
    \frac{1}{\afiveip{7}} + \frac{1}{\afiveip{8}}
    \notag \\
    &=
    \frac{1}{s_{14}} \left( \frac{1}{s_{23}} + \frac{1}{s_{134}} \right) +
    \frac{1}{s_{34}} \left( \frac{1}{s_{134}} + \frac{1}{s_{234}} \right)+
    \frac{1}{s_{23}s_{234}} \, ,
    \label{eq:a5_ppm_elements}
\end{align}
where $\afiveip{i}$ stands for the product of inverse propagators of the $i$-th diagram in Figure~\ref{fig:trphi2_5pt}.
$\ns(\afiveppm)$ gives corresponding BCJ relations
\begin{align}
    (s_{14}+s_{34}) \catracdbq +
    s_{14} \catradcbq
    &=
    \notag \\
    & \hspace{-4cm}
    (s_{123} - m^{2}) \catracbdq \,,
    \notag \\
    -s_{13}s_{24}\catracdbq +
    s_{14}(m^{2}-s_{12}-s_{23})\catradcbq
    &=
    \notag \\
    & \hspace{-4cm}
    (s_{12}-m^{2})(s_{123} - m^2) \catrabcdq \,,
    \label{eq:a5_phi2_bcj}
\end{align}
and another two by exchanging $3 \leftrightarrow 4$.

\subsubsection*{Bleach to $\ff_4(1^\phi, 2^\phi, 3^g, 4^g)$}

\newcommand{\fFourppm}{\Theta^{\ff_4}_{2 \times 2}}

Analogous to the $\ff_3$ case,
applying the bleaching procedure to $\amp_5$
gives a new CK-dual representation for $\ff_4$.
In these solutions,
numerators of topologies $(9)$ to $(15)$ are non-zero,
while their color factors are annihilated by $\blcBH$.
These numerators allow the double copy to be performed
without the introduction of spurious poles.

Through an analysis of the planar diagrams
related to partial amplitudes,
we identify two bleaching-invariant partial amplitudes from $\amp_5$,
\begin{equation}
    \catracdbq =\cf_4(1^\phi, 3^g, 4^g, 2^\phi)\, ,
    \quad
    \catradcbq =\cf_4(1^\phi, 4^g, 3^g, 2^\phi)\, .
    \label{eq:a5_f4_mapping}
\end{equation}
With these two partial amplitudes,
we confirm that the propagator matrix of the four-point FF is a submatrix of the five-point dyed amplitude's propagator matrix,
\begin{gather}
    \fFourppm \colorxy{1,3,4,2,q}{1,3,4,2,q} =
    \afiveppm \colorxy{1,3,4,2,q}{1,3,4,2,q} \, ,
    \notag \\
    \fFourppm \colorxy{1,3,4,2,q}{1,4,3,2,q} =
    \fFourppm \colorxy{1,4,3,2,q}{1,3,4,2,q} =
    \afiveppm\colorxy{1,3,4,2,q}{1,4,3,2,q}\,,
    \notag \\
    \fFourppm \colorxy{1,4,3,2,q}{1,4,3,2,q} =
    \afiveppm \colorxy{1,4,3,2,q}{1,4,3,2,q} \, .
\end{gather}
Equivalently, this block gives the inverse KLT kernel in the chosen
KLT basis.
Therefore, the KLT forms of $\ff_4$ and $\amp_5$
are identical and give the same resulting gravity amplitude.

\subsubsection*{Hidden factorization relations from on-shell limits of BCJ relations}

We now explain the hidden factorization relations of $\ff_4$
in (\ref{eq:f4_hidden_fac}).
Consider the first BCJ relation in
(\ref{eq:a5_phi2_bcj}).
Taking $(s_{123} - m^2)$ to be on-shell gives
\begin{align}
    \left[
        (s_{14}+s_{34}) \catracdbq +
        s_{14} \catradcbq
    \right]_{s_{123} = m^2}
    \hspace{-4cm} &
    \notag \\
    &=
    \ca_4(1^{\phi},3^{g}, 2^{\phi}, -\mathbf{P}_{123}^{\bHBlue}) \times
    \ca_3(\mathbf{P}_{123}^{\bHBlue}, 4^g, q^{\bHBlue})
    \hspace{-4cm} &
    \notag \\
    &=
    \cf_3(1^{\phi},3^{g}, 2^{\phi}) \times
    \ca_3(\mathbf{P}_{123}^{\bHBlue}, 4^g, q^{\bHBlue})
    \,,
    \label{eq:a5_bcj_on_shell_s123}
\end{align}
where the equivalence
$
    \ca_4(1^{\phi},3^{g}, 2^{\phi}, -\mathbf{P}_{123}^{\bHBlue})
    =
    \cf_3(1^{\phi},3^{g}, 2^{\phi})
$
is used.
Using the kinematic conditions and the mapping (\ref{eq:a5_f4_mapping}),
the expression becomes
\begin{align}
    -\left[
        s_{24}\cf_4(1^{\phi},3^g,4^g,2^{\phi})+
        (s_{24}+s_{34})\cf_4(1^{\phi},4^g,3^g,2^{\phi})
    \right]_{s_{123}=q^2}
    &
    \notag \\
    &\hspace{-3cm}
    =\cf_3(1^\phi,3^g,2^\phi)
    \times \ca_3(\mathbf{P}_{123}^{\bH},4^g,q^{\bH})\,,
\end{align}
which reproduces the hidden factorization relation of $\ff_4$ at the
pole $s_{123}=q^2$ in (\ref{eq:f4_hidden_fac}).

The second relation in (\ref{eq:f4_hidden_fac}) follows from taking the on-shell limit of the second BCJ
relation in (\ref{eq:a5_phi2_bcj}) by enforcing
$(s_{12} - m^2) \rightarrow 0$.
The right-hand side factorizes,
\begin{align}
    \lim_{s_{12} \rightarrow m^2}
    (s_{12}-m^{2})(s_{123} - m^2) \catrabcdq
    \hspace{-3cm}&
    \notag \\
    &=
        (s_{123} - m^2)
        \ca_3(1^\phi, 2^\phi, -\mathbf{P}_{12}^{\bHBlue})
        \times
        \ca_4(\mathbf{P}_{12}^{\bHBlue}, 3^g, 4^g, q^{\bHBlue})
    \notag \\
    &=
        (s_{123} - m^2)
        \cf_2(1^\phi, 2^\phi)
        \times
        \ca_4(\mathbf{P}_{12}^{\bHBlue}, 3^g, 4^g, q^{\bHBlue})
    \, .
\end{align}
Therefore,
the relation can be rearranged as
\begin{align}
    -\left[
        s_{13}s_{24}\catracdbq +
        s_{14}s_{23}\catradcbq
    \right]_{s_{12} = m^2}
    \hspace{-7cm} &
    \notag \\
    &=
    \left[
        (s_{123} - m^2)
        \cf_2(1^\phi, 2^\phi)
        \times
        \ca_4(\mathbf{P}_{12}^{\bHBlue}, 3^g, 4^g, q^{\bHBlue})
    \right]_{s_{12} = m^2}
    \, .
    \label{eq:a5_bcj_on_shell_s12}
\end{align}
With $s_{123} - m^2 = s_{13} + s_{23}$ under $s_{12} = m^2$ and the mapping (\ref{eq:a5_f4_mapping}),
the expression can be written as
\begin{align}
    -\left[
        \frac{s_{13}s_{24}}{s_{13}+s_{23}}
        \cf_{4}(1^{\phi},3^g,4^g,2^{\phi})+
        \frac{s_{14}s_{23}}{s_{13}+s_{23}}
        \cf_{4}(1^{\phi},4^g,3^g,2^{\phi})
    \right]_{s_{12}=m^2}
    &
    \notag \\
    &\hspace{-3cm}
    =\cf_{2}(1^\phi,2^\phi)
    \times
    \ca_{4}(\mathbf{P}_{12}^{\bHBlue},3^g,4^g,q^{\bHBlue})\,,
\end{align}
which exactly reproduces the second hidden factorization relation of $\ff_4$ in (\ref{eq:f4_hidden_fac}).
These two examples demonstrate that this on-shell-limit procedure extends to higher-point cases.


  \section{Dyed amplitudes with multiple Higgs particles}
\label{sec:multi_higgs}

In this section,
we extend the dyeing construction to FFs with multiple
local operator insertions,
\begin{align}
  \ff_{\mathcal{O}_1\cdots\mathcal{O}_{n_h},\,n}
  (1,\ldots,n; q_1,\ldots,q_{n_h})
  &=
  \int \left[\prod_{i = 1}^{n_h} d^D x_i\, e^{i q_i \cdot x_i}\right]
  \left\langle 1,\ldots,n \left|
  \mathcal{O}_1(x_1)\cdots\mathcal{O}_{n_h}(x_{n_h})
  \right|0\right\rangle
  \, ,
  \label{eq:multi_operator_ff_def}
\end{align}
which is noted as $\ffnh{n_h}{n}$.
Such multi-operator FFs are a natural extension of ordinary FFs and
also appear as building blocks in on-shell constructions of correlation
functions~\cite{Engelund:2012re,Gao:2013dza,Koster:2016fna,Ahmed:2019yjt}.
The vertical bar separates the on-shell external states from
the off-shell momenta $q_i$ carried by the operator insertions.

In the Higgs-EFT description,
each operator insertion is represented by a color-singlet Higgs
particle $H_i$ carrying momentum $q_i$.
Dyeing then promotes the $n_h$ Higgs fields $H_i$ to adjoint fields $\bH_i$.
The resulting colored amplitudes with $n_h$ dyed Higgs particles are denoted by $\ampnh{n_h}{n + n_h}$.
After bleaching all dyed Higgs legs,
these amplitudes reduce to the corresponding multi-operator FFs,
$\ffnh{n_h}{n}$.
The dyed amplitude $\ampnh{n_h}{n + n_h}$ gives a consistent gravity result and
provides a systematic way to derive
generalized hidden factorization relations for
$\ffnh{n_h}{n}$.
Furthermore,
we identify a new structural feature called
``scalar-ordering'' within $\ampnh{n_h}{n + n_h}$,
which appears at the gauge level and
is inherited on the gravity side through
the double copy.

\subsection{Difficulty of double copy without dyeing}
\label{sec:naive_double_copy_ff}

\newcommand{\ffnhppqq}{\ffnh{2}{2}(1^{\phi}, 2^{\phi} | q_1^{H_1}, q_2^{H_2})}

In this subsection, we perform the double copy construction \emph{without} using the dyeing method, which will identify the difficulties in this picture. 

\begin{figure}[t]
  \begin{center}
    \includegraphics[width=0.4\textwidth]{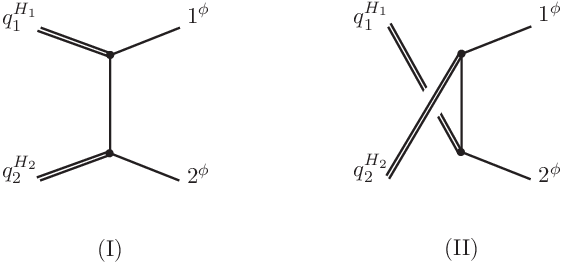}
  \end{center}
  \caption{
    Trivalent topologies of $\ffnhppqq$.
  }
  \label{fig:f2_cubic}
\end{figure}

Consider first the simple two-point FF $\ffnhppqq$.
Its trivalent topologies are shown in Figure~\ref{fig:f2_cubic}.
The FF is given by
\begin{equation}
  \ffnhppqq = \frac{C_\clsA N_\clsA}{s_{1q_1}} + \frac{C_\clsB
  N_\clsB}{s_{1q_2}}\, ,
  \label{eq:F_2_q1_q2}
\end{equation}
with the notation $s_{iq_j} = (p_i + q_j)^2$,
and
\begin{equation}
  \quad C_\clsA = C_\clsB = \delta_{a_\clsA a_\clsB},\ N_\clsA = N_\clsB = 1 \, .
\end{equation}
These numerators inherently satisfy CK duality,
so a direct double-copy operation gives a gravity result.
This result has correct factorization properties and is a consistent gravitational amplitude.
However,
the construction quickly encounters obstructions at higher-point cases.

\subsubsection*{The three-point case}

\newcommand{\ffnhppgqq}{\ffnh{2}{3}(1^\phi, 2^\phi, 3^g | q_1^{H_1}, q_2^{H_2})}
\newcommand{\cffnhacbAB}{\cffnh{2}{3}(1^\phi, 3^g, 2^\phi | q_1^{H_1}, q_2^{H_2})}
\newcommand{\ganhpphqq}{\ganh{2}{5}(1^\phi, 2^\phi, 3^h, q_1^{H_1}, q_2^{H_2})}

\begin{figure}[t]
  \begin{center}
    \includegraphics[width=0.5\textwidth]{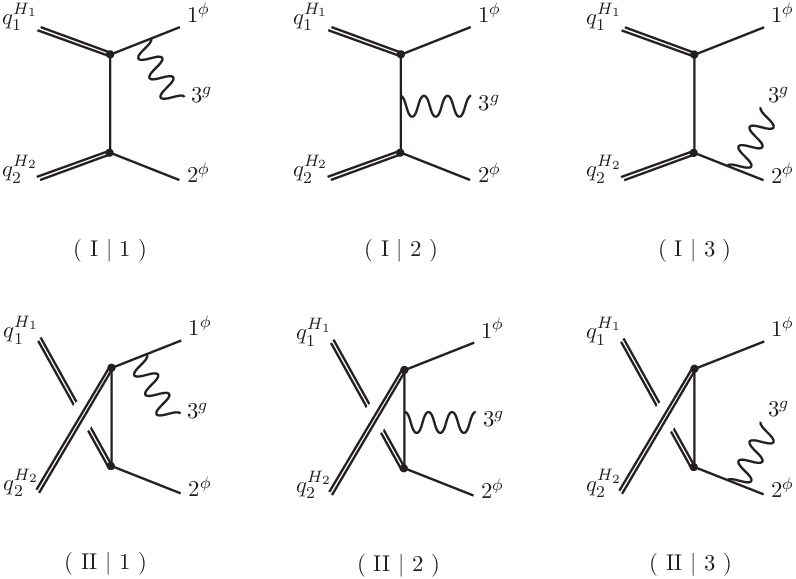}
  \end{center}
  \caption{
    Trivalent topologies of $\ffnhppgqq$.
    The labels class-$\clsA$ and class-$\clsB$ denote distinct topological classes.
  }
  \label{fig:f3_trivalent}
\end{figure}

Figure~\ref{fig:f3_trivalent} illustrates the six Feynman diagrams
contributing to $\ffnhppgqq$,
partitioned into two distinct classes, class-$\clsA$ and class-$\clsB$.
Each class contains three diagrams, indexed by $\{1, 2, 3\}$,
and is mapped to the other class under the exchange of $q_1 \leftrightarrow q_2$.
The full-color FF is written as
\begin{equation}
  \ffnhppgqq = \sum_{S,\, i}
  \frac{C_{S|i} N_{S|i}}{D_{S|i}},\quad S \in \{\clsA, \clsB\},\ i
  \in \{1, 2, 3\} \, .
  \label{eq:f3_q1_q2}
\end{equation}
Quantities associated with the $i$-th diagram of class $S$ are denoted by
the subscript `$S\, |\, i$'.

Because the operator insertions are color-singlets,
the color factors are all identical in this case:
\begin{equation}
  C_{\clsA|1} = C_{\clsA|2} = C_{\clsA|3} = C_{\clsB|1} = C_{\clsB|2}
  = C_{\clsB|3} = f^{a_1 a_3 a_2}
  \equiv \tilde{C} \, .
  \label{eq:f3_cfs}
\end{equation}
Imposing naive CK duality requires only a single master numerator:
\begin{equation}
  N_{\clsA|1}^{\CK} = N_{\clsA|2}^{\CK} = N_{\clsA|3}^{\CK} = N_{\clsB|1}^{\CK}
  = N_{\clsB|2}^\CK = N_{\clsB|3}^{\CK} \equiv \tilde{N}^\CK \, .
  \label{eq:f3_ns}
\end{equation}
With (\ref{eq:f3_cfs}) and (\ref{eq:f3_ns}),
the FF can be written as
\begin{equation}
  \ffnhppgqq = \tilde{C} \tilde{N}^\CK
  \sum_{S,\, i} \frac{1}{D_{S|i}} = \tilde{C} \tilde{N}^\CK
  \frac{\mathcal{D}}{\prod_{S,\, i} D_{S|i}}\, ,
  \label{eq:f3_expanded}
\end{equation}
where $\mathcal{D}$ is a sum over products of $D_i$,
\begin{equation}
  \mathcal{D} = \sum_{S, i} \frac{D_{\clsA|1} D_{\clsA|2}
  D_{\clsA|3}D_{\clsB|1} D_{\clsB|2} D_{\clsB|3}}{D_{S|i}}\, .
  \label{eq:f3_spurious_pole}
\end{equation}
with $D_{\clsA|1} = s_{13}s_{2q_2}$, etc.
Alternatively, (\ref{eq:f3_q1_q2}) can be formulated in terms of color-ordered FFs,
\begin{equation}
  \ffnhppgqq = \tilde{C} \cffnhacbAB \, ,
  \label{eq:f3_color_ordered}
\end{equation}
where $\cffnhacbAB$ is the color-ordered
FF associated with the color factor $\tilde C$,
which is identical to the color-ordered
FF in the single-trace basis
$\tr(T^{a_1}T^{a_3} T^{a_2})$.
Comparing (\ref{eq:f3_expanded}) with (\ref{eq:f3_color_ordered}) gives
the master numerator,
\begin{equation}
  \tilde{N}^\CK = \cffnhacbAB
  \frac{\prod_{S,\, i} D_{S|i}}{\mathcal{D}} \, .
  \label{eq:f3_nck_sol}
\end{equation}
The CK-satisfying numerator $\tilde{N}^{\CK}$ introduces $\mathcal{D}$
as a spurious pole,
which cancels in the full-color expression for $\ffnh{2}{3}$ (\ref{eq:f3_expanded}).
However, after double copy,
this spurious pole manifests as a real singularity in the result,
\begin{equation}
  \mathcal{G}_3 = (\tilde{N}^\CK)^2
  \frac{\mathcal{D}}{\prod_{S,\, i} D_{S|i}}
  = \left(\cffnhacbAB\right)^2
  \frac{\prod_{S,\, i} D_{S|i}}{\mathcal{D}}\, ,
\end{equation}
which is unphysical.
Moreover, for a consistent gravity amplitude $\ganhpphqq$,
it must correctly factorize on the new physical poles
$(s_{3q_1} - q_1^2)$ and $(s_{3q_2} - q_2^2)$.
However, $\mathcal{G}_3$ fails to factorize correctly on these poles.
Therefore, it is clear that $\mathcal{G}_3$ is inconsistent with the expected gravity amplitude $\ganhpphqq$.

\subsubsection*{The three-point case with extended CK masters}

\newcommand{\ffnhppgA}{\ffnh{2}{3,\clsA}}
\newcommand{\ffnhppgB}{\ffnh{2}{3,\clsB}}
\newcommand{\cffnhpgpBAI}{\cffnh{2}{3}(1^\phi, 3^g, 2^\phi, q_2^{H_2}, q_1^{H_1})}
\newcommand{\cffnhpgpABII}{\cffnh{2}{3}(1^\phi, 3^g, 2^\phi, q_1^{H_1}, q_2^{H_2})}
\newcommand{\vecffnhThree}{\vec{\cf}_{3}^{\,[2]}}
\newcommand{\ppmffnhThree}{\Theta_{2 \times 2}^{\ffnh{2}{3}}}
\newcommand{\kltffnhThree}{\klt_{2 \times 2}^{\ffnh{2}{3}}}

A refined analysis is to decompose $\ffnh{2}{3}$ into two distinct
gauge-invariant subsectors,
$\ffnhppgA$ and $\ffnhppgB$,
\begin{align}
  \ffnhppgA = \sum_{i = 1}^3 \frac{C_{\clsA|i}
  N_{\clsA|i}}{D_{\clsA|i}}, \quad
  \ffnhppgB = \sum_{i = 1}^3 \frac{C_{\clsB|i}
  N_{\clsB|i}}{D_{\clsB|i}} \, .
  \label{eq:f3_cls_AB}
\end{align}
This classification is validated by independently evaluating $\ffnhppgA$ and $\ffnhppgB$
via standard Feynman rules and confirming their individual gauge invariance.

Correspondingly, we relax the color relations by dividing color factors into two sectors,
\begin{equation}
  C_{\clsA|1} = C_{\clsA|2} = C_{\clsA|3} \equiv \tilde{C}_\clsA\, ,
  \quad
  C_{\clsB|1} = C_{\clsB|2} = C_{\clsB|3} \equiv \tilde{C}_\clsB \, .
  \label{eq:f3_cfs_AB}
\end{equation}
By imposing CK duality,
two CK masters $\set{\tilde{N}^{\CK}_\clsA, \tilde{N}^{\CK}_\clsB}$
are needed instead of only one master in (\ref{eq:f3_ns}),
\begin{equation}
  N_{\clsA|1}^{\CK} = N_{\clsA|2}^{\CK} = N_{\clsA|3}^{\CK} \equiv
  \tilde{N}^{\CK}_\clsA\, ,
  \quad
  N_{\clsB|1}^{\CK} = N_{\clsB|2}^{\CK} = N_{\clsB|3}^{\CK} \equiv
  \tilde{N}^{\CK}_\clsB \, .
  \label{eq:f3_numerator_class}
\end{equation}
Combining (\ref{eq:f3_cfs_AB}), (\ref{eq:f3_numerator_class}) and
(\ref{eq:f3_cls_AB}),
and comparing with color-ordered FFs gives
\begin{equation}
  \vecffnhThree = \ppmffnhThree \cdot \vec{N}_3\, ,
\end{equation}
with
\begin{equation}
  \vec{\cf}^{[2]}_3 =
  \begin{bmatrix}
    \cffnhpgpBAI \\ \cffnhpgpABII
  \end{bmatrix}
  ,\quad
  \vec{N}_3 =
  \begin{bmatrix}
    \tilde{N}^{\CK}_\clsA\\
    \tilde{N}^{\CK}_\clsB
  \end{bmatrix}\, .
  \label{eq:f3_vec_f3_n3}
\end{equation}
Here,
$\cffnhpgpBAI$ is the color-ordered FF in the trace basis
$\tr(T^{a_1} T^{a_3} T^{a_2})$.
It belongs to the class-$\clsA$ sector.
We omit the vertical bar ``$|$'' notation to explicitly show
the integrated planar ordering of external states and operators,
matching the topologies in Figure~\ref{fig:f3_trivalent}.
The propagator matrix assumes a diagonal, full-rank form,
\begin{equation}
  \ppmffnhThree =
  \begin{bmatrix}
    \frac{1}{D_{\clsA|1}} + \frac{1}{D_{\clsA|2}} + \frac{1}{D_{\clsA|3}} & 0 \\
    0 & \frac{1}{D_{\clsB|1}} + \frac{1}{D_{\clsB|2}} + \frac{1}{D_{\clsB|3}}
  \end{bmatrix}\, .
  \label{eq:f3_propagator_matrix}
\end{equation}
The KLT kernel $\kltffnhThree$ is its inverse,
\begin{equation}
  \kltffnhThree = \left( \ppmffnhThree \right)^{-1} =
  \begin{bmatrix}
    \frac{D_{\clsA|1} D_{\clsA|2} D_{\clsA|3}}{D_{\clsA|1}
      D_{\clsA|2} + D_{\clsA|1} D_{\clsA|3}
    + D_{\clsA|2} D_{\clsA|3}} & 0 \\
    0 & \frac{D_{\clsB|1} D_{\clsB|2} D_{\clsB|3}}{D_{\clsB|1}
      D_{\clsB|2} + D_{\clsB|1}
    D_{\clsB|3} + D_{\clsB|2} D_{\clsB|3}}
  \end{bmatrix}\, .
\label{eq:f3_klt_kernel}
\end{equation}
Implementing the KLT double copy generates the associated gravity quantity,
\begin{equation}
  \tilde{\mathcal{G}}_3 =
  \left( \vecffnhThree \right)^T \cdot
  \kltffnhThree \cdot
  \vecffnhThree \, .
\label{eq:f3_klt}
\end{equation}
Inserting (\ref{eq:f3_vec_f3_n3}) and (\ref{eq:f3_klt_kernel}) into
(\ref{eq:f3_klt}) shows that $\tilde{\mathcal{G}}_3$ can also be divided into
two parts,
\begin{equation}
  \tilde{\mathcal{G}}_3 = \tilde{\mathcal{G}}_{3,\clsA} +
  \tilde{\mathcal{G}}_{3,\clsB}\, ,
\end{equation}
where
\begin{align}
  \tilde{\mathcal{G}}_{3,\clsA} &=
  \left( \cffnhpgpBAI \right)^2
  \frac{D_{\clsA|1} D_{\clsA|2} D_{\clsA|3}}{D_{\clsA|1} D_{\clsA|2} +
    D_{\clsA|1} D_{\clsA|3} +
  D_{\clsA|2} D_{\clsA|3}} \, ,
  \notag \\
  \tilde{\mathcal{G}}_{3,\clsB} &=
  \left( \cffnhpgpABII \right)^2 \frac{D_{\clsB|1} D_{\clsB|2}
  D_{\clsB|3}}{D_{\clsB|1} D_{\clsB|2} +
    D_{\clsB|1} D_{\clsB|3} +
  D_{\clsB|2} D_{\clsB|3}} \, .
  \label{eq:g3_ab}
\end{align}
However,
$\tilde{\mathcal{G}}_3$
contains two unphysical poles
\begin{equation}
  D_{\clsA|1} D_{\clsA|2} + D_{\clsA|1} D_{\clsA|3} + D_{\clsA|2}
  D_{\clsA|3},\quad D_{\clsB|1}
  D_{\clsB|2} + D_{\clsB|1} D_{\clsB|3} + D_{\clsB|2} D_{\clsB|3}  \, ,
  \label{eq:f3_double_pole}
\end{equation}
and $\tilde{\mathcal{G}}_3$ also fails to factorize on the expected new poles $(s_{3q_1} - q_1^2)$ and $(s_{3q_2} - q_2^2)$.
Hence, $\tilde{\mathcal{G}}_3$
does not reproduce the correct physical gravity amplitude associated with $\ffnh{2}{3}$.

 \subsection{Dyeing Higgs-EFT}

\newcommand{\YMSHi}[1]{\text{YMsH}_{#1}}
\newcommand{\dYMSHi}[1]{\overline{\text{YMsH}}_{#1}}
\newcommand{\dHiggsi}[1]{\overline{\text{Higgs}}_{#1}}
\newcommand{\LaYMSHi}[1]{\La^{\YMSHi{#1}}}
\newcommand{\LaHiggsi}[1]{\La^{\text{Higgs}_{#1}}}
\newcommand{\LaHppi}[1]{\La^{H_{#1} \phi \phi}}
\newcommand{\LadYMSHi}[1]{\La^{\dYMSHi{#1}}}
\newcommand{\LadHiggsi}[1]{\La^{\dHiggsi{#1}}}
\newcommand{\LadHppi}[1]{\La^{\bar{H}_{#1} \phi \phi}}
\newcommand{\mH}[1]{m_{H_{#1}}}
\newcommand{\lH}[1]{\lambda_{H_{#1}}}
\newcommand{\LaMixdYMSH}[2]{\La^{\dYMSHi{#1}^{#2}}}

The above discussions show that it is difficult to generalize the picture of \cite{Lin:2021pne} to form factors with multi-operator insertions, due to the appearance of complicated unphysical poles.
As we will show, the dyeing theory naturally solve these problems.
Before presenting explicit constructions, we first discuss the setup of the dyed theory. This is a straightforward generalization of Section~\ref{sec:dyedphi2}, except that now we have multiple Higgs particles with different masses.

FFs with multiple $\tr(\phi^2)$ operator insertions can be interpreted as amplitudes
of a generalized Higgs-EFT $\YMSHi{n_h}$ theory,
\begin{equation}
  \LaYMSHi{n_h}_{\tr(\phi^2)} = \LaYMS +
  \sum_{i=1}^{n_h} \left(\LaHiggsi{i} + \LaHppi{i}\right) \, ,
\end{equation}
with
\begin{equation}
  \LaHiggsi{i} =
  \frac{1}{2} (\partial_\mu H_i)(\partial^\mu H_i) - \frac{1}{2} \mH{i}^{2} H_i^2 \, ,
  \quad
  \LaHppi{i} = \lH{i} H_i \phi^a \phi^a \,.
\end{equation}
Since different operators in general carry different off-shell momenta $q_i$, we introduce multiple Higgs fields with distinct masses $\mH{i}^2 = q_i^2$. Here $n_h$ counts the number of inserted operators.

In the dyed theory, one promotes all Higgs fields $H_i$ to
the adjoint representation of the gauge group,
\begin{equation}
  \bH_i = H_i^a T^a \, ,
\end{equation}
giving the $\dYMSHi{n_h}$ theory,
\begin{equation}
  \LadYMSHi{n_h}_{\tr(\phi^2)} = \LaYMS +
  \sum_{i = 1}^{n_h} \left(\LadHiggsi{i} + \LadHppi{i}\right)\, ,
\end{equation}
with the dyed components defined as
\begin{equation}
  \LadHiggsi{i} =
  \frac{1}{2} (D_\mu H_i)(D^\mu H_i)
  - \frac{1}{2} \mH{i}^{2} H_i^a H_i^a \, ,
  \quad
  \LadHppi{i} = \lH{i} d^{abc} H_{i}^{a} \phi^b \phi^c \,.
\end{equation}
The bleaching procedure maps them back to color singlets
\begin{equation}
  \blc_{\bH_i} \left( \LadHiggsi{i} \right) = \LaHiggsi{i}\, ,
  \quad
  \blc_{\bH_i} \left( \LadHppi{i} \right) = \LaHppi{i}\, .
\end{equation}
To recover the original $\LaYMSHi{n_h}_{\tr(\phi^2)}$ theory,
the bleaching procedure can be applied to the complete set of
dyed Higgs fields $\set{\bH_i}:$\footnote{
  One may also bleach only a subset $\Lambda \subseteq \set{\bH_i}$,
  which gives an intermediate partially mixed theory.
}
\begin{equation}
  \left( \prod_{i = 1}^{n_h} \blc_{\bH_i} \right)
  \left( \LadYMSHi{n_h}_{\tr(\phi^2)} \right) =
  \LaYMSHi{n_h}_{\tr(\phi^2)}\, .
\end{equation}

A corresponding BAS theory can be used to calculate the
propagator matrix in amplitudes with multiple Higgs external states,
where we need kinetic and cubic terms that match the amplitude structures,
\begin{equation}
  \La^{\text{BAS}}_{\dYMSHi{n_h},{\tr(\phi^2)}} =
  \LaBAS_{\text{YMs}}
  + \sum_{i = 1}^{n_h}
  \left(
    \LaBAS_{\varphi_{\bH_i}} +
    \LaBAS_{\bH_i\phi\phi}
  \right)\, ,
\end{equation}
with
\begin{equation}
    \LaBAS_{\varphi_{\bH_i}} =
    \frac{1}{2}(\partial_\mu \varphi_{\bH_i}^{a\ba}) (\partial^\mu \varphi_{\bH_i}^{a\ba})
    - \frac{1}{2} \mH{i}^2 \varphi_{\bH_i}^{a\ba}\varphi_{\bH_i}^{a\ba} \, ,
    \quad
    \LaBAS_{\bH_i\phi\phi} =
    \frac{\lambda_{H_i\phi\phi}}{3!} d^{abc}d^{\ba\bb\bc}
    \pp^{a\ba} \pp^{b\bb} \varphi_{\bH_i}^{c\bc} \, .
\end{equation}
 
\subsection{The two-Higgs five-point dyed amplitude}
\label{sub:2_higgs_5pt} 

\newcommand{\dampPPGHH}{\ampnh{2}{5}({1^\phi, 2^\phi, 3^g, q_1^{\bHBlueIdx{1}}, q_2^{\bHOrangeIdx{2}}})}
\newcommand{\cddmAppghh}{\set{C_{\text{DDM}}^{\ampnh{2}{5}}}}

Now we apply the dyed framework to the five-point amplitude $\dampPPGHH$ of
the $\LadYMSHi{2}_{\tr(\phi^2)}$ theory, which is the dyeing version of $\ffnhppgqq$.

\subsubsection*{The propagator matrix}

\newcommand{\ppmAppghh}{\Theta_{6 \times 6}^{\ampnh{2}{5}}}
\newcommand{\ppmAppghhI}{\Theta_{3 \times 3}^{\ampnh{2}{5, \clsA}}}
\newcommand{\ppmAppghhII}{\Theta_{3 \times 3}^{\ampnh{2}{5, \clsB}}}

\begin{figure}[t]
  \begin{center}
    \includegraphics[width=0.85\textwidth]{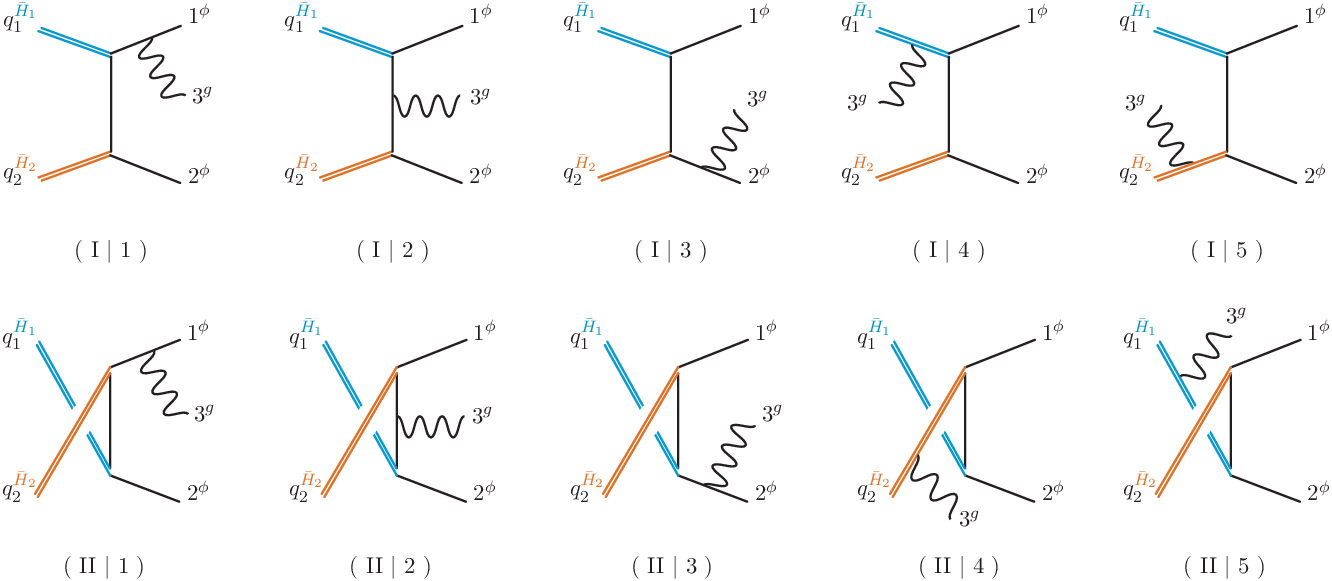}
  \end{center}
  \caption{
    Trivalent topologies of $\dampPPGHH$.
    To distinguish different Higgs,
    $\bHBlueIdx{1}$ is depicted with a blue double line,
    while $\bHOrangeIdx{2}$ is shown in orange.}
  \label{fig:a5_trivalent}
\end{figure}

The ten relevant Feynman diagrams are presented in Figure~\ref{fig:a5_trivalent}.
We organize the ten diagrams into two gauge-invariant groups,
class-$\clsA$ and class-$\clsB$.
We will comment on this grouping from the color factor analysis below.
The amplitude is written as a sum over all diagrams,
\begin{equation}
  \dampPPGHH =
  \sum_{i = 1}^{5} \frac{C_{\clsA|i} N_{\clsA|i}}{D_{\clsA|i}}
  + \sum_{i = 1}^{5} \frac{C_{\clsB|i} N_{\clsB|i}}{D_{\clsB|i}} \, .
  \label{eq:a5_diag}
\end{equation}
The color factors for the class-$\clsA$ diagrams are
\begin{gather}
  C_{\clsA|1} = f^{a_1 a_3 b_1} d^{b_1 b_2 q_1} d^{b_2 a_2 q_2},\
  C_{\clsA|2} = f^{b_1 a_3 b_2} d^{a_1 b_1 q_1} d^{b_2 a_2 q_2},\
  C_{\clsA|3} = f^{b_2 a_3 a_2} d^{a_1 b_1 q_1} d^{q_2 b_1 b_2}, \notag \\
  C_{\clsA|4} = f^{a_3 q_1 b_1} d^{a_1 b_2 b_1} d^{b_2 a_2 q_2},\
  C_{\clsA|5} = f^{a_3 b_2 q_2} d^{a_1 b_1 q_1} d^{b_2 b_1 a_2}\, ,
\end{gather}
where $b_{1, 2}$ represent the color indices of the internal lines.
The color factors for the class-$\clsB$ diagrams are obtained under
the exchange $q_1 \leftrightarrow q_2$.
Using the corresponding color relations,
the color factors within each group satisfy
\begin{gather}
  C_{\clsA|4} = C_{\clsA|1} - C_{\clsA|2}, \quad C_{\clsA|5} =
  -C_{\clsA|2} + C_{\clsA|3}\, , \notag \\
  C_{\clsB|4} = C_{\clsB|1} - C_{\clsB|2}, \quad C_{\clsB|5} =
  -C_{\clsB|2} + C_{\clsB|3}\, .
  \label{eq:a5_color_relations}
\end{gather}
A key observation is that the color relations
in~(\ref{eq:a5_color_relations}) respect the above grouping:
the relations for class-$\clsA$ diagrams involve only class-$\clsA$ color factors,
and likewise for class-$\clsB$.
In other words,
the ten color factors decompose into two \emph{closed} subsets under these color relations.
Indeed, explicit computation via Feynman rules confirms
that each class is independently gauge invariant,
\begin{equation}
  \ampnh{2}{5, \clsA} \equiv \sum_{i = 1}^{5} \frac{C_{\clsA|i}
  N_{\clsA|i}}{D_{\clsA|i}}\, , \quad
  \ampnh{2}{5, \clsB} \equiv \sum_{i = 1}^{5} \frac{C_{\clsB|i}
  N_{\clsB|i}}{D_{\clsB|i}} \, .
\end{equation}

Selecting the independent elements from each class,
we define the color basis as
\begin{equation}
  \cddmAppghh =
  \set{
    C_{\clsA|1},\ C_{\clsA|2},\ C_{\clsA|3}, \
    C_{\clsB|1},\ C_{\clsB|2},\ C_{\clsB|3}
  }\, .
  \label{eq:a52_ddmbasis}
\end{equation}
The propagator matrix is extracted from
the dual BAS theory with respect to the ordered basis in (\ref{eq:a52_ddmbasis}),
resulting in a block-diagonal matrix,
\begin{gather}
  \ppmAppghh =
  \begin{bmatrix}
    \ppmAppghhI & 0 \\
    0 & \ppmAppghhII
  \end{bmatrix}\, ,\\
  \ppmAppghhI =
  \begin{bmatrix}
    \frac{1}{D_{\clsA|1}} + \frac{1}{D_{\clsA|4}} &
    -\frac{1}{D_{\clsA|4}} & 0 \\
    -\frac{1}{D_{\clsA|4}} & \frac{1}{D_{\clsA|2}} + \frac{1}{D_{\clsA|4}} +
    \frac{1}{D_{\clsA|5}} & -\frac{1}{D_{\clsA|5}} \\
    0 & -\frac{1}{D_{\clsA|5}} & \frac{1}{D_{\clsA|3}} + \frac{1}{D_{\clsA|5}}
  \end{bmatrix}, \quad
  \ppmAppghhII =
  \left. \ppmAppghhI \right|_{\clsA \rightarrow \clsB}\, .
\end{gather}
Crucially, $\ppmAppghh$ is singular.
Its rank is reduced to
\begin{equation}
  \rank \ppmAppghh =
    \rank \ppmAppghhI +
    \rank \ppmAppghhII
    = 2 + 2 = 4\, .
\end{equation}
This rank reduction has two important consequences at the gauge and gravity levels:

First, at the gauge-theory level,
the matrix singularity shows the existence of two independent BCJ relations
among the partial amplitudes ---
one governing $\ampnh{2}{5,\clsA}$ and the other governing $\ampnh{2}{5,\clsB}$.
Second, at the gravity level,
the vanishing determinant of $\ppmAppghh$ eliminates
the complicated unphysical poles that previously obstructed the naive double copy procedure.

\subsubsection*{KLT double copy}

\newcommand{\GAppghh}{\ganh{2}{5}}
\newcommand{\GAppghhI}{\ganh{2}{5,\clsA}}
\newcommand{\GAppghhII}{\ganh{2}{5,\clsB}}
\newcommand{\vecAppghh}{\vec{\ca}_{5}^{\, [2]}}
\newcommand{\kltAppghh}{\klt^{\ampnh{2}{5}}_{4 \times 4}}
\newcommand{\kltAppghhI}{\klt^{\ampnh{2}{5, \clsA}}_{2 \times 2}}
\newcommand{\kltAppghhII}{\klt^{\ampnh{2}{5, \clsB}}_{2 \times 2}}

The KLT double copy of $\ampnh{2}{5}$ generates a gravity amplitude $\ganhpphqq$:
\begin{equation}
  \GAppghh =
  \left( \vecAppghh \right)^T \cdot
  \kltAppghh \cdot
  \vecAppghh \, ,
\end{equation}
with
\begin{equation}
  \vecAppghh =
  \begin{bmatrix}
      \campnh{2}{5,\clsA|1},\ &\ \campnh{2}{5,\clsA|2},\
    &\ \campnh{2}{5,\clsB|1},\ &\ \campnh{2}{5,\clsB|2}
  \end{bmatrix}^T \, ,
  \label{eq:vecA52}
\end{equation}
where $\campnh{2}{5, S|i}$ represents the partial amplitude
associated with the color basis element $C_{S|i}$ from (\ref{eq:a52_ddmbasis}).
The KLT kernel is constructed by
inverting a full-rank $4 \times 4$ submatrix of $\ppmAppghh$,
\begin{equation}
  \kltAppghh =
  \begin{bmatrix}
    \kltAppghhI & 0 \\
    0 & \kltAppghhII
  \end{bmatrix}\, ,
  \quad
  \kltAppghhI =
  \begin{bmatrix}
    \frac{1}{D_{\clsA|1}} + \frac{1}{D_{\clsA|4}} & -\frac{1}{D_{\clsA|4}} \\
    -\frac{1}{D_{\clsA|4}} & \frac{1}{D_{\clsA|2}} + \frac{1}{D_{\clsA|4}} +
    \frac{1}{D_{\clsA|5}}
  \end{bmatrix}^{-1}\, ,
  \quad
  \kltAppghhII = \left. \kltAppghhI \right|_{
      \clsA \rightarrow \clsB} \, .
\end{equation}
The block-diagonalization of $\kltAppghh$,
divides the resulting gravity amplitude into
two diffeomorphism invariant subsectors:
\begin{equation}
  \GAppghh
  = \GAppghhI + \GAppghhII \,,
\end{equation}
where
\begin{equation}
  \GAppghhI=
    \begin{bmatrix}
      \campnh{2}{5,\clsA|1}, & \campnh{2}{5,\clsA|2}
    \end{bmatrix} \cdot \kltAppghhI \cdot
    \begin{bmatrix}
      \campnh{2}{5,\clsA|1} \\ \campnh{2}{5,\clsA|2}
    \end{bmatrix} , \qquad
    \GAppghhII = \GAppghhI \Big|_{ (\clsA \rightarrow \clsB)} \, .
\end{equation}
We have verified that this result has consistent poles and factorization properties,
confirming its validity as the tree-level gravity amplitude.

\subsubsection*{BCJ relations and hidden factorizations}

The rank of the propagator matrix implies the existence of
two BCJ relations of the color-ordered amplitudes $\campnh{2}{5}$.
Because the class-$\clsA$ and class-$\clsB$ sectors are symmetric under $\clsA \leftrightarrow \clsB$,
we restrict our analysis to the class-$\clsA$ sector.
Analyzing the left null space of the propagator matrix gives the BCJ relation
\begin{align}
-s_{13} \campnh{2}{5,\clsA|1}
  + (-s_{3q_1} - s_{13} + \mH{1}^2) \campnh{2}{5,\clsA|2}
  + s_{23} \campnh{2}{5,\clsA|3} = 0 \, .
  \label{eq:a5_bcj}
\end{align}
To connect this relation to the hidden factorizations,
we map the color-ordered amplitudes in (\ref{eq:a5_bcj}) to the single-trace basis,
\begin{gather}
  \campnh{2}{5,\clsA|1} =
  - \campnh{2}{5}(3^{g}, 1^\phi, 2^\phi, q_2^{\bHOrangeIdx{2}}, q_1^{\bHBlueIdx{1}})\, ,
  \quad
  \campnh{2}{5, \clsA|3} = - \campnh{2}{5}(1^\phi, 2^\phi, 3^{g}, q_2^{\bHOrangeIdx{2}},
  q_1^{\bHBlueIdx{1}}) \, ,
  \notag \\
  \campnh{2}{5,\clsA|2} = \campnh{2}{5}(1^\phi, 2^\phi, 3^{g}, q_2^{\bHOrangeIdx{2}},
  q_1^{\bHBlueIdx{1}}) + \campnh{2}{5}(1^\phi, 3^{g}, 2^\phi, q_2^{\bHOrangeIdx{2}} ,q_1^{\bHBlueIdx{1}})
  + \campnh{2}{5}(3^{g}, 1^\phi, 2^\phi, q_2^{\bHOrangeIdx{2}}, q_1^{\bHBlueIdx{1}})
  \, .
\end{gather}
Therefore (\ref{eq:a5_bcj}) can be rewritten as
\begin{align}
  - (s_{3q_1} - \mH{1}^2)
  \campnh{2}{5}(3^g, 1^\phi, 2^\phi, q_2^{\bHOrangeIdx{2}}, q_1^{\bHBlueIdx{1}}) +
  (s_{3q_2} - \mH{2}^2)
  \campnh{2}{5}(1^\phi, 2^\phi, 3^g, q_2^{\bHOrangeIdx{2}}, q_1^{\bHBlueIdx{1}})
  \hspace{-7.5cm} &
  \notag \\
  & - (s_{3q_1} + s_{13} - \mH{1}^2)
  \campnh{2}{5}(1^\phi, 3^g, 2^\phi, q_2^{\bHOrangeIdx{2}}, q_1^{\bHBlueIdx{1}}) = 0\, .
  \label{eq:a5_bcj_tr}
\end{align}
As in the single-operator FF cases, we find a direct
equivalence between the color-ordered amplitude and FF (i.e., bleaching invariance),
\begin{equation}
  \campnh{2}{5}(1^\phi, 3^g, 2^\phi, q_2^{\bHOrangeIdx{2}}, q_1^{\bHBlueIdx{1}}) =
  \cffnh{2}{3}(1^\phi, 3^g, 2^\phi, q_2^{H_2}, q_1^{H_1})\, .
  \label{eq:a5_eq_f3}
\end{equation}

Now we take $(s_{3q_1} - \mH{1}^2)$ to be on-shell, which factorizes the first term of (\ref{eq:a5_bcj_tr})
into two tree blocks,
\begin{equation}
  \lim_{s_{3q_1} \rightarrow \mH{1}^2}
  (s_{3q_1} - \mH{1}^2)
  \times
  \campnh{2}{5}(
    3^g, 1^\phi, 2^\phi, q_2^{\bHOrangeIdx{2}},
    q_1^{\bHBlueIdx{1}})
  =
  \ca_3(q_1^{\bHBlueIdx{1}}, 3^g, -\mathbf{P}_{3q_1}^{\bHBlueIdx{1}}) \times
  \campnh{2}{4}(1^\phi, 2^\phi,
  q_2^{\bHOrangeIdx{2}}, \mathbf{P}_{3q_1}^{\bHBlueIdx{1}})\,.
\end{equation}
Under the limit, (\ref{eq:a5_bcj_tr}) can be written as
\begin{align}
  & \quad\ \ca_3(q_1^{\bHBlueIdx{1}}, 3^g, -\mathbf{P}_{3q_1}^{\bHBlueIdx{1}}) \times
  \campnh{2}{4}(1^\phi, 2^\phi,
  q_2^{\bHOrangeIdx{2}}, \mathbf{P}_{3q_1}^{\bHBlueIdx{1}}) \notag \\
  & =
  \left[(s_{3q_2} - \mH{2}^2) \campnh{2}{5}(1^\phi, 2^\phi, 3^g, q_2^{\bHOrangeIdx{2}}, q_1^{\bHBlueIdx{1}})
    - s_{13} \campnh{2}{5}(1^\phi, 3^g, 2^\phi, q_2^{\bHOrangeIdx{2}}, q_1^{\bHBlueIdx{1}})
  \right]_{s_{3q_1} = \mH{1}^2}\, .
  \label{eq:a5_first_on_shell_limit}
\end{align}
Or graphically,
\begin{equation}
  \begin{gathered} {\includegraphics[height=3.2cm]{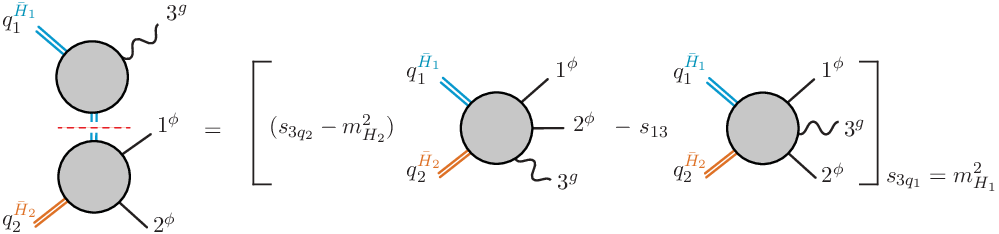} }
  \end{gathered} \,.
\end{equation}
The red dotted line marks the on-shell channel, and the gray blobs denote planar amplitudes.
Although another pole remains in (\ref{eq:a5_first_on_shell_limit}), it still exhibits
the hidden factorization structure.
Interpreting the planar subamplitudes in the corresponding bleached channels,
(\ref{eq:a5_first_on_shell_limit}) can be written as
\begin{align}
  & \quad\ \ca_3(q_1^{\bHBlueIdx{1}}, 3^g, -\mathbf{P}_{3q_1}^{\bHBlueIdx{1}})
  \times
  \cffnh{2}{2}(1^\phi, 2^\phi, q_2^{H_2}, \mathbf{P}^{H_1}_{3q_1}) \notag \\
  & =
  \left[(s_{3q_2} - \mH{2}^2)
    \cffnh{\set{2}}{3}(1^\phi, 2^\phi, 3^g, q_2^{\bHOrangeIdx{2}} | q_1^{H_1})
    - s_{13} \cffnh{2}{3}(1^\phi, 3^g, 2^\phi, q_2^{H_2}, q_1^{H_1})
  \right]_{s_{3q_1} = \mH{1}^2}\, ,
\end{align}
which is a mixed factorization relation between
multi-operator and single-operator FFs.

Imposing the second on-shell condition
$(s_{3q_2} - \mH{2}^2) \to 0$
in addition to the first one $(s_{3q_1} - \mH{1}^2) \to 0$,
the expression further reduces to\footnote{
In this example the two limits commute, so the ordering is not important.
However, the ordering of taking on-shell limits will become relevant in the higher-point case discussed later.
}
\begin{align}
  & \quad\ \left.s_{13} \campnh{2}{5}(1^\phi, 3^g, 2^\phi, q_2^{\bHOrangeIdx{2}}, q_1^{\bHBlueIdx{1}})
  \right|_{s_{3q_2} = \mH{2}^2}^{s_{3q_1} = \mH{1}^2}
  \notag \\
  & = - \left[
    \ca_3(q_1^{\bHBlueIdx{1}}, 3^g, -\mathbf{P}_{3q_1}^{\bHBlueIdx{1}})
    \times
    \campnh{2}{4}(1^\phi, 2^\phi, q_2^{\bHOrangeIdx{2}}, \mathbf{P}_{3q_1}^{\bHBlueIdx{1}})
  \right]_{s_{3q_2} = \mH{2}^2} \notag \\
  & \qquad \qquad +
  \left[
    \ca_3(-\mathbf{P}^{\bHOrangeIdx{2}}_{3q_2}, 3^g, q_2^{\bHOrangeIdx{2}})
    \times
    \campnh{2}{4}(1^\phi, 2^\phi, \mathbf{P}^{\bHOrangeIdx{2}}_{3q_2}, q_1^{\bHBlueIdx{1}})
  \right]_{s_{3q_1} = \mH{1}^2}\, .
  \label{eq:a5_second_on_shell_limit}
\end{align}
Here and throughout, when multiple on-shell conditions are stacked as superscript and subscript, the limit on top is taken first.
The result is represented as
\begin{equation}
  \begin{gathered} {\includegraphics[height=3.2cm]{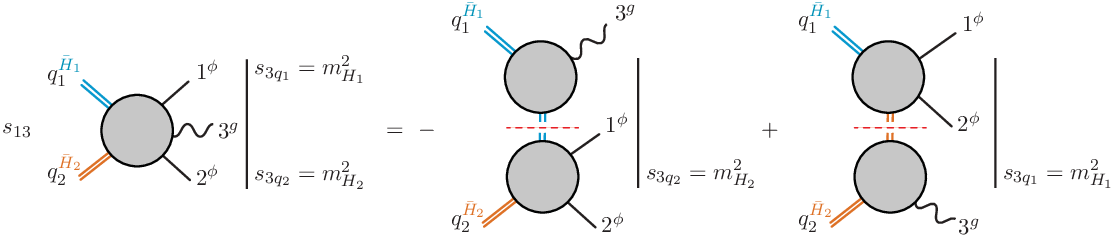} }
  \end{gathered} \,.
\end{equation}
Using the bleaching-invariant amplitudes,
we get the hidden factorization relation of
$\ffnh{2}{3}$:
\begin{align}
  & \quad\ \left.s_{13}
    \cffnh{2}{3}(1^\phi, 3^g, 2^\phi, q_2^{H_2}, q_1^{H_1})
  \right|_{s_{3q_2} = \mH{2}^2 }^{s_{3q_1} = \mH{1}^2 }
  \notag \\
  & = -\left[
    \ca_3(q_1^{\bHBlueIdx{1}}, 3^g, -\mathbf{P}_{3q_1}^{\bHBlueIdx{1}})
    \times
    \cffnh{2}{2}(1^\phi, 2^\phi, q_2^{H_2}, \mathbf{P}_{3q_1}^{H_1})
  \right]_{s_{3q_2} = \mH{2}^2} \notag \\
  & \qquad \qquad +
  \left[
    \ca_3(-\mathbf{P}_{3q_2}^{\bHOrangeIdx{2}}, 3^g, q_2^{\bHOrangeIdx{2}})
    \times
    \cffnh{2}{2}(1^\phi, 2^\phi, \mathbf{P}^{H_2}_{3q_2}, q_1^{H_1})
  \right]_{s_{3q_1} = \mH{1}^2}\, .
  \label{eq:f3_hidden_fac}
\end{align}

\subsubsection*{Revisit the naive double copy}

\newcommand{\nckff}[2]{N^{\CK}_{\ff, #1|#2}}
\newcommand{\nckamp}[2]{N^{\CK}_{\amp, #1|#2}}

We can review the failure of the naive double copy in Section~\ref{sec:naive_double_copy_ff}. By
applying bleaching $\blc_{\bH_{1,2}}$ to the full-color dyed amplitude, we have
\begin{equation}
  \blc_{\bHBlueIdx{1}}\blc_{\bHOrangeIdx{2}}
  \left( \dampPPGHH \right) =
  \ffnhppgqq\, .
\end{equation}
After bleaching,
the color factors of $\ampnh{2}{5}$ collapse to match those of $\ffnh{2}{3}$.
Specifically,
the color factors for topologies
with labels $4$ and $5$ in the class-$\clsA$ and class-$\clsB$ sectors are annihilated to zero.
Introducing these two graphs as auxiliary graphs,
a set of CK-satisfying numerators for $\ffnh{2}{3}$ is
generated directly from the corresponding numerators of $\ampnh{2}{5}$,
\begin{equation}
  \nckff{S}{i} = \nckamp{S}{i} \, ,
\end{equation}
with $S \in \set{\clsA,\ \clsB}$ and $i \in \set{1,2,3,4,5}$.
These numerators have no unphysical spurious poles.

As established, $\rank(\ppmAppghh) = 4$,
which indicates that two kinematic numerators remain unconstrained as free parameters.
However,
in the naive construction,
auxiliary topologies such as $(S|4)$ and $(S|5)$ do not exist,
even at the kinematic level,
which imposes the strict algebraic constraints
\begin{equation}
  \nckff{\clsA}{4} = \nckff{\clsA}{5} =
  \nckff{\clsB}{4} = \nckff{\clsB}{5} = 0 \, ,
\end{equation}
forcing four numerators to zero, which over-constrains the system and makes the double-copy result inconsistent.

 \subsection{Scalar-ordered amplitudes} \label{sec:scalar_ordered_amplitudes}

The amplitude $\dampPPGHH$ admits a decomposition into two gauge-invariant subsectors,
labeled class-$\clsA$ and class-$\clsB$.
Not only are $\campnh{2}{5,\clsA}$ and $\campnh{2}{5,\clsB}$ independently gauge-invariant,
but their associated color factors and propagator matrices map symmetrically into one another under the exchange of the Higgs legs,
$q_1 \leftrightarrow q_2$.
This classification corresponds to two distinct planar topologies composed
only of scalar fields ($\bH$ and $\phi$),
which are isomorphic to the fundamental topologies of $\ffnhppqq$ depicted in Figure~\ref{fig:f2_cubic}.

\begin{figure}[t]
  \centering
  \subfloat[Full-color]{\includegraphics[width=.18\linewidth]{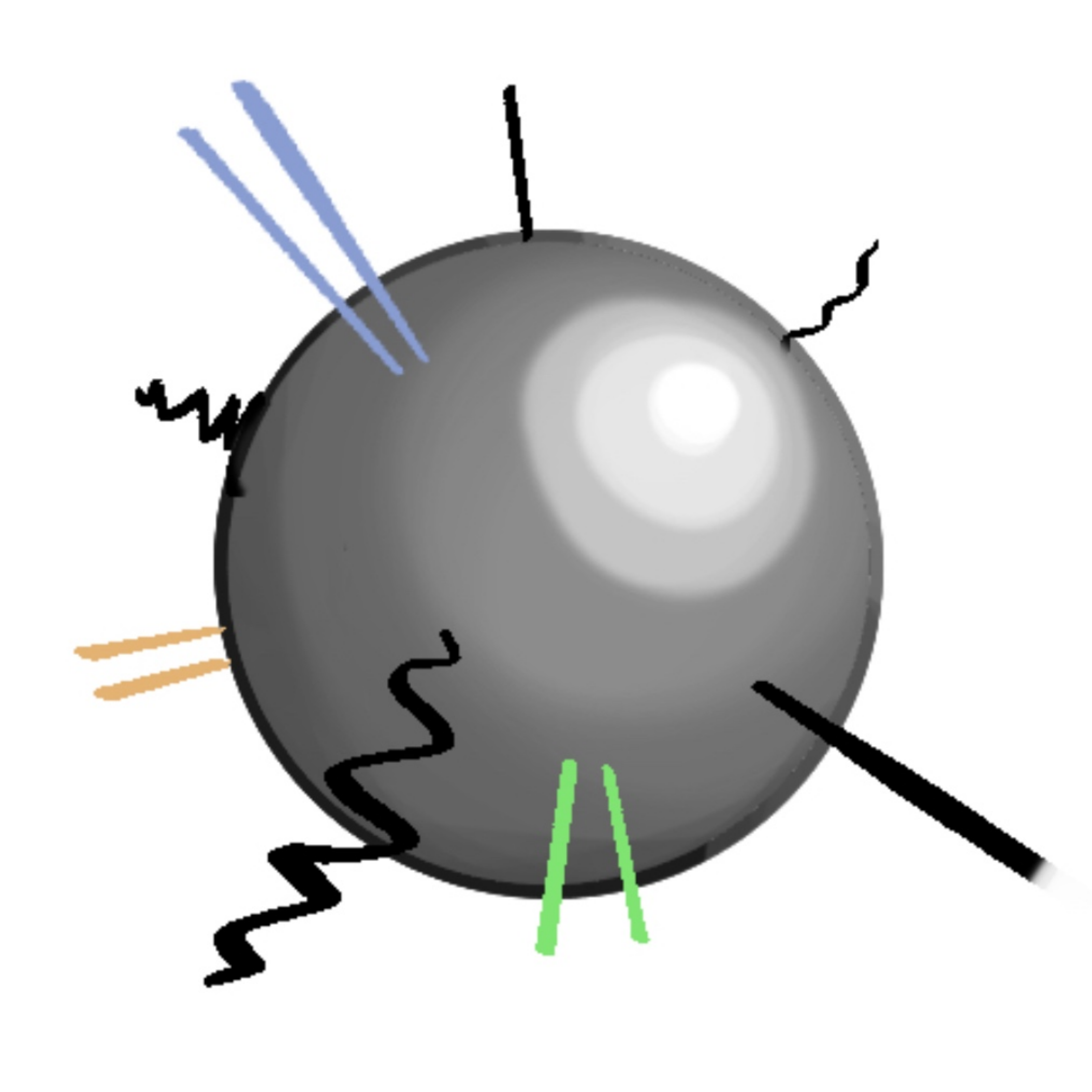}}
  \hspace{2cm}
  \subfloat[Color-ordered]{\includegraphics[width=.18\linewidth]{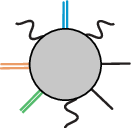}}
  \hspace{2cm}
  \subfloat[Scalar-ordered]{\includegraphics[width=.18\linewidth]{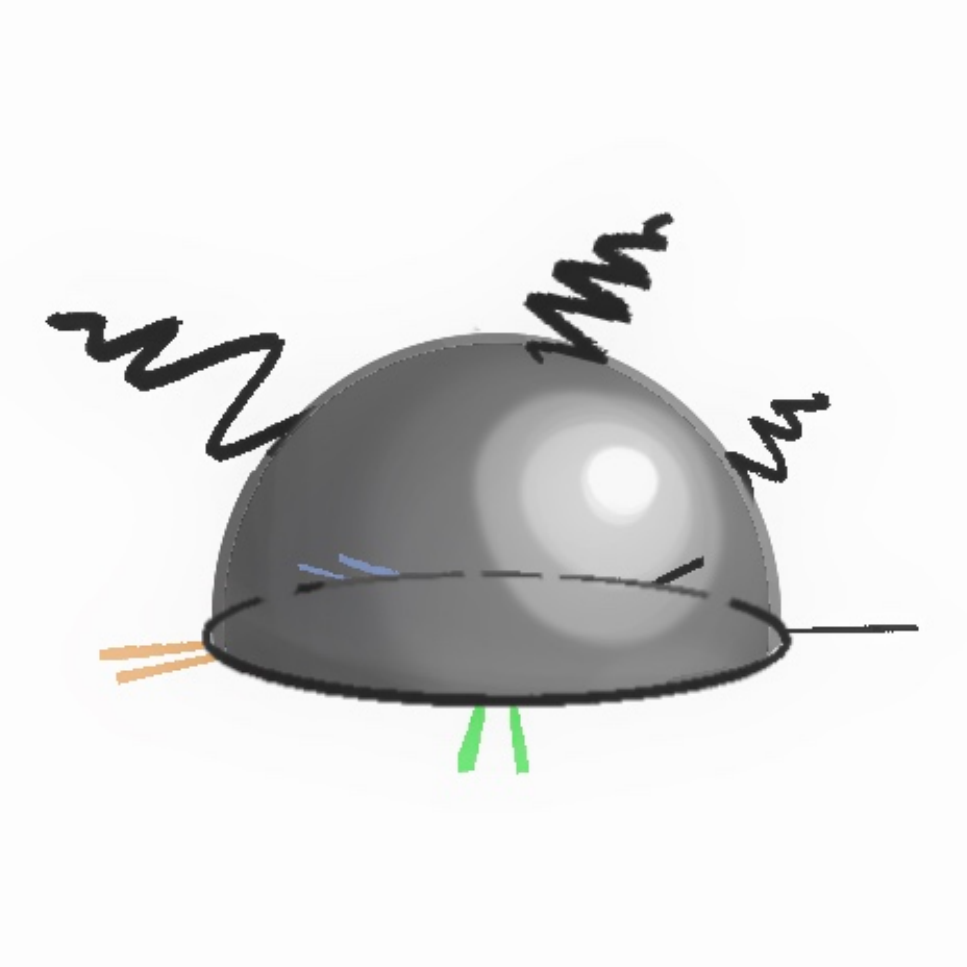}}
  \caption{
    Illustrations for full-color, color-ordered and scalar-ordered amplitudes.
    The full-color amplitude is represented by a sphere,
    where external states are attached to the surface without a defined order.
The color-ordered amplitude is plotted as a planar disk with
    particles fixed on the boundary in a prescribed order.
    In the scalar-ordered amplitude,
    scalars are projected onto the boundary of
    the hemisphere to show the scalar order
    $\trs({\color{green!70!black} \bH}
    {\color{orange} \bH}
    {\color{cyan!70!black} \bH} \phi \phi)$,
    which leaves the gluons unordered on the surface.
  }
  \label{fig:scalar_decompose}
\end{figure}

To make this classification precise,
we isolate the part composed only of scalar fields.
Consider the pure-scalar amplitude
\begin{equation}
  \ampnh{n_h}{n_h + 2}
    \left(1^{\bHBlueIdx{1}}, \ldots, n_h^{\bHGreenIdx{n_h}},
    (n_h + 1)^\phi, (n_h + 2)^\phi \right)\, .
\end{equation}
We refer to each distinct planar topology of this amplitude as a
``scalar skeleton'',
and denote the collection by $\sk(n_h)$.

A scalar ordering $\trs(O_s)$ is the cyclic ordering of the color labels
carried by the external scalar lines along such a topology.
We fix $(n_h + 1)^\phi$ and $(n_h + 2)^\phi$ on the
right and left sides of the half-ladder diagrams,
and place the other particles above the $\phi$-$\phi$ line.
The set $\sk(n_h)$ is generated by performing
permutations $\sigma \in S_{n_h}$ among Higgs particles.
The half-ladder representative with scalar order $\trs(O_s^\sigma)$,
where $O_s^\sigma = ((n_h + 2), \sigma(1), \ldots, \sigma(n_h), (n_h + 1))$,
is depicted as
\begin{equation}
  \begin{gathered}
    \includegraphics[width=0.55\linewidth]{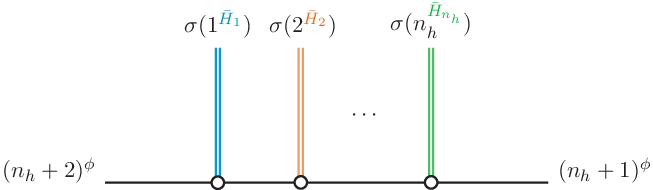}
  \end{gathered} \,.
  \label{fig:half_ladder}
\end{equation}

At this purely scalar level,
there is no independent $dd$ relation that exchanges two adjacent scalar vertices.
Thus one skeleton cannot be mapped to another by color relations.
In this sense,
a fixed planar scalar topology already defines a scalar-ordered sector.

\subsubsection*{Scalar decomposition and color closure}

We now return to general amplitude by including gluons,
\begin{equation}
  \ampnh{n_h}{n} \left(1^{\bHBlueIdx{1}}, \ldots, n_h^{\bHGreenIdx{n_h}},
    (n_h + 1)^\phi, (n_h + 2)^\phi,
  (n_h + 3)^{g}, \ldots ,n^{g} \right)\, .
  \label{eq:full_color_an}
\end{equation}
For a fixed scalar ordering $\trs(O_s)$,
we take the corresponding scalar skeleton and insert the gluons
$\set{(n_h + 3)^g, \ldots, n^g}$
in all possible ways.
This generates all Feynman diagrams whose external scalars
preserve the order $\trs(O_s)$,
with no ordering imposed on the gluons.
The sum over these diagrams is referred to as ``scalar-ordered amplitude"
associated with $\trs(O_s)$,
denoted by $\soa{n_h}{n}{O_s}$.
With this definition,
the full-color amplitude (\ref{eq:full_color_an}) can be written as
\begin{equation}
  \ampnh{n_h}{n}
  =
  \sum_{\sigma \in S_{n_h}}
  \soa{n_h}{n}{O_{s}^\sigma}
  \, .
  \label{eq:amp_scalar_decomposition}
\end{equation}
This expression is referred to as the ``scalar decomposition''.

In multi-Higgs amplitudes,
the color decomposition has the following property.
Consider a scalar skeleton with $\trs(2^\phi, q_1, q_2, q_3, 1^\phi)$.
For a product involving one $f$ and one $d$, the relevant color relation takes the form
\begin{equation}
  \begin{gathered}
    \includegraphics[width=0.9\linewidth]{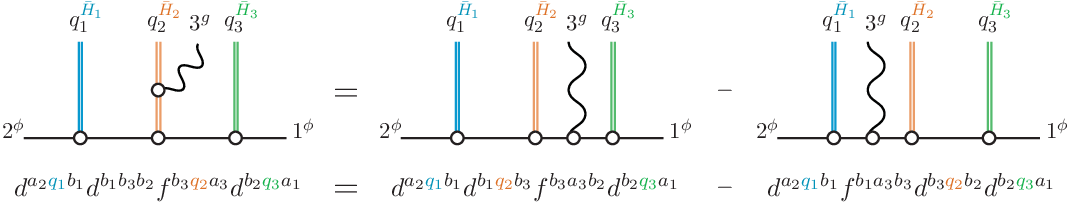}
  \end{gathered} \,,
  \label{fig:ddm_fd_dec}
\end{equation}
and the Jacobi relation takes the form
\begin{equation}
  \begin{gathered}
    \includegraphics[width=0.9\linewidth]{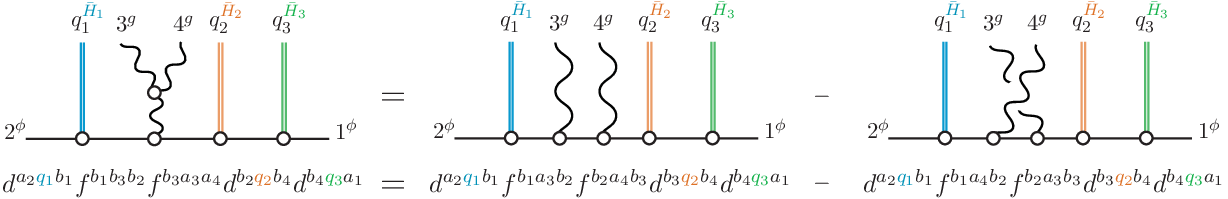}
  \end{gathered} \,.
  \label{fig:ddm_ff_dec}
\end{equation}
The relations shown in (\ref{fig:ddm_fd_dec}) and (\ref{fig:ddm_ff_dec})
can move gluons,
but they cannot permute the ordering of the external scalar colors.
The obstruction already present in the scalar skeletons persists after gluons are added.
Once the skeleton is fixed,
the application of identities (\ref{fig:ddm_fd_dec}) and (\ref{fig:ddm_ff_dec})
cannot map the color structure into a skeleton with a different scalar order.
Thus color relations close within each scalar-ordering sector.
The above color relations also illustrate that a DDM color basis can be obtained by inserting gluons only to the bottom line of the scalar-skeleton graphs.

The five-point example in Section~\ref{sub:2_higgs_5pt}
illustrates this decomposition.
Class-$\clsA$ and class-$\clsB$ correspond to the two scalar orders
\begin{equation}
  \trs(O_s^{\clsA}) = \trs(2^\phi, q_2^{\bHOrangeIdx{2}},
    q_1^{\bHBlueIdx{1}}, 1^\phi), \qquad
  \trs(O_s^{\clsB}) = \trs(2^\phi, q_1^{\bHBlueIdx{1}},
    q_2^{\bHOrangeIdx{2}}, 1^\phi)\, .
\end{equation}
The color relations in (\ref{eq:a5_color_relations}) only relate color factors
inside class-$\clsA$ or inside class-$\clsB$,
and do not mix the two classes.
The two closed gauge-invariant blocks are the two scalar-ordered partial amplitude:
\begin{equation}
  \ampnh{2}{5,\clsA} = \soa{2}{5}{O_s^{\clsA}}, \qquad
  \ampnh{2}{5,\clsB} = \soa{2}{5}{O_s^{\clsB}}\, .
\end{equation}

\subsubsection*{Gravity}

\begin{figure}
  \begin{center}
    \includegraphics[width=0.74\textwidth]{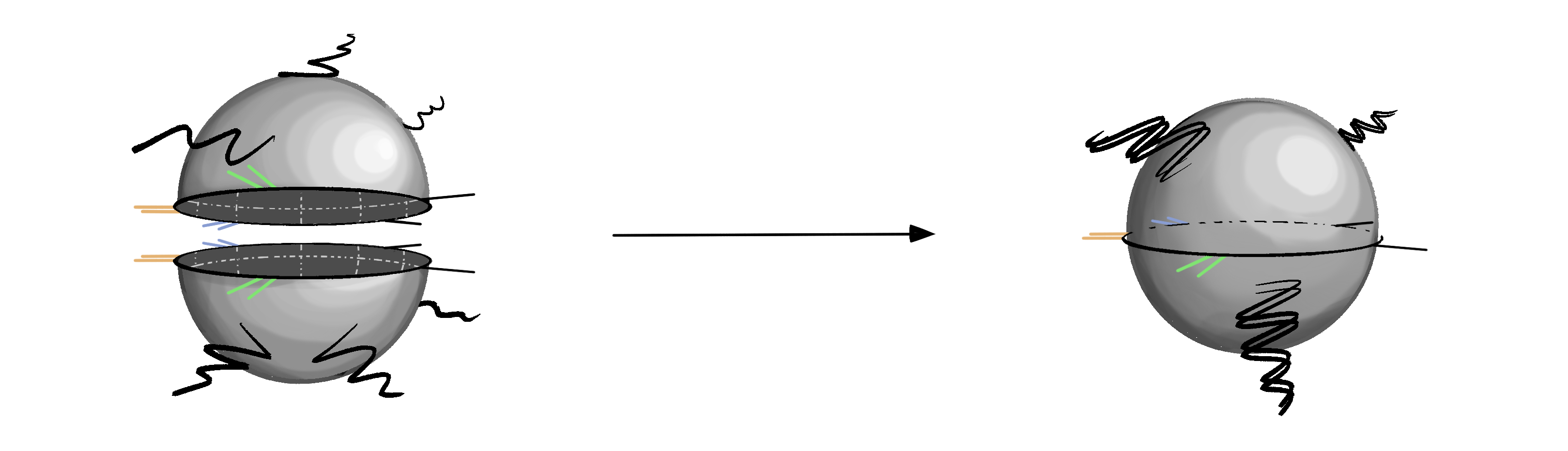}
  \end{center}
  \caption{
    The double copy of a scalar-ordered gauge amplitude generates a
    scalar-ordered gravity amplitude.
Gravitons are denoted by double curly lines.
A nonzero gravity result requires that the scalar orders of
    the constituent gauge amplitudes are compatible.
  }
  \label{fig:scalar_ordered_dc}
\end{figure}

The scalar-ordered structure is present not only at the
gauge level, but also at the gravity level, giving
the scalar-ordered gravity amplitude $\soga{n_h}{n}{O_s}$.

At the gauge-theory level,
the construction above shows that a scalar-ordered amplitude is
gauge invariant,
as it is composed of a complete sum over a closed color sector.
At the gravity level, since all particles are colorless,
there are no ``color-ordered gravity amplitudes''.
However, the scalar-ordering property is preserved by the KLT double-copy formalism.
This can be verified by calculating a scalar-ordered gravity amplitude
using standard Feynman rules,
and independently confirming its diffeomorphism invariance.

Another argument follows from the KLT kernel.
This kernel is derived from the block-diagonalized propagator matrix of
$\ampnh{n_h}{n}$.
Matrix elements of the KLT kernel coupling distinct scalar orders
$\set{O_{{s}}, O_{{s}}'}$ are zero,
\begin{equation}
  \klt^{[n_h]}_{n} \colorxy{C_{O_{{s}}}}{C_{O_{{s}}'}} = 0
  \, .
\end{equation}
This disentangles the kinematic contributions of $\soa{n_h}{n}{O_{{s}}}$
and $\soa{n_h}{n}{O_{{s}}'}$,
permitting the gravity amplitude to be expressed via a scalar decomposition
\begin{equation}
  \ganh{n_h}{n} =
  \sum_{\sigma \in S_{n_h}}
  \soga{n_h}{n}{O_{s}^\sigma}
  \, .
  \label{eq:gravity_scalar_decomposition}
\end{equation}
This establishes the double-copy correspondence
\begin{equation}
  \soa{n_h}{n}{O_{s}^\sigma}
  \xrightarrow{\text{Double copy}}
  \soga{n_h}{n}{O_{s}^\sigma}
  \,.
\end{equation}

 \subsection{High-point cases} 

\newcommand{\dampPPGGHH}{\ampnh{2}{6}({1^\phi, 2^\phi, 3^g, 4^g, q_1^{\bHBlueIdx{1}}, q_2^{\bHOrangeIdx{2}}})}
\newcommand{\dampPPGHHH}{\ampnh{3}{6}({1^\phi, 2^\phi, 3^g, q_1^{\bHBlueIdx{1}}, q_2^{\bHOrangeIdx{2}}, q_3^{\bHGreenIdx{3}}})}
\newcommand{\dampPPGGGH}{\ampnh{1}{6}({1^\phi, 2^\phi, 3^g, 4^g, 5^g, q_1^{\bHBlueIdx{1}}})}
\newcommand{\dampPPHHHH}{\ampnh{4}{6}({1^\phi, 2^\phi, q_1^{\bHBlueIdx{1}}, q_2^{\bHOrangeIdx{2}}, q_3^{\bHGreenIdx{3}}, q_4^{\bHPurpleIdx{4}}})}
\newcommand{\ppmos}{\Theta_{4 \times 4}^{O_s^1}}

To verify the scalar-ordered amplitude strategy
and provide more examples,
we turn to higher-point amplitudes.
We discuss a six-point case in detail.

\subsubsection*{The six-point case}

There are four kinds of six-point dyed amplitudes ---
\begin{gather}
  \dampPPGGGH\,,\ \dampPPGGHH\,, \notag\\
  \dampPPGHHH\,,\ \dampPPHHHH\,.
\end{gather}
$\dampPPGGGH$ reduces to the single-Higgs scattering already studied in Section~\ref{sec:dyedphi2},
and $\dampPPHHHH$ involves no gluon scattering and is the scalar skeleton.
We discuss $\dampPPGHHH$ in detail below, which corresponds to
the three-point FF $\ffnh{3}{3}$ and has $42$ trivalent topologies.

\begin{figure}
  \begin{center}
    \includegraphics[width=0.98\textwidth]{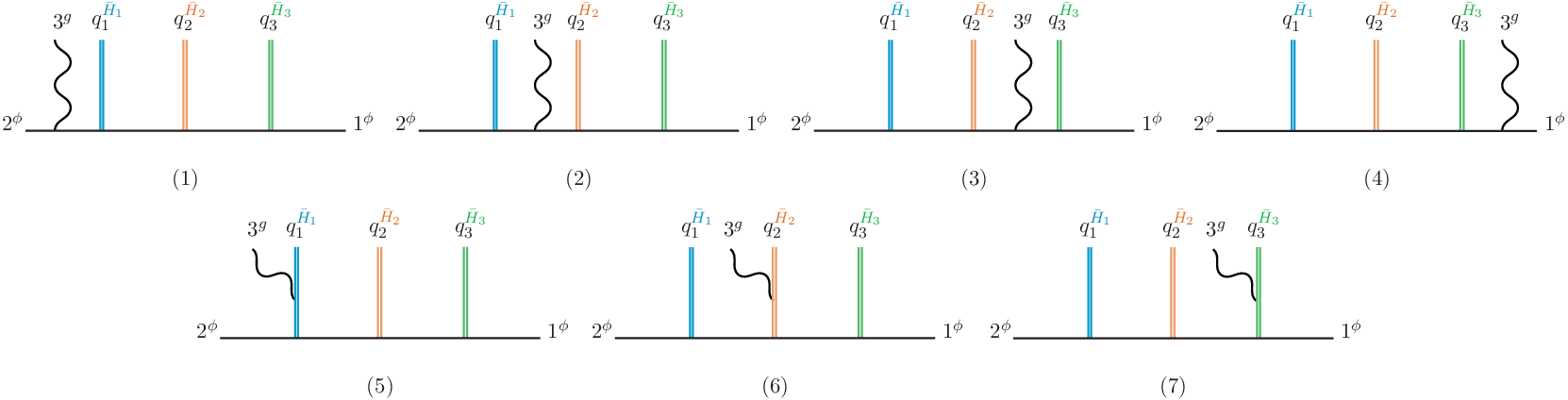}
  \end{center}
  \caption{
    Trivalent topologies related to $\soa{3}{6}{O_s^1}$.
    The first four can be interpreted as half-ladder
    diagrams of DDM color basis,
    while the last three can be expanded by the corresponding color relations.
  }\label{fig:a36_os1_cubic}
\end{figure}

There are 6 scalar skeletons corresponding to the permutations of the three Higgs. 
The full-color amplitude can be decomposed as
\begin{equation}
  \ampnh{3}{6}
  =
  \sum_{\sigma \in S_3}
  \soa{3}{6}{O_{s}^\sigma}
  \, .
  \label{eq:amp_scalar_decomposition6pt}
\end{equation}
We select a single scalar ordering and focus on the scalar-ordered amplitude $\soa{3}{6}{O_s^1}$ with
\begin{align}
  \trs(O_s^1) = \trs \left(
    2^\phi, q_1^{\bHBlueIdx{1}}, q_2^{\bHOrangeIdx{2}}, q_3^{\bHGreenIdx{3}}, 1^\phi
  \right)
  \, .
\end{align}
The cubic topologies associated with $O_s^1$ are depicted in Figure~\ref{fig:a36_os1_cubic}.
The DDM color basis is determined by inserting the single gluon directly into the bottom lines of the scalar skeleton graph,
which contains four elements given by the first $4$ graphs in Figure~\ref{fig:a36_os1_cubic},
and their color factors are \begin{align}
  \co(O_a^{1|1}) =
  \co \left(
        2^\phi, 3^g, q_1^{\bHBlueIdx{1}}, q_2^{\bHOrangeIdx{2}}, q_3^{\bHGreenIdx{3}}, 1^\phi
      \right)
      = f^{a_2 a_3 b_1} d^{b_1 q_1 b_2} d^{b_2 q_2 b_3}
        d^{b_3 q_3 a_1} \, ,
  \notag \\
  \co(O_a^{1|2}) =
  \co \left(
        2^\phi, q_1^{\bHBlueIdx{1}}, 3^g, q_2^{\bHOrangeIdx{2}}, q_3^{\bHGreenIdx{3}}, 1^\phi
      \right)
      = d^{a_2 q_1 b_1} f^{b_1 a_3 b_2} d^{b_2 q_2 b_3}
        d^{b_3 q_3 a_1} \, ,
  \notag \\
  \co(O_a^{1|3}) =
  \co \left(
        2^\phi, q_1^{\bHBlueIdx{1}}, q_2^{\bHOrangeIdx{2}}, 3^g, q_3^{\bHGreenIdx{3}}, 1^\phi
      \right)
      = d^{a_2 q_1 b_1} d^{b_1 q_2 b_2} f^{b_2 a_3 b_3}
        d^{b_3 q_3 a_1} \, ,
  \notag \\
  \co(O_a^{1|4}) =
  \co \left(
        2^\phi, q_1^{\bHBlueIdx{1}}, q_2^{\bHOrangeIdx{2}}, q_3^{\bHGreenIdx{3}}, 3^g, 1^\phi
      \right)
      = d^{a_2 q_1 b_1} d^{b_1 q_2 b_2} d^{b_2 q_3 b_3}
        f^{b_3 a_3 a_1} \, .
  \label{eq:}
\end{align}
$\set{\co (O_a^{1|i})}$ defines the DDM color basis of
$\soa{3}{6}{O_s^1}$,
leading to the scalar-ordered propagator matrix
\begin{equation}
  \ppmos =
  \begin{bmatrix}
    \frac{1}{D_1} + \frac{1}{D_5} & -\frac{1}{D_5} & 0 & 0 \\
    -\frac{1}{D_5} &
    \frac{1}{D_2} + \frac{1}{D_5} + \frac{1}{D_6} &
    -\frac{1}{D_6} & 0 \\
    0 & -\frac{1}{D_6} &
    \frac{1}{D_3} + \frac{1}{D_6} + \frac{1}{D_7}  &
    - \frac{1}{D_7} \\
    0 & 0 & - \frac{1}{D_7} &
    \frac{1}{D_4} + \frac{1}{D_7}
  \end{bmatrix} \, .
\end{equation}
The matrix is singular, with
\begin{equation}
  \rank \ppmos = 3 \, .
\end{equation}
KLT double copy is implemented by extracting a $3 \times 3$
submatrix from $\ppmos$.
For example,
\begin{equation}
  \soga{3}{6}{O_s^1} =
  \left(\vec{\ca}^{\, O_a^{1}}\right)^T \cdot
  \klt_{3 \times 3}^{O_s^1} \cdot
  \vec{\ca}^{\, O_a^{1}} \, ,
\end{equation}
with
\begin{equation}
  \klt_{3 \times 3}^{O_s^1} =
  \begin{bmatrix}
    \frac{1}{D_1} + \frac{1}{D_5} & -\frac{1}{D_5} & 0 \\
    -\frac{1}{D_5} &
    \frac{1}{D_2} + \frac{1}{D_5} + \frac{1}{D_6} &
    -\frac{1}{D_6} \\
    0 & -\frac{1}{D_6} &
    \frac{1}{D_3} + \frac{1}{D_6} + \frac{1}{D_7}
  \end{bmatrix}^{-1},
  \quad
  \vec{\ca}^{\, O_a^{1}} =
  \begin{bmatrix}
    \ca^{O_a^{1|1}} \\ \ca^{O_a^{1|2}} \\ \ca^{O_a^{1|3}}
  \end{bmatrix}
  \, .
\end{equation}
where $\ca^{O_a^{1|i}}$ is the partial amplitude of the
color $\co(O_a^{1|i})$.
The null space of $\ppmos$
gives the BCJ relation
\begin{equation}
  s_{23} \ca^{O_a^{1|1}} +
  (s_{23} + s_{3q_1} - \mH{1}^2) \ca^{O_a^{1|2}} -
  (s_{13} + s_{3 q_3} - \mH{3}^2) \ca^{O_a^{1|3}} -
  s_{13} \ca^{O_a^{1|4}} = 0 \, .
  \label{eq:a36_ddm_bc}
\end{equation}
To deduce the hidden factorization,
project (\ref{eq:a36_ddm_bc}) into single-trace basis
\begin{align}
  (s_{3q_3} - \mH{3}^2 + s_{13}) &\campnh{3}{6}(1^\phi,3^g,2^\phi,q_1^{\bHBlueIdx{1}},q_2^{\bHOrangeIdx{2}},q_3^{\bHGreenIdx{3}})
  \notag \\
  + (s_{3q_3} - \mH{3}^2) &\campnh{3}{6}(1^\phi,2^\phi,q_1^{\bHBlueIdx{1}},q_2^{\bHOrangeIdx{2}},q_3^{\bHGreenIdx{3}},3^g)
  \notag \\
  - [(s_{3q_1} - \mH{1}^2) + (s_{3q_2} - \mH{2}^2)] & \campnh{3}{6}(1^\phi,2^\phi,3^g,q_1^{\bHBlueIdx{1}},q_2^{\bHOrangeIdx{2}},q_3^{\bHGreenIdx{3}})
  \notag \\
  - (s_{3q_2} - \mH{2}^2) & \campnh{3}{6}(1^\phi,2^\phi,q_1^{\bHBlueIdx{1}},3^g,q_2^{\bHOrangeIdx{2}},q_3^{\bHGreenIdx{3}}) = 0
  \, ,
  \label{eq:a36_tr_bc}
\end{align}
where the relevant poles appear explicitly as coefficients of color-ordered amplitudes.

Taking the relevant poles in (\ref{eq:a36_tr_bc}) on shell,
namely $(s_{3q_1} - \mH{1}^2)$, $(s_{3q_2} - \mH{2}^2)$
and $(s_{3q_3} - \mH{3}^2)$,
we extract the hidden factorization relation of $\ffnh{3}{3}$.
A subtlety here is that
$\campnh{3}{6}(1^\phi,2^\phi,q_1^{\bHBlueIdx{1}},3^g,q_2^{\bHOrangeIdx{2}},q_3^{\bHGreenIdx{3}})$
contains both $(s_{3q_1} - \mH{1}^2)$ and $(s_{3q_2} - \mH{2}^2)$ as propagators,
while its coefficient in (\ref{eq:a36_tr_bc}) is only $(s_{3q_2} - \mH{2}^2)$.
Therefore, the limit $s_{3q_2} = \mH{2}^2$ must be taken first to avoid the divergence.
Following the convention stated after (\ref{eq:a5_second_on_shell_limit})
that stacked on-shell conditions are taken from top to bottom,
we impose $(s_{3q_2} - \mH{2}^2) \to 0$ first, then the remaining poles, giving
\begin{align}
  &\quad \left[
    s_{13} \campnh{3}{6}(1^\phi,3^g,2^\phi,q_1^{\bHBlueIdx{1}},q_2^{\bHOrangeIdx{2}},q_3^{\bHGreenIdx{3}})
  \right] \hspace{-0.15cm}
  {\tiny \begin{array}{c} s_{3q_2} = \mH{2}^2 \\ s_{3q_1} = \mH{1}^2 \\ s_{3q_3} = \mH{3}^2 \end{array}}
  \notag \\
  &=\ \ \
  \ca_3(-\mathbf{P}_{3q_1}^{\bHBlueIdx{1}}, 3^g, q_1^{\bHBlueIdx{1}}) \times
  \campnh{3}{5}(1^\phi, 2^\phi, \mathbf{P}_{3q_1}^{\bHBlueIdx{1}}, q_2^{\bHOrangeIdx{2}}, q_3^{\bHGreenIdx{3}})
  \notag \\
  &\quad +
  \ca_3(-\mathbf{P}_{3q_2}^{\bHOrangeIdx{2}}, 3^g, q_2^{\bHOrangeIdx{2}}) \times
  \campnh{3}{5}(1^\phi, 2^\phi, q_1^{\bHBlueIdx{1}},\mathbf{P}_{3q_2}^{\bHOrangeIdx{2}}, q_3^{\bHGreenIdx{3}})
  \notag \\
  &\quad
  - \ca_3(q_3^{\bHGreenIdx{3}}, 3^g, -\mathbf{P}_{3q_3}^{\bHGreenIdx{3}}) \times
  \campnh{3}{5}(1^\phi, 2^\phi, q_1^{\bHBlueIdx{1}}, q_2^{\bHOrangeIdx{2}}, \mathbf{P}_{3q_3}^{\bHGreenIdx{3}})
  \, .
\end{align}
Using the bleaching-invariant relations,
\begin{align}
  \campnh{3}{6}(1^\phi,3^g,2^\phi,q_1^{\bHBlueIdx{1}},q_2^{\bHOrangeIdx{2}},q_3^{\bHGreenIdx{3}})
  &= \cffnh{3}{3}(1^\phi,3^g,2^\phi,q_1^{H_1},q_2^{H_2},q_3^{H_3})\, ,
  \notag \\
  \campnh{3}{5}(1^\phi, 2^\phi, \mathbf{P}_{3q_1}^{\bHBlueIdx{1}}, q_2^{\bHOrangeIdx{2}}, q_3^{\bHGreenIdx{3}})
  &= \cffnh{3}{2}(1^\phi, 2^\phi, \mathbf{P}_{3q_1}^{H_1}, q_2^{H_2}, q_3^{H_3})\, ,
  \notag \\
  \campnh{3}{5}(1^\phi, 2^\phi, q_1^{\bHBlueIdx{1}},\mathbf{P}_{3q_2}^{\bHOrangeIdx{2}}, q_3^{\bHGreenIdx{3}})
  &= \cffnh{3}{2}(1^\phi, 2^\phi, q_1^{H_1},\mathbf{P}_{3q_2}^{H_2}, q_3^{H_3}) \, ,
  \notag \\
  \campnh{3}{5}(1^\phi, 2^\phi, q_1^{\bHBlueIdx{1}}, q_2^{\bHOrangeIdx{2}}, \mathbf{P}_{3q_3}^{\bHGreenIdx{3}})
  &= \cffnh{3}{2} (1^\phi, 2^\phi, q_1^{H_1}, q_2^{H_2}, \mathbf{P}_{3q_3}^{H_3})  \, ,
\end{align}
the factorization can be written as
\begin{align}
  &\quad \left[
    s_{13} \cffnh{3}{3}(1^\phi,3^g,2^\phi,q_1^{H_1},q_2^{H_2},q_3^{H_3})
  \right] \hspace{-0.15cm}
  {\tiny \begin{array}{c} s_{3q_2} = \mH{2}^2 \\ s_{3q_1} = \mH{1}^2 \\ s_{3q_3} = \mH{3}^2 \end{array}}
  \notag \\
  &=\ \ \
  \ca_3(-\mathbf{P}_{3q_1}^{\bHBlueIdx{1}}, 3^g, q_1^{\bHBlueIdx{1}}) \times
  \cffnh{3}{2}(1^\phi, 2^\phi, \mathbf{P}_{3q_1}^{H_1}, q_2^{H_2}, q_3^{H_3})
  \notag \\
  &\quad +
  \ca_3(-\mathbf{P}_{3q_2}^{\bHOrangeIdx{2}}, 3^g, q_2^{\bHOrangeIdx{2}}) \times
  \cffnh{3}{2}(1^\phi, 2^\phi, q_1^{H_1},\mathbf{P}_{3q_2}^{H_2}, q_3^{H_3})
  \notag \\
  &\quad
  - \ca_3(q_3^{\bHGreenIdx{3}}, 3^g, -\mathbf{P}_{3q_3}^{\bHGreenIdx{3}}) \times
  \cffnh{3}{2} (1^\phi, 2^\phi, q_1^{H_1}, q_2^{H_2}, \mathbf{P}_{3q_3}^{H_3})
  \, ,
\end{align}
which is a hidden factorization relation for the form factor. This factorization relation can be represented graphically as
\begin{equation}
  \begin{gathered}
    \includegraphics[width=0.95\linewidth]{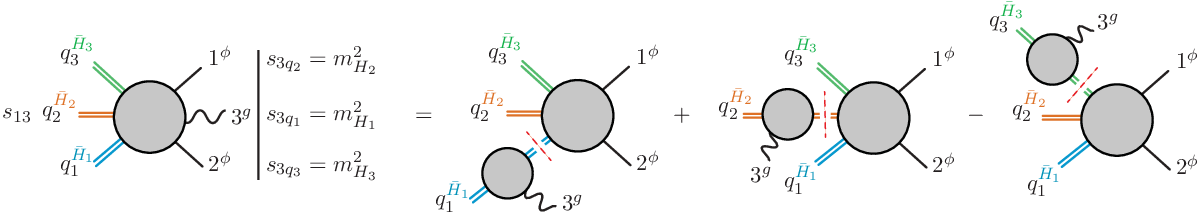}
  \end{gathered} \,.
  \label{fig:a36_fac}
\end{equation}

To obtain the amplitudes of alternative scalar orders --- at both the gauge and gravity levels --- one simply applies permutations of the Higgs fields to the result derived above.
For example, combing all scalar orderings, the full propagator matrix of $\ampnh{3}{6}$
has size $4 \times (3!) = 24$ and rank $3 \times (3!) = 18$.

\subsubsection*{Counting}

With the strategy proposed previously,
scalar-ordered amplitudes $\soa{n_h}{n}{O_s}$ can be systematically constructed to arbitrary multiplicities.
We have performed explicit constructions up to
the eight-point case with four dyed Higgs legs.

\begin{table}[t]
  \centering
  \begin{tabular}{|c|c|c|c|c|c|}
\hline
    external legs & Higgs legs & cubic diagrams & \#SKLT & \#SIZE & \#RANK \\
    \hline
    \hline
    3 & \multirow{4}{*}{1} & 1 & \multirow{4}{*}{1} & 1 & 1 \\
    \cline{1-1} \cline{3-3} \cline{5-6}
    4 & & 3 & & 2 & 1 \\
    \cline{1-1} \cline{3-3} \cline{5-6}
    5 & & 15 & & 6 & 2 \\
    \cline{1-1} \cline{3-3} \cline{5-6}
    6 & & 105 & & 24 & 6 \\
    \hline
    \hline
    4 & \multirow{4}{*}{2} & 2 & \multirow{4}{*}{2} & 1 & 1 \\
    \cline{1-1} \cline{3-3} \cline{5-6}
    5 & & 10 & & 3 & 2 \\
    \cline{1-1} \cline{3-3} \cline{5-6}
    6 & & 70 & & 12 & 6 \\
    \cline{1-1} \cline{3-3} \cline{5-6}
    7 & & 630 & & 60 & 24 \\
    \hline
    \hline
    5 & \multirow{4}{*}{3} & 6 & \multirow{4}{*}{6} & 1 & 1 \\
    \cline{1-1} \cline{3-3} \cline{5-6}
    6 & & 42 & & 4 & 3 \\
    \cline{1-1} \cline{3-3} \cline{5-6}
    7 & & 378 & & 20 & 12 \\
    \cline{1-1} \cline{3-3} \cline{5-6}
    8 & & 4158 & & 120 & 60 \\
    \hline
    \hline
    6 & \multirow{3}{*}{4} & 24 & \multirow{3}{*}{24} & 1 & 1 \\
    \cline{1-1} \cline{3-3} \cline{5-6}
    7 & & 216 & & 5 & 4 \\
    \cline{1-1} \cline{3-3} \cline{5-6}
    8 & & 2376 & & 30 & 20 \\
    \hline
    \hline
    $n$ & $n_h$ & $n_{h}! \dfrac{(2n-5)!!}{(2n_{h}
    - 1)!!}$ & $n_h!$ & $\dfrac{(n-2)!}{n_{h}!}$ &
    $\dfrac{(n-3)!}{(n_h - 1)!}$ \\
    \hline
  \end{tabular}
  \caption{
	    Counting data for amplitudes $\ampnh{n_h}{n}$ of $\LadYMSHi{n_h}_{\tr(\phi^2)}$,
    comprising $n_h$ Higgs particles, two $\phi$ scalar fields and $(n - n_h - 2)$ gluons.
The third column gives the number of cubic diagrams.
    The last three columns are counts related to the scalar-ordered amplitude $\soa{n_h}{n}{O_s}$.
    \#SKLT is the number of scalar skeletons.
	    \#SIZE and \#RANK give the dimension and rank of
    each scalar-ordered propagator matrix.
}
  \label{tab:multi_higgs}
\end{table}
A general count of the number of cubic diagrams, as well as the size and rank of propagator matrices, are summarized in Table~\ref{tab:multi_higgs}.
Several observations can be drawn from the table.
First, the counts with $n_h = 1$ is the same as that of pure-gluon scattering,
as expected from the unique scalar order $\trs(2q_11)$.
Second,
for $\ampnh{n_h}{n}$ with $n_g$ gluons, the ratio between the rank and size scales as
\begin{equation}
  \frac{\text{\#RANK}^{n_h}_n}{\text{\#SIZE}^{n_h}_n}
  = \frac{n_h \times (n-3)!}{(n-2)!}
  = \frac{n_h}{n - 2}
  = \frac{n - n_g - 2}{n - 2}
  \, .
\end{equation}
This reveals two different limits.
	If $n_g$ is fixed and $n \rightarrow \infty$,
i.e., Higgs dominance, the ratio approaches $1$.
When the number of Higgs particles is fixed and $n_g$ grows, the ratio again approaches $0$, in agreement with $\amp^g_n$.
Since the matrix rank is directly related to the number of BCJ relations, the ratio quantifies the redundancies among partial amplitudes.


  \section{Dyed theory: general operators}
\label{sec:dyedgeneral}
So far, the dyeing procedures have been illustrated primarily for the $\tr(\phi^2)$ operator.
In this section, we apply the framework to more general operators.
We begin with high-length scalar operators $\tr(\phi^m)$, where the color structure involves generalized fully symmetric $d$-symbols, and then turn to fermion-related operators $\trpp$ and $\trpGp$.
Since the construction is similar, we will mainly focus on the results and new features.

\subsection[High-length \protect\texorpdfstring{${\rm tr}(\phi^m)$}{tr(phi^m)} operator]{High-length ${\rm tr}(\phi^m)$ operator}
\label{sec:trphin}

\newcommand{\trm}{\tr(\phi^m)}
\newcommand{\LaHpm}{\La^{H \phi^m}}
\newcommand{\LadHpm}{\La^{\bHBlue \phi^m}}

In this subsection, we study the high-length operator $\trm$ in the YMs theory, focusing on the FF $\ff_{n, \trm}(1^\phi, \ldots, m^\phi, (m+1)^g, \ldots n^g)$, which can be interpreted as a Higgs amplitude of the $\trm$ YMsH theory,
\begin{equation}
  \LaYMSH_{\trm} = \LaYMSH + \LaHpm,
  \quad
  \LaHpm = \lambda_{H\phi^m}
  d^{a_1 \ldots a_m} H \phi^{a_1} \ldots \phi^{a_m} \, ,
\end{equation}
with the fully symmetric constant
\begin{equation}
  d^{a_1 \ldots a_k} =
  \sum_{\sigma \in S_{k-1}}
  \tr(T^{a_{\sigma(1)}} \ldots T^{a_{\sigma(k-1)}} T^{a_k}) \, .
\end{equation}
The Lagrangian is dyed as
\begin{equation}
  \LadYMSH_{\trm} = \LadYMSH + \LadHpm,
  \quad
  \LadHpm = \lambda_{H\phi^m}
  d^{a_1 \ldots a_{m+1}}
  \phi^{a_1} \cdots \phi^{a_{m}} H^{a_{m+1}} \, .
\end{equation}
Hence the amplitude is bleached to the FF as
\begin{align}
  \blcqBH \left(
    \amp_{n+1}^{\bar{\phi}^m} (1^\phi, \ldots, m^\phi, (m+1)^g, \ldots n^g, q^{\bHBlue})
  \right)
  &=
    \amp_{n+1}^{\phi^m} (1^\phi, \ldots, m^\phi, (m+1)^g, \ldots n^g, q^{H})
  \notag \\
  &= \ff_{n, \trm} (1^\phi, \ldots, m^\phi, (m+1)^g, \ldots n^g)\, .
\end{align}

\subsubsection*{The five-point example}

\newcommand{\amppfive}{\amp_5^{\bar{\phi}^3}(1^\phi, 2^\phi, 3^\phi, 4^g, q^{\bHBlue})}
\newcommand{\cfap}[1]{C_{#1}^{\bar{\phi}^3}}
\newcommand{\nsap}[1]{N_{#1}^{\bar{\phi}^3}}

\begin{figure}
  \begin{center}
    \includegraphics[width=0.75\textwidth]{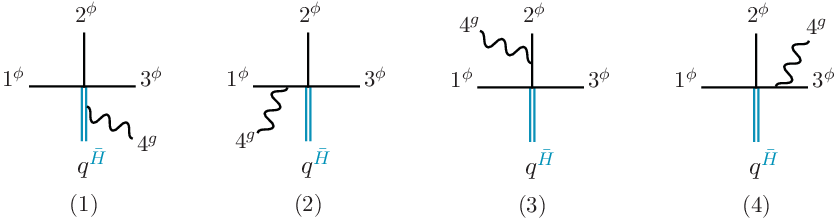}
  \end{center}
  \caption{Feynman diagrams of $\amppfive$.}\label{fig:amp5_diagrams}
\end{figure}

To illustrate the basic idea, we use the amplitude $\amppfive$ of $\LadYMSH_{\tr(\phi^3)}$ as an example. It can be written as a sum over the four topologies shown in Figure~\ref{fig:amp5_diagrams},
\begin{equation}
  \amppfive =
  \frac{\cfap{1}\nsap{1}}{s_{4q} - m^2} + \frac{\cfap{2}\nsap{2}}{s_{14}} + \frac{\cfap{3}\nsap{3}}{s_{24}} + \frac{\cfap{4}\nsap{4}}{s_{34}} \, ,
\end{equation}
with color factors
\begin{equation}
  \cfap{1} = d^{a_1a_2a_3b} f^{ba_4q}\,, \
  \cfap{2} = d^{ba_2a_3q} f^{ba_4a_1}\,, \
  \cfap{3} = d^{a_1ba_3q} f^{ba_4a_2}\,, \
  \cfap{4} = d^{a_1a_2bq} f^{ba_4a_3}\,.
\end{equation}
These color factors satisfy the relation
\begin{equation}
  \cfap{1} + \cfap{2} + \cfap{3} + \cfap{4} = 0 \,.
  \label{eq:a5p3_color_relation}
\end{equation}
This relation shifts the gluon emission from the $\bHBlue$-leg to other $\phi$-legs, and thus allows $\set{\cfap{2}, \cfap{3}, \cfap{4}}$ to form a DDM basis of $\amp_5^{\bar{\phi}^3}$.
The kinematic numerators obtained from the Feynman diagrams satisfy the corresponding dual $\CK$ relation, so no spurious pole is introduced.
An analogous relation holds for higher-length operators $\tr(\phi^m)$ with $m > 3$.
Based on this feature, the dimension of the DDM color basis for a general case with a single dyed $\tr(\phi^m)$ vertex and $(n - m - 1)$ gluons is $(n-2)! / (m-1)!$.

The propagator matrix in the basis $\set{\cfap{2}, \cfap{3}, \cfap{4}}$ is given by
\begin{equation}
  \Theta_{3 \times 3}^{\amp_5^{\bar{\phi}^3}} =
  \begin{bmatrix}
    \frac{1}{s_{14}} + \frac{1}{s_{4q} - m^2} & \frac{1}{s_{4q} - m^2} & \frac{1}{s_{4q} - m^2} \\
    \frac{1}{s_{4q} - m^2} & \frac{1}{s_{24}} + \frac{1}{s_{4q} - m^2} & \frac{1}{s_{4q} - m^2} \\
    \frac{1}{s_{4q} - m^2} & \frac{1}{s_{4q} - m^2} & \frac{1}{s_{34}} + \frac{1}{s_{4q} - m^2}
  \end{bmatrix}\, ,
\end{equation}
which is singular with rank $2$ using $-(s_{4q} - m^2) = s_{14} + s_{24} + s_{34}$.
After translating into the single-trace basis, the BCJ relation is
\begin{align}
  - (s_{4q} - m^2) \ca_5^{\bar{\phi}^3}(1^\phi, 2^\phi, 3^\phi, q^{\bHBlue}, 4^g)
  &+
  (s_{24} + s_{34}) \ca_5^{\bar{\phi}^3}(1^\phi, 4^g, 2^\phi, 3^\phi, q^{\bHBlue})
\notag \\
  & \qquad \qquad \qquad + s_{34} \ca_5^{\bar{\phi}^3}(1^\phi, 2^\phi, 4^g, 3^\phi, q^{\bHBlue}) = 0
  \, .
  \label{eq:a5p_bcj_tr}
\end{align}
Taking the pole $(s_{4q} - m^2)$ on shell in (\ref{eq:a5p_bcj_tr}), we obtain
\begin{align}
  \left[
    (s_{24} + s_{34}) \ca_5^{\bar{\phi}^3}(1^\phi, 4^g, 2^\phi, 3^\phi, q^{\bHBlue}) +
    s_{34} \ca_5^{\bar{\phi}^3}(1^\phi, 2^\phi, 4^g, 3^\phi, q^{\bHBlue})
  \right]_{s_{4q} = m^2}
  = \hspace{-5cm}
  &
  \notag \\
  &
  \ca_3^{\bar{\phi}^3}(q^{\bHBlue}, \mathbf{P}_{123}^{\bHBlue}, 4^g)
  \times
  \ca_4^{\bar{\phi}^3}(1^\phi, 2^\phi, 3^\phi, - \mathbf{P}_{123}^{\bHBlue}) \, .
\end{align}
With the bleaching invariant amplitudes,
\begin{gather}
  \ca_5^{\bar{\phi}^3}(1^\phi, 4^g, 2^\phi, 3^\phi, q^{\bHBlue}) =
  \cf_4^{\phi^3}(1^\phi, 4^g, 2^\phi, 3^\phi)\, ,
  \quad
  \ca_5^{\bar{\phi}^3}(1^\phi, 2^\phi, 4^g, 3^\phi, q^{\bHBlue}) =
  \cf_4^{\phi^3}(1^\phi, 2^\phi, 4^g, 3^\phi)\, ,
  \notag \\
  \ca_4^{\bar{\phi}^3}(1^\phi, 2^\phi, 3^\phi, - \mathbf{P}_{123}^{\bHBlue}) =
  \cf_3^{\phi^3}(1^\phi, 2^\phi, 3^\phi) \, ,
\end{gather}
the expression is rewritten as
\begin{align}
  \left[
    (s_{24} + s_{34}) \cf_4^{{\phi}^3}(1^\phi, 4^g, 2^\phi, 3^\phi) +
    s_{34} \cf_4^{{\phi}^3}(1^\phi, 2^\phi, 4^g, 3^\phi)
  \right]_{s_{4q} = m^2}
  = \hspace{-3cm}
  &
  \notag \\
  &
  \ca_3^{\bar{\phi}^3}(q^{\bHBlue}, \mathbf{P}_{123}^{\bHBlue}, 4^g)
  \times
  \cf_3^{{\phi}^3}(1^\phi, 2^\phi, 3^\phi) \, ,
\end{align}
which is the corresponding FF factorization.

\newcommand{\HQCD}{\text{HQCD}}
\newcommand{\PQCD}{\text{PQCD}}
\newcommand{\LaHQCD}{\La^{\HQCD}}
\newcommand{\LaQCD}{\La^{\text{QCD}}}
\newcommand{\dHQCD}{\overline{\HQCD}}
\newcommand{\LadHQCD}{\La^{\dHQCD}}
\newcommand{\LaPQCD}{\La^{\PQCD}}
\newcommand{\Fgamma}{F^{(\photon)}_{\mu\nu}}
\newcommand{\ampPQCDFour}{\amp_{4}^{\overline{\PQCD}}}
\newcommand{\ampPQCDFourFull}{\ampPQCDFour(1^{\bar{\psi}}, 2^\psi, 3^g, q^{\bCPurple})}

\subsection[Operator \protect\texorpdfstring{$\trpp$}{bar psi psi}]{Operator $\trpp$}

\newcommand{\ampHQCDThree}{\amp_{3}^{\dHQCD}}
\newcommand{\ampHQCDFour}{\amp_{4}^{\dHQCD}}
\newcommand{\ampHQCDFourFull}{\ampHQCDFour(1^{\bar{\psi}}, 2^\psi, 3^g, q^{\bHBlue})}
\newcommand{\ampTrPhi}{\amp_{4}^{\tr(\phi^2)}}
\newcommand{\ampHQCDThreeFull}{\ampHQCDThree(1^{\bar{\psi}}, 2^\psi, q^{\bHBlue})}

In this and the following subsection, we apply the dyeing and bleaching strategy to FFs involving fermion-related operators, based on the Lagrangian
\begin{equation}
    \LaQCD =
    - \frac{1}{4} \tr(F_{\mu \nu} F^{\mu \nu})
    + i \bp \gamma^\mu D_\mu \psi \, .
\end{equation}
The color indices of the fundamental fermions in $\trpp$ and $\trpGp$
are implicitly contracted.
FFs of $\trpp$ can be interpreted as amplitudes of the Higgs QCD (HQCD) theory,
\begin{equation}
    \LaHQCD = \LaQCD + \La^{\text{Higgs}} + \La^{H \psibarpsi}\, ,
    \quad
    \La^{H\psibarpsi} =
    \lambda_{\psibarpsi} H \psibarpsi\, ,
\end{equation}
which is dyed as
\begin{equation}
    \LadHQCD = \LaQCD + \LadHiggs + \La^{\bHBlue \psibarpsi}\, ,
    \quad
    \La^{\bHBlue \psibarpsi} =
    \lambda_{\psibarpsi} H^a \bar{\psi}^i(T^a)^{ij} \psi^j \,.
\end{equation}
The indices $i,j$ are fundamental color indices.
The bleaching is applied using
\begin{equation}
    \blcBH\left[(T^a)^{ij}\right] = \idd^{ij} = \delta^{ij} \,,
    \label{eq:bleach_Ta}
\end{equation}
which leads to
\begin{equation}
    \blcBH \left(\La^{\bHBlue \psibarpsi}\right) =
\La^{H\psibarpsi}
    \, .
\end{equation}
At the amplitude level,
\begin{align}
    \blcqBH
    \left(
        \amp_{n + 1}^{\dHQCD}(1^{\bp}, 2^\psi, 3^g, \ldots, n^g, q^{\bHBlue})
    \right)
    & =
    \amp_{n + 1}^{\HQCD}(1^{\bp}, 2^\psi, 3^g, \ldots, n^g, q^H)
    \notag \\
    &=
    \ff_{n, \trpp}(1^{\bp}, 2^\psi, 3^g, \ldots, n^g)
    \, .
\end{align}

\subsubsection*{The three-point case}

The new element in this theory is the spinor.
Under the double copy, keeping only the symmetric representation, $\psi$ produces a massless vector boson $\photon_\mu$, which we refer to as a photon.

To verify this interpretation, we examine the simplest case of the three-point amplitude $\ampHQCDThreeFull$.
This amplitude contains only one Feynman diagram with the cubic vertex $\bar{\psi}$-$\psi$-$\bHBlue$,
\begin{equation}
    \ampHQCDThreeFull = (T^q)^{i_1 i_2} \bar{u}_1 v_2 \, .
\end{equation}
$\bar{u}_1$ is the spinor of $1^{\bar{\psi}}$, and $v_2$ is the spinor of $2^{\psi}$.
Using the spinor helicity formalism with the configuration $\bar{\psi}^+$ and $\psi^+$, the amplitude can be rewritten as
\begin{equation}
    \amp_3^{\dHQCD}(1^{\bar{\psi}^+}, 2^{\psi^+}, q^{\bHBlue}) =
    (T^q)^{i_1i_2} \spb{1}{2} \, .
\end{equation}
Performing the double copy by replacing the color factor $(T^q)^{i_1 i_2}$ with the kinematic numerator, we obtain
\begin{equation}
    \mathcal{G}_3^{++} = \spb{1}{2}^2 \,.
\end{equation}
The factor $\spb{1}{2}^2$ corresponds to the anti-self-dual part of the photon kinetic term $\frac{1}{4} \Fgamma F^{(\photon)\,\mu\nu}$.
The opposite helicity configuration $(\bar{\psi}^-, \psi^-)$ gives the self-dual part $\spa{1}{2}^2$.
This confirms that $\mathcal{G}_3^{++}$ corresponds to the three-point gravitational amplitude $\gA_3(1^{\photon}, 2^{\photon}, q^H)$ for the given helicity.

\subsubsection*{The four-point case}

\begin{figure}
    \begin{center}
        \includegraphics[width=0.75\textwidth]{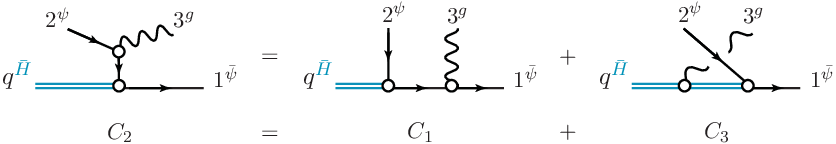}
    \end{center}
    \caption{
        The color relation of $\ampHQCDFourFull$.
        Colors are represented by half-ladder diagrams.
    }\label{fig:color_relation_psipsi}
\end{figure}

The amplitude $\ampHQCDFourFull$ has the same topologies as $\ampppgq$ of $\LadYMSH_{\tr(\phi^2)}$ studied in Section~\ref{sec:dyedphi2}, with the scalar $\phi$ lines in Figure~\ref{fig:a4_cubic} replaced by quark lines.
We denote the latter amplitude by $\ampTrPhi$ below.
The color factors of $\ampHQCDFour$ involve both the generators $T^a$ and the structure constant $f^{abc}$, which are related by
\begin{equation}
    f^{abc}T^c = T^a T^b - T^b T^a \, .
\end{equation}
For the three trivalent topologies, the color factors read
\begin{equation}
    C_1^{\psibarpsi} = (T^{a_3}T^q)^{i_1 i_2},\quad
    C_2^{\psibarpsi} = (T^q T^{a_3})^{i_1 i_2},\quad
    C_3^{\psibarpsi} = f^{qa_3 b}(T^b)^{i_1 i_2}\, ,
\end{equation}
satisfying the color relation $C_1^{\psibarpsi} - C_2^{\psibarpsi} + C_3^{\psibarpsi} = 0$.
Using the color relation to project out $C_2^{\psibarpsi}$, we work in the basis $\{C_1^{\psibarpsi}, C_3^{\psibarpsi}\}$, shown in Figure~\ref{fig:color_relation_psipsi}.
In the half-ladder representation, this basis corresponds to fixing $q$ and $1^{\bar{\psi}}$ at the two endpoints.

The propagator matrix and BCJ relations of $\{C_1^{\psibarpsi}, C_3^{\psibarpsi}\}$ then match those of $\ampTrPhi$ in the basis $\{\tr(T^{a_1}T^{a_3}T^{a_2}T^{q}),\ \tr(T^{a_1}T^{a_2}T^{a_3}T^{q})\}$, so the KLT kernel remains identical.
The partial amplitude associated with $C_1^{\psibarpsi}$ is bleaching invariant, which ensures that the double copy of the dyed amplitude agrees with the FF result.

\subsection[Operator \protect\texorpdfstring{$\trpGp$}{bar psi gamma mu psi}]{Vector operator $\trpGp$}

Compared with $\trpp$, the situation for $\trpGp$ is more subtle because the operator insertion carries a Lorentz vector index.
The operator is mapped to a massive vector field $V_\mu$, and the corresponding effective theory is a Proca extension of QCD, which we abbreviate as Proca QCD (PQCD) in this subsection (see e.g.~\cite{Zerwekh:2012cy}),
\begin{gather}
    \LaPQCD = \LaQCD + \La^{\text{Proca}} + \La^{\psiGabarpsi}\, ,
    \quad
    \La^{\text{Proca}} =
    -\frac{1}{4} \FV_{\mu \nu} \FV^{\mu \nu} + \frac{1}{2} m^2 V_\mu V^\mu \, ,
    \notag \\
    \La^{\psiGabarpsi} = \lambda_{\gamma} V_\mu \psiGabarpsi \, ,
\end{gather}
with $\FV_{\mu\nu} = \partial_\mu V_\nu - \partial_\nu V_\mu$.
Dyeing $V_\mu$ into the adjoint representation, $\bCPurple_\mu = V_\mu^a T^a$, gives the dyed PQCD Lagrangian
\begin{gather}
    \La^{\overline{\PQCD}} = \LaQCD + \La^{\overline{\text{Proca}}} + \La^{\bCPurple \psiGabarpsi} + \La^{F \bCPurple \bCPurple}
    \, ,
\end{gather}
with
\begin{gather}
    \La^{\overline{\text{Proca}}} =
    -\frac{1}{4} \FC_{\mu \nu}^a \FC^{a\,\mu \nu} + \frac{1}{2} m^2 V_\mu^a V^{a,\mu} \, ,
    \notag \\
    \La^{\bCPurple \psiGabarpsi} =
    \lambda_\gamma V_\mu^a \bar{\psi}^i (T^a)^{ij} \gamma^\mu \psi^j \,,
    \quad
    \La^{F \bCPurple \bCPurple} = - \frac{1}{2} g f^{abc} F^{a,\mu\nu} V^b_\mu V^c_\nu\,.
\end{gather}
The kinetic term is dyed with $\FC_{\mu\nu}^a = D_\mu V^a_\nu - D_\nu V^a_\mu$.
The dyed vector field $\bCPurple_\mu$ is a massive vector boson in the
adjoint representation.\footnote{
    A massive adjoint vector also appears in phenomenological models as
    a coloron, emerging from the spontaneous breaking of an extended color
    gauge symmetry $SU(N)\times SU(N)$ down to its diagonal $SU(N)$
    subgroup~\cite{Hill:1991sq,Simmons:1996ws}.
}
Unlike the gluon, whose gauge transformation involves a derivative term,
the dyed vector transforms covariantly as
$\bCPurple_\mu \to U \, \bCPurple_\mu \, U^\dagger$.\footnote{
There are also studied of color-kinematics duality for off-shell 
vector currents~\cite{Mastrolia:2015maa,Mafra:2015vca,Ben-Shahar:2021zww} which are gauge dependent.}
As a consequence, the mass term in the Proca Lagrangian remains gauge
invariant.

Unlike the dyeing constructions in previous sections, the $\overline{\text{PQCD}}$ contains an additional interaction $\La^{F \bCPurple \bCPurple}$.
This term is required by $\CK$ duality: without it, no consistent set of kinematic numerators satisfying the $\CK$ relations exists.
With $\La^{F \bCPurple \bCPurple}$, the $\bCPurple$-$g$-$\bCPurple$ and $\bCPurple$-$g$-$g$-$\bCPurple$ Feynman vertices take the same form as the pure gluon three- and four-point vertices.
Moreover, since its color factor is $f^{abc}$, bleaching annihilates it as
\begin{equation}
    \blc_{\bCPurple} \left(\La^{F \bCPurple \bCPurple} \right) = 0 \,,
\end{equation}
while bleaching the remaining terms gives
\begin{equation}
    \blc_{\bCPurple} \left(\La^{\bCPurple\psiGabarpsi}\right) = \La^{\psiGabarpsi} \, ,
    \quad
    \blc_{\bCPurple} \left(\La^{\overline{\text{Proca}}}\right) = \La^{\text{Proca}} \, .
\end{equation}
At the amplitude level,
\begin{align}
    \blc_{q^{\bCPurple}}
    \left(
        \amp_{n + 1}(1^{\bp}, 2^\psi, 3^g, \ldots, n^g, q^{\bCPurple})
    \right)
    &=
    \amp_{n + 1}(1^{\bp}, 2^\psi, 3^g, \ldots, n^g, q^V)
    \notag \\
    &=
    \ff_{n, \trpGp}(1^{\bp}, 2^\psi, 3^g, \ldots, n^g)
    \, .
\end{align}

Under the double copy, the massive vector $\bCPurple_\mu$ generates a massive spin-2 field $G_{\mu\nu}$ coupled to the electromagnetic energy-momentum tensor $T^{(\photon),\mu\nu}$:
\begin{equation}
    \bCPurple_\mu \, \psiGabarpsi \quad \xrightarrow{\text{double copy}} \quad G_{\mu\nu} \, T^{(\photon),\mu\nu} \, ,
\end{equation}
where
\begin{equation}
    T^{(\photon),\mu\nu} =
    - F^{(\photon),\mu\rho} F^{(\photon),\nu}{}_{\rho}
    + \frac{1}{4} \eta^{\mu\nu} F^{(\photon), \rho\sigma} F^{(\photon)}_{\rho\sigma}
    \, .
\end{equation}

To verify this prediction, we examine the three-point amplitude.
The dyed three-point amplitude contains a single Feynman diagram with the cubic vertex $V_\mu^a$-$\bar\psi$-$\psi$,
\begin{equation}
    \amp_{3}^{\overline{\text{PQCD}}}(1^{\bp}, 2^\psi, q^{\bCPurple}) = (T^a)^{i_1 i_2} \, (\bar{u}_1 \gamma^\mu  v_2) \varepsilon_{q,\mu} \, .
\end{equation}
In the spinor helicity formalism, choosing the configuration $\bar{u}_1^+$ and $v_2^-$, the amplitude becomes
\begin{equation}
    \amp_{3}^{\overline{\text{PQCD}}}(1^{\bp^+}, 2^{\psi^-}, q^{\bCPurple}) =
    (T^a)^{i_1 i_2}  \, [1 | \gamma^\mu | 2 \rangle \varepsilon_{q,\mu}
    \, .
\end{equation}
Performing the double copy by replacing the color factor $(T^a)^{i_1 i_2}$ with the kinematic numerator, we obtain
\begin{equation}
    \mathcal{G}_3^{+-} =
    \, [1 | \gamma^\mu | 2 \rangle [1 | \gamma^\nu | 2 \rangle \varepsilon_{q,\mu} \varepsilon_{q,\nu}
    = \sigma^\mu_{\alpha \dot{\alpha}} \sigma^\nu_{\beta\dot{\beta}} \varepsilon_{q,\mu} \varepsilon_{q,\nu}
    (\tilde{\lambda}_1^{\dot{\alpha}} \tilde{\lambda}_1^{\dot{\beta}} \lambda_2^{\alpha} \lambda_2^\beta )
    \, .
\end{equation}
Here $\lambda_i^\alpha$ and $\tilde{\lambda}_i^{\dot{\alpha}}$ are the Weyl spinors of particle $i$.
The tensor structure in $\mathcal{G}_3^{+-}$ is symmetric, traceless, and conserved with respect to the momentum $q$.
This result exactly matches $T^{(\photon),\mu\nu}$ for the given configuration.
The opposite helicity configuration, $\bar{u}_1^-$ and $v_2^+$, provides the piece related by $1 \leftrightarrow 2$ symmetry.
Together, they reconstruct the fully symmetric tensor $T^{(\photon),\mu\nu}$, confirming the double copy prediction.

The four-point amplitude $\ampPQCDFourFull$ shares the same topologies, color factors, and color relations as $\ampHQCDFourFull$ discussed above, and we do not repeat the analysis here.
Using the same color basis, the partial amplitude associated with $C_1^{\psibarpsi}$ is again bleaching invariant, ensuring that the double copy of the dyed amplitude agrees with the FF result.

The double copy maps $\psi \to \photon$ and the $q$-leg to the massive spin-2 field $q^H \to G$, giving
\begin{equation}
    \gA_4(1^{\photon}, 2^{\photon}, 3^h, q^G)\, .
\end{equation}
We have verified that it passes all factorization tests on the physical poles $s_{13}$, $s_{23}$, and $(s_{12} - m^2)$, and is diffeomorphism invariant, confirming it as a valid gravitational scattering amplitude.
  

\section{Summary and discussion}
\label{sec:discussion}

In this paper, we have unified form factors and amplitudes via a dyeing mechanism. By ``dyeing'' the color-singlet operator --- assigning it an adjoint color index --- we lift the problem to the level of fully colored amplitudes.
The form factor (FF) is then the ``bleached'' (color-singlet) limit of an amplitude involving a colored massive particle (the ``dyed'' Higgs or vector field).
The main results are summarized as follows:
\begin{itemize}
  \item
  The dyed amplitude generates additional topologies beyond those of the FF. These contributions have zero color factors after bleaching.
  They contribute to the double copy without introducing ``spurious'' poles in the previous FF construction.

  \item
  The hidden factorization relations of the FFs acquire a clear interpretation in the dyed theory: they emerge as on-shell limits of BCJ relations among dyed amplitudes, which in turn follow from CK duality in the dyed theory.

  \item
The framework generalizes straightforwardly to multi-Higgs dyed amplitudes (corresponding to multiple operator insertions) and yields consistent gravity results via double copy.
We identify a new ``scalar-ordering'' structure,
  which appears at the gauge level and is inherited by gravity through the double copy.
  Generalized hidden factorization relations are also derived. 

  \item
The mechanism extends to more general types of operators.
For $\tr(\phi^m)$, the dyed theory involves color relations based on fully symmetric $d^{a_1 \cdots a_{m+1}}$ tensors.
Notably, the operator $\trpGp$, associated with an effective massive vector field, double copies to a stress-tensor operator.
This suggests a potential route toward higher-spin operator insertions.
\end{itemize}
Several directions remain open for future work:
\begin{itemize}
  \item A natural next step is to extend the dyeing framework to pure gluon operators.
     For instance, the operator $\tr(F^2)$
    appears as the effective Higgs--gluon coupling in the heavy-top
    limit~\cite{Wilczek:1977zn, Shifman:1979eb, Dawson:1990zj, Kniehl:1995tn}, and the corresponding form factor is equivalent to Higgs-plus-gluons amplitudes \cite{Dixon:2004za,Brandhuber:2011tv}.
    In \cite{Lin:2023rwe}, a double-copy picture for the $\tr(F^2)$ form factor was proposed, based on a decomposition into scalar-operator form factors.
    In the dyed picture, the dyed vertex would read
    \begin{equation}
      H \tr(F^2)
      \quad \xrightarrow{\ \text{dyeing}\ } \quad
      d^{abc} H^a F^{b, \mu\nu} F^c_{\mu\nu}
      \,.
    \end{equation}
    But preliminary checks show that
    the resulting gravity object does not factorize correctly on all
    physical poles. This indicates that a naive dyeing construction is insufficient for
    $\tr(F^2)$ compared to the operators
    studied in this work.

  \item Extending the dyeing framework
    to loop order is another important direction.
Color-kinematics duality has been used to
    construct loop form factors in both ${\cal N}=4$ SYM theory~\cite{Boels:2012ew,Yang:2016ear,Lin:2021kht,Lin:2021lqo} and pure YM theory~\cite{Li:2022tir,Li:2024wzj}.
    In those constructions, the Jacobi relations involving internal lines associated with the
    operator insertion are generally neglected, and the corresponding gravity double-copy
    quantity remains obscure.
The dyed theory offers a new perspective on this problem:
    once the operator is promoted to a colored field, it can circulate
    inside the loop as a massive internal line, enlarging the set of topologies; these additional topologies 
    possess relatively few symmetries, providing more freedom for constructing
    CK-satisfying numerators.
We have given a preliminary one-loop example illustrating this mechanism
    in Appendix~\ref{sec:loop} and leave a complete construction to
    future work.

  \item The dyeing procedure promotes the original theory
    into a special EFT --- the dyed theory --- by assigning a gauge
    group index to the operator.
    In this work we have focused on its amplitude-level aspects.
    One can systematically explore its EFT
    structure and possible extensions, in connection with the
    broader study of color-kinematics duality and double copy for
    higher-dimensional
    operators~\cite{Broedel:2012rc,Bonnefoy:2021qgu}.
    Related CK-dual constructions involving symmetric color tensors and
    mixed $d$-$f$ color structures have also appeared in EFT
    double-copy studies~\cite{Carrasco:2021ptp,Carrasco:2022jxn,Carrasco:2026hxf}.
    Another interesting direction is to determine whether dyed amplitudes provide a natural embedding into the CHY formalism, extending previous four-dimensional connected descriptions to general dimensions \cite{He:2016jdg,Brandhuber:2016xue,He:2016dol}.
\end{itemize}


\acknowledgments

We would like to thank Zhiming Cai, Zeyu Li and Guanda Lin for discussions.
This work is supported by the National Natural Science Foundation of
China (Grants No.~12425504, 12447101, 12247103) and by the Chinese
Academy of Sciences (Grant No. YSBR-101).
We also acknowledge the support of the HPC Cluster of ITP-CAS.


\appendix
\section[A preliminary dyed one-loop integrand: \protect\texorpdfstring{$H \to \phi\phi$}{H to phi phi}]{A preliminary dyed one-loop integrand: $H \to \phi\phi$}
\label{sec:loop}

\newcommand{\dampThree}{\mathcal{I}^{\tr(\phi^2)}_{3,(1)}(1^\phi, 2^\phi, q^{\bHBlue})}
\newcommand{\ffTwo}{\ff^{\tr(\phi^2)}_{2,(1)}(1^\phi, 2^\phi)}
\newcommand{\gIthree}{\mathcal{I}^{\text{grav}}_{3,(1)}(1^\phi, 2^\phi, q^H)}

In this appendix we consider a
one-loop integrand for the simplest case:
the $\trphitwo$ form factor $\ffTwo$ and its dyed counterpart $\dampThree$.

\begin{figure}
  \begin{center}
    \includegraphics[width=0.85\textwidth]{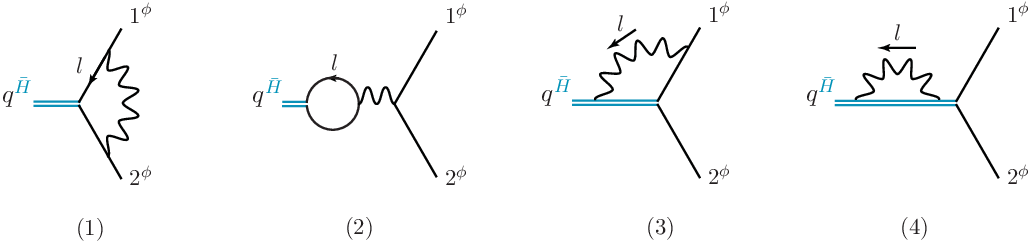}
  \end{center}
  \caption{Trivalent topologies of the dyed integrand $\dampThree$. }
  \label{fig:topo_1loop_H_phi_phi}
\end{figure}

The dyed integrand is organized by the
$4$ cubic topologies shown in Figure~\ref{fig:topo_1loop_H_phi_phi}.
For compactness, we write
\begin{equation}
  \dampThree =
  \sum_{\sigma_2}  \sum_{i=1}^{4}
  \int \frac{d^D \ell}{(2\pi)^D}
  \frac{1}{S_i} \frac{C_i N_i}{D_i} \, ,
  \label{eq:loop_decomp}
\end{equation}
where $\sum_{\sigma_2}$ sums over permutations of identical external $\phi$ states.
$S_i$ is the symmetry factor of the $i$-th topology, with $S_1 = 2$, $S_2 = 4$, $S_3 = 1$, $S_4 = 2$.
$D_i$ denotes the product of inverse propagators of the corresponding topology,
and $\Nt{i}{p_1}{p_2}{\ell}$ is the kinematic numerator
with $p_{1,2}$ denoting the external momenta and $\ell$ the loop momentum.
We use the shorthand $N_i \equiv \Nt{i}{p_1}{p_2}{\ell}$ when the arguments are in the canonical ordering.
The color factors are
\begin{gather}
  C_1 = \td^{q b_1 b_3} \tf^{b_1 a_1 b_2} \tf^{b_2 a_2 b_3}, \quad
  C_2 = \td^{q b_1 b_3} \tf^{b_1 b_2 b_3} \tf^{b_2 a_1 a_2} \, ,
  \notag \\
  C_3 = \tf^{q b_1 b_3} \tf^{b_2 b_1 a_1} \td^{b_3 b_2 a_2}, \quad
  C_4 = \tf^{q b_1 b_3} \tf^{b_3 b_1 b_2} \td^{b_2 a_1 a_2} \, .
  \label{eq:loop_cf_34}
\end{gather}
Under the color algebra, all color factors are proportional to $\tilde{C} = N_c\, \td^{q a_1 a_2}$, with coefficients $\set{1,0,1,2}$ for $C_1$ through $C_4$ respectively.
The factor $N_c$ is the number of colors of the gauge group $SU(N_c)$.

To express the integrand of each term, we define two propagator bases:
\begin{equation}
  \Pbas^{\text{A}} = \set{\ell^2,\; (\ell + p_1)^2,\; (\ell + p_1 + p_2)^2} \, ,
  \quad
  \Pbas^{\text{B}} = \set{\ell^2,\; (\ell + p_1)^2,\; (\ell + p_1 + p_2)^2 - m^2} \, .
\end{equation}
$\Pbas^{\text{A}}$ is used for topologies $(1)$ and $(2)$, whose internal lines are all massless, while $\Pbas^{\text{B}}$ is used for topologies $(3)$ and $(4)$, which involve the massive dyed Higgs propagator.
We denote by $\Pbas^{\text{X}}[\alpha_1, \alpha_2, \alpha_3]$ the product of propagators from basis $\Pbas^{\text{X}}$, where the integer $\alpha_i$ is the power of the $i$-th propagator in the denominator.
The integrand of each term in (\ref{eq:loop_decomp}) can be written as
\begin{equation}
  \mathbb{I}_1 = N_1 \times \Pbas^{\text{A}}[1,1,1] \, ,
  \quad
  \mathbb{I}_2 = N_2 \times \frac{\Pbas^{\text{A}}[1,0,1]}{s_{12}} \, ,
  \quad
  \mathbb{I}_3 = N_3 \times \Pbas^{\text{B}}[1,1,1] \, ,
  \quad
  \mathbb{I}_4 = N_4 \times \frac{\Pbas^{\text{B}}[1,0,1]}{s_{12} - m^2} \, .
\end{equation}
With the color factors and symmetry factors, this can be expressed as
\begin{equation}
  \dampThree = \tilde{C} \int \frac{d^D \ell}{(2\pi)^D} \left[\frac{1}{2}\,\mathbb{I}_1 + \mathbb{I}_3 + \mathbb{I}_4 + (p_1 \leftrightarrow p_2) \right]\, .
  \label{eq:loop_expand}
\end{equation}
Here $C_2 = 0$ eliminates topology $(2)$.
The term $(p_1 \leftrightarrow p_2)$ accounts for the permutation of $\phi$ states.

We do not construct the ordinary amputated on-shell amplitude.
The CK relations in the next subsection are imposed only within the chosen truncated
set of topologies.
The construction is restricted as follows:
\begin{itemize}
  \item Scaleless sectors, such as massless tadpoles and massless
    bubbles attached to on-shell massless external legs, are removed
    in dimensional regularization.
  \item Massive tadpoles are not scaleless, but are left outside
    in this example.
\end{itemize}
The finite expressions below should therefore be read as the
tadpole-subtracted part of the dyed integrand.
This example serves only as an initial illustration of the dyed
mechanism at loop level, and a complete one-loop construction is
beyond our scope.

Topology $(4)$ is a one-particle-reducible self-energy topology of the
dyed massive leg and carries the pole $s_{12} - m^2$.
Because of this pole, we keep the dyed leg partially off shell,
treating $s_{12}$ and $m^2$ as independent variables until the pole
is separated.

\subsubsection*{CK duality}

\begin{figure}
  \begin{center}
    \includegraphics[width=0.6\textwidth]{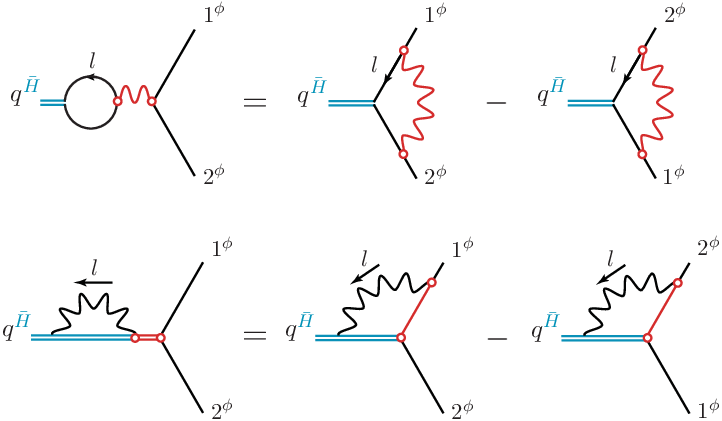}
  \end{center}
  \caption{
    Two color relations of $\dampThree$.
    The red line with hollow circles at both ends represents the propagator involved in the CK operation.
  }\label{fig:CK_example_1loop_H_phi_phi}
\end{figure}

We now impose the dual CK relations on the numerators.
Within the restricted topology set, these relations reduce the cubic topologies to a smaller independent subset,
whose numerators are called CK masters.
For this case, the two CK relations shown in Figure~\ref{fig:CK_example_1loop_H_phi_phi} express all numerators in terms of topologies $(1)$ and $(3)$,
\begin{equation}
  N_2 = N_1 - \Nt{1}{p_2}{p_1}{\ell}\, ,
  \quad
  N_4 = N_3 + \Nt{3}{p_2}{p_1}{\ell} \, .
  \label{eq:loop_masters}
\end{equation}
Hence $N_1$ and $N_3$ are chosen as masters.

The numerator $N_i$ has mass dimension $2$.
Each master numerator can be written as a linear combination of $5$ independent scalar products:
\begin{equation}
  \Nt{i}{p_1}{p_2}{\ell}
  = c_{i,1} \, (p_1 \cdot p_2)
  + c_{i,2} \, (p_1 \cdot \ell)
  + c_{i,3} \, (p_2 \cdot \ell)
  + c_{i,4} \, \ell^2
  + c_{i,5} \, m^2 \, ,
  \quad i \in \set{1,3} \, ,
  \label{eq:loop_ansatz}
\end{equation}
where $c_{i,j}$ are undetermined coefficients
and $m$ is the mass of the dyed Higgs.
Since $s_{12}$ and $m^2$ are treated as independent variables,
both $p_1 \cdot p_2$ and $m^2$ appear in the ansatz.
The total number of parameters is $10$.

\subsubsection*{Symmetry}

\begin{figure}
  \begin{center}
    \includegraphics[width=0.35\textwidth]{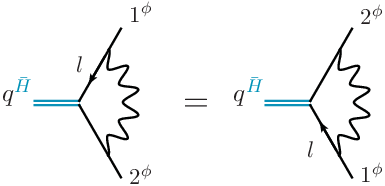}
  \end{center}
  \caption{
    Symmetry of the topology (1).
  }
  \label{fig:symmetry_1loop_H_phi_phi}
\end{figure}

Numerators are required to obey the same symmetry as their corresponding topologies.
For topology $(1)$, it is invariant under a vertical flip, which is shown in Figure~\ref{fig:symmetry_1loop_H_phi_phi},
\begin{equation}
  N_1 = \Nt{1}{p_2}{p_1}{-\ell - p_1 - p_2} \, .
  \label{eq:loop_sym_1}
\end{equation}
Substituting the ansatz (\ref{eq:loop_ansatz}) into (\ref{eq:loop_sym_1})
yields $c_{1,2} + c_{1,3} = 2 c_{1,4}$,
which is used to reduce one of the parameters.
The remaining symmetries,
\begin{gather}
  N_2 = -\Nt{2}{p_1}{p_2}{-\ell - p_1 - p_2} \, , \quad
  N_2 = -\Nt{2}{p_2}{p_1}{\ell} \, ,
  \notag \\
  N_4 = \Nt{4}{p_2}{p_1}{\ell} \, ,
  \label{eq:loop_sym_234}
\end{gather}
are automatically satisfied once both the CK relations~(\ref{eq:loop_masters})
and the symmetry~(\ref{eq:loop_sym_1}) of topology $(1)$ are imposed,
and do not further constrain the master numerators.

\subsubsection*{Unitarity cuts}
\label{sub:loop_unitarity}

\begin{figure}
  \begin{center}
    \includegraphics[width=0.8\textwidth]{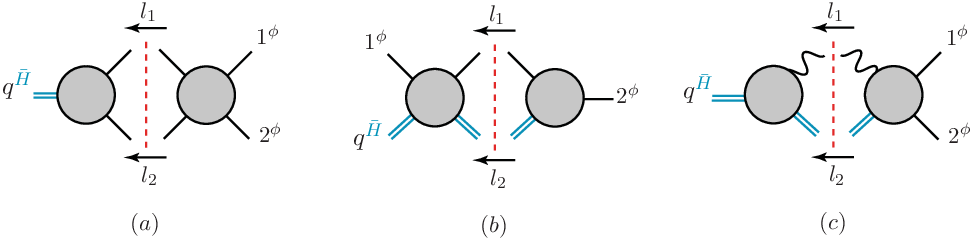}
  \end{center}
  \caption{
    Three planar cuts considered for $\dampThree$.
  }\label{fig:planar_cuts_1loop_H_phi_phi}
\end{figure}

We impose three planar unitarity cuts shown in Figure~\ref{fig:planar_cuts_1loop_H_phi_phi} on the integrand to constrain the remaining unknown coefficients, which suffices to fix the residue of the integrand at the pole $s_{12} = m^2$ at the gauge level.
Cut $(a)$ puts two massless internal lines on-shell, while cuts $(b)$ and $(c)$ each put one massless line and one massive dyed Higgs internal line on-shell.

We start with planar cut $(a)$.
This cut receives contributions from topologies $(1)$ and $(2)$, giving
\begin{equation}
  \mathrm{Cut}_{(a)} = \frac{\Nt{1}{p_1}{p_2}{\ell_1}}{(\ell_1 + p_1)^2} + \frac{\Nt{2}{p_1}{p_2}{\ell_1}}{s_{12}} \, ,
\end{equation}
with the conditions $\ell_1^2 = \ell_2^2 = 0$.
On the other hand, the cut can be computed using tree-level amplitudes.
Since the internal lines crossing the cut are scalars, no polarization sum is needed, and the cut factorizes as
\begin{equation}
  \mathrm{Cut}_{(a)} = \ca_3^{\tr(\phi^2)}(q^{\bHBlue}, -\ell_1^\phi, -\ell_2^\phi) \times \ca_4^{\tr(\phi^2)}(1^\phi, 2^\phi, \ell_2^\phi, \ell_1^\phi) \, .
\end{equation}
Comparing the two expressions constrains the ansatz coefficients.
For this case, $2$ additional constraints are found.

Cut $(b)$ receives contribution from topology $(3)$ only, giving
\begin{equation}
  \mathrm{Cut}_{(b)} = \frac{\Nt{3}{p_1}{p_2}{\ell_1 - p_1}}{(\ell_1 - p_1)^2} \, ,
\end{equation}
with the on-shell conditions $\ell_1^2 = 0$ and $\ell_2^2 = m^2$.
The cut factorizes as
\begin{equation}
  \mathrm{Cut}_{(b)} = \ca_3^{\tr(\phi^2)}(q^{\bHBlue}, -\ell_1^\phi, -\ell_2^{\bHBlue}) \times \ca_3^{\tr(\phi^2)}(2^\phi, \ell_2^{\bHBlue}, \ell_1^\phi) \, .
\end{equation}
This gives $2$ new constraints on the $N_3$ ansatz coefficients.

Cut $(c)$ involves the subtlety of topology $(4)$.
For cuts $(a)$ and $(b)$, since no pole $s_{12} - m^2$ appears, one can safely set $q^2 = m^2$ on-shell.
For cut $(c)$ this is not possible, as the propagator $s_{12} - m^2$ would diverge.
This cut receives contributions from two topologies,
\begin{equation}
  \mathrm{Cut}_{(c)} = \frac{\Nt{3}{p_1}{p_2}{\ell_1}}{(\ell_1 + p_1)^2} + \frac{\Nt{4}{p_1}{p_2}{\ell_1}}{s_{12} - m^2} \, ,
\end{equation}
with $\ell_1^2 = 0$ and $\ell_2^2 = m^2$.
The cut factorizes as
\begin{equation}
  \mathrm{Cut}_{(c)} = \sum_{\lambda = \pm} \ca_3^{\tr(\phi^2)}(q^\bHBlue, -\ell_1^{g,\lambda}, -\ell_2^{\bHBlue}) \times \ca_4^{\tr(\phi^2)}(1^\phi, 2^\phi, \ell_2^{\bHBlue}, \ell_1^{g,-\lambda}) \, ,
\end{equation}
where $\ell_1$ is a gluon, requiring a polarization sum.
To match the Feynman diagram result, we use
\begin{equation}
  \sum_{\lambda = \pm} \epsilon^\lambda_\mu(\ell) \epsilon^{*, -\lambda}_\nu(-\ell) = \eta_{\mu \nu}\, ,
\end{equation}
which corresponds to Feynman gauge.
This provides one additional constraint.

The above three planar cuts provide $5$ constraints, which leaves $4$ free parameters in total.
The resulting numerators are
\begin{align}
  N_1 &=
    -\tfrac{1}{2}(c_{1,1} - c_{1,2} + 2)\,m^2
    + (c_{1,2} + 1)\,\ell^2
    + c_{1,2}\,(\ell \cdot p_1)
    + (c_{1,2} + 2)\,(\ell \cdot p_2)
    + c_{1,1}\,(p_1 \cdot p_2) \, , \notag \\
  N_2 &=
    -2\,(\ell \cdot p_1)
    + 2\,(\ell \cdot p_2) \, , \notag \\
  N_3 &=
    -\tfrac{1}{2}(c_{3,1} - 4)\,m^2
    + \tfrac{1}{2}(c_{3,2} - 2)\,\ell^2
    + c_{3,2}\,(\ell \cdot p_1)
    + (c_{3,1} - 3)\,(\ell \cdot p_2)
    + c_{3,1}\,(p_1 \cdot p_2) \, , \notag \\
  N_4 &=
    \begin{aligned}[t]
      &-(c_{3,1} - 4)\,m^2
      + (c_{3,2} - 2)\,\ell^2 \\
      &+ (c_{3,1} + c_{3,2} - 3)\,(\ell \cdot p_1)
      + (c_{3,1} + c_{3,2} - 3)\,(\ell \cdot p_2) \\
      &+ 2\,c_{3,1}\,(p_1 \cdot p_2) \, .
    \end{aligned}
\end{align}

\subsubsection*{Integrand reduction}

\begin{figure}
  \begin{center}
    \includegraphics[width=0.4\linewidth]{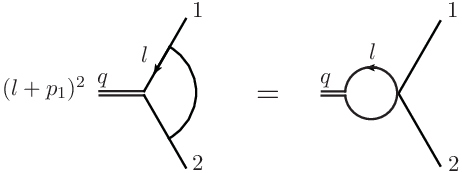}
  \end{center}
  \caption{Reducing a triangle topology to a bubble by cancelling an internal propagator.}
  \label{fig:shrink}
\end{figure}

The CK solution above contains $4$ free parameters, reflecting generalized gauge freedom.
We display the result in a reduced propagator-basis representation.
For example, the full propagator product of topology $(1)$ is
\begin{equation}
  \Pbas^{\text{A}}[1, 1, 1] = \frac{1}{\ell^2 \, (\ell + p_1)^2 \, (\ell + p_1 + p_2)^2} \, .
\end{equation}
If the numerator contains a factor $(\ell + p_1)^2$, it cancels one propagator in the denominator, reducing the topology
\begin{equation}
  (\ell + p_1)^2 \times \Pbas^{\text{A}}[1, 1, 1] = \Pbas^{\text{A}}[1, 0, 1] \, .
\end{equation}
This is plotted in Figure~\ref{fig:shrink}.

Decomposing each numerator into the propagator basis, the integrands are reduced to
\begin{gather}
  \mathbb{I}_1 = \tfrac{1}{2}\Big[(c_{1,2} + 2)\,\Pbas^{\text{A}}[0,1,1] - 2\,\Pbas^{\text{A}}[1,0,1] + (c_{1,2} + 2)\,\Pbas^{\text{A}}[1,1,0] - 4\,m^2\,\Pbas^{\text{A}}[1,1,1]\Big] \, , \notag \\
  \mathbb{I}_2 = \tfrac{1}{m^2}\Big[\Pbas^{\text{A}}[0,0,1] - 2\,\Pbas^{\text{A}}[1,-1,1] + \Pbas^{\text{A}}[1,0,0] - m^2\,\Pbas^{\text{A}}[1,0,1]\Big] \, , \notag \\
  \mathbb{I}_3 = \tfrac{1}{2}\Big[-2\,\Pbas^{\text{B}}[0,1,1] + (c_{3,2} - c_{3,1} + 3)\,\Pbas^{\text{B}}[1,0,1] + (c_{3,1} - 3)\,\Pbas^{\text{B}}[1,1,0] + 4\,m^2\,\Pbas^{\text{B}}[1,1,1]\Big] \, .
\end{gather}
For topology $(4)$, the integrand carries a pole at $s_{12} = m^2$.
We separate $\mathbb{I}_4$ into a part containing this pole and a finite
remainder:
\begin{equation}
  \mathbb{I}_4 = \mathbb{I}_4^{\text{div}} + \mathbb{I}_4^{\text{fin}} \,,
\end{equation}
with
\begin{gather}
  \mathbb{I}_4^{\text{fin}} = \tfrac{1}{2}(c_{3,1} - c_{3,2} - 5)\,\Pbas^{\text{B}}[1,0,1] \, , \notag \\
  \mathbb{I}_4^{\text{div}} = \frac{1}{2(s_{12} - m^2)}\Big[(c_{3,2} - c_{3,1} - 1)\,\Pbas^{\text{B}}[0,0,1] + (c_{3,1} + c_{3,2} - 3)\,\Pbas^{\text{B}}[1,0,0] + 8\,m^2\,\Pbas^{\text{B}}[1,0,1]\Big] \, .
\end{gather}
Therefore, the finite reduced part of the integrand is
\begin{equation}
  \begin{gathered}
    \includegraphics[width=0.97\linewidth]{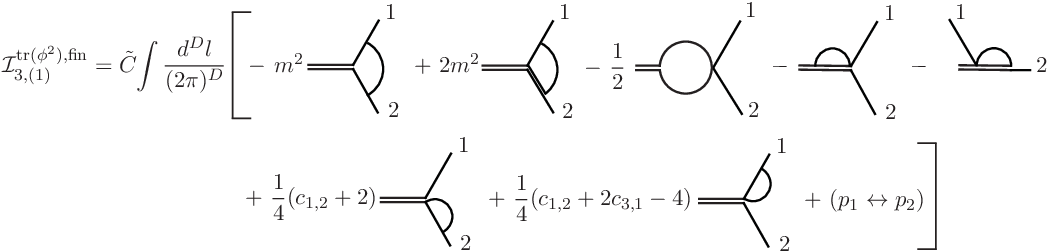}
  \end{gathered} \,.
  \label{eq:finite_integrand_res_1loop_H_phi_phi}
\end{equation}
Here $\Pbas$ is represented graphically as triangles and bubbles.
The first row contains integrands whose free parameters have already canceled, whereas the second row still carries residual parameters.
The reason is that the corresponding integrals have not been constrained by unitarity cuts: they are massless bubble integrals, which vanish after integration.
The final integrated result is therefore free of residual gauge parameters.

\subsubsection*{Comparison with Feynman diagrams}

The Feynman rule computation gives the numerators
\begin{gather}
  N_1^{\text{feyn}} =
    \ell^2
    + 2\,(\ell \cdot p_2)
    - 2\,(p_1 \cdot p_2) \, , \notag \\
  N_2^{\text{feyn}} =
    -2\,(\ell \cdot p_1)
    + 2\,(\ell \cdot p_2) \, , \notag \\
  N_3^{\text{feyn}} =
    \ell^2
    + 4\,(\ell \cdot p_1)
    + 2\,(\ell \cdot p_2)
    + 4\,(p_1 \cdot p_2) \, , \notag \\
  N_4^{\text{feyn}} =
    \ell^2
    + 4\,(\ell \cdot p_1)
    + 4\,(\ell \cdot p_2)
    + 8\,(p_1 \cdot p_2) \, .
\end{gather}
After reduction, we have
\begin{gather}
  \mathbb{I}_1^{\text{feyn}} =
    \Pbas^{\text{A}}[0,1,1]
    - \Pbas^{\text{A}}[1,0,1]
    + \Pbas^{\text{A}}[1,1,0]
    - 2\,m^2\,\Pbas^{\text{A}}[1,1,1] \, , \notag \\
  \mathbb{I}_2^{\text{feyn}} = \mathbb{I}_2 \, , \notag \\
  \mathbb{I}_3^{\text{feyn}} =
    -\Pbas^{\text{B}}[0,1,1]
    + \Pbas^{\text{B}}[1,0,1]
    + \Pbas^{\text{B}}[1,1,0]
    + 2\,m^2\,\Pbas^{\text{B}}[1,1,1] \, .
\end{gather}
For topology $(4)$, we again separate the pole part and the finite remainder,
\begin{equation}
  \mathbb{I}_4^{\text{feyn}}
  =
  \mathbb{I}_4^{\text{feyn},\text{div}}
  + \mathbb{I}_4^{\text{feyn},\text{fin}} \, ,
\end{equation}
with
\begin{gather}
  \mathbb{I}_4^{\text{feyn},\text{fin}} =
    -2\,\Pbas^{\text{B}}[1,0,1] \, , \notag \\
  \mathbb{I}_4^{\text{feyn},\text{div}} =
    \frac{1}{s_{12} - m^2}
    \Big[
      -\Pbas^{\text{B}}[0,0,1]
      + 2\,\Pbas^{\text{B}}[1,0,0]
      + 4\,m^2\,\Pbas^{\text{B}}[1,0,1]
    \Big] \, .
\end{gather}
Comparing with the CK result, $\mathbb{I}_1$, $\mathbb{I}_2$,
$\mathbb{I}_3$, and $\mathbb{I}_4^{\text{fin}}$ agree at
$c_{1,2} = 0$, $c_{3,1} = 5$, and $c_{3,2} = 4$, with $c_{1,1}$ remaining
arbitrary.
Therefore, $\mathbb{I}^{\text{feyn},\text{fin}}$ is a special representative of the CK-dual solution.
The remaining
difference appears only in $\mathbb{I}_4^{\text{div}}$, where the two
representations differ by a term proportional to $\Pbas^{\text{B}}[1,0,0]$,
which vanishes after integration.

\subsubsection*{Bleaching}

Following bleaching, one expects $\blcqBH$ to recover the FF integrand
$\ffTwo$ from the dyed integrand $\dampThree$,
\begin{equation}
  \blcqBH \Big( \dampThree \Big) = \ffTwo \, .
\end{equation}
At the loop level, however, bleaching must be performed
before the internal color indices are contracted.
Before the contraction of color factors, bleaching removes the color
factors of topologies $(3)$ and $(4)$. The surviving part then gives the
usual FF color factor after internal color contraction.
We denote by $\cint$ the operator that contracts the internal color indices.

To see why the order matters, consider the finite integrand result~(\ref{eq:finite_integrand_res_1loop_H_phi_phi}), which is already expressed after applying $\cint$.
Applying $\blcqBH$ directly to this result produces integrals with massive internal propagators, inconsistent with the structure of $\ffTwo$.
The reason is that bleaching modifies the color factor at each vertex before $\cint$ acts, so the contracted result changes.

Consider topology $(3)$ as an example.
If we contract first,
\begin{equation}
  \blcqBH \circ \cint (C_3) = \blcqBH(\tilde{C}) = N_c \delta^{a_1a_2}\,.
\end{equation}
If, instead, $\blcqBH$ is applied first, the color factor would vanish,
\begin{equation}
  \cint \circ \blcqBH(C_3) = \cint \circ \blcqBH\Big(\tf^{q b_1 b_3} \, \tf^{b_2 b_1 a_1} \, \td^{b_3 b_2 a_2}\Big) = \cint(0) = 0 \, .
\end{equation}
In summary, $\blcqBH$ and $\cint$ do not commute at loop level.
To obtain the correct FF integrand, bleaching must be performed before contracting the internal color indices.

\subsubsection*{Double copy}

As in the tree-level cases, replacing $C_i \rightarrow N_i$ in
(\ref{eq:loop_decomp}) gives a gravity integrand,
\begin{equation}
  \gIthree  =
  \sum_{\sigma_2}  \sum_{i=1}^{4}
  \int \frac{d^D \ell}{(2\pi)^D}
  \frac{1}{S_i} \frac{N_i^2}{D_i} \, .
  \label{eq:loop_dc_gravity}
\end{equation}
We comment on some subtleties here. At the gauge level, possible massive-tadpole data are
confined to $\mathbb{I}_4^{\text{div}}$. After the double copy, cross
terms between $\mathbb{I}_4^{\text{div}}$ and
$\mathbb{I}_4^{\text{fin}}$ can feed such data into the finite
gravity-side integrand. Since the
corresponding single cut was not included, these coefficients are not
fixed here.
In general, using Feynman rules, or via single-cut and loop-tree duality methods may be used to determine such tadpole
coefficients~\cite{Catani:2008xa,NigelGlover:2008ur,Britto:2009wz,Britto:2010um}.
We leave this to future work.


\providecommand{\href}[2]{#2}\begingroup\raggedright\endgroup
 
\end{document}